\documentclass[a4paper,11pt]{article}
\usepackage{jcappub} 
\usepackage{lineno}
\usepackage{subcaption}
\usepackage{booktabs}
\usepackage{bm}

\title{\boldmath Safe Phantom Divide Crossing from Unscreened Non-Minimal Coupling to Gravity}

\author{Joel Argudo-Panes,}
\author{Alex Gonz\'alez-Fuentes}
\author{and Adri\`a G\'omez-Valent}
\affiliation{Departament de F\'isica Qu\`antica i Astrof\'isica (FQA) and Institut
de Ciències del Cosmos (ICCUB), Universitat de Barcelona (UB), c. Martí i Franqués 1, 08028 Barcelona, Catalonia, Spain}

\emailAdd{jargudpa7@alumnes.ub.edu}
\emailAdd{agonzalezfuentes@icc.ub.edu}
\emailAdd{agomezvalent@icc.ub.edu}

\abstract{Recent hints of dynamical dark energy, including a possible crossing of the phantom divide inferred from the latest cosmological data, have motivated the search for viable theoretical explanations. Scalar fields non-minimally coupled to gravity with a coupling of the form $F(\varphi)=1+\alpha\varphi^2$ are physically motivated and have emerged as promising candidates, providing a significantly improved fit to the data compared to $\Lambda$CDM and even outperforming the $w_0w_a$CDM parametrization. However, in the absence of a screening mechanism, the regions of parameter space examined exhaustively so far in the literature (with $\alpha<0$) violate stringent local constraints on the strength and time variation of gravity and enhance the matter growth rate at low redshifts, in tension with galaxy clustering observations, thereby compromising the actual viability of these solutions. In this work, we investigate a largely unexplored region of parameter space (with $\alpha>0$) that realizes a crossing of the phantom divide while remaining consistent with Cassini–Huygens bounds and the most recent (and tightest) Lunar Laser Ranging measurements, without introducing higher-derivative terms in the action. This setup, which admittedly is subject to some degree of fine-tuning, fully respects the BBN constraints and leads only to mild deviations of the effective gravitational coupling in the past, reaching an extremum in the late universe that differs from Newton's constant by no more than approximately $0.3\%$. We review the background and linear perturbation dynamics of the model, deriving illustrative analytical results that provide physical insight into its behavior, and update the constraints on the model using the latest cosmological data, including the DES-Dovekie supernova sample. We perform for the first time a full-fledged Monte Carlo analysis with $\alpha>0$, studying also the impact of the variation of $G$ on the supernovae absolute luminosity, and compare the results with those obtained for $\alpha<0$, considering in both cases a linear potential. In the context of the model with non-minimal coupling and $\alpha<0$, the standard model is excluded at the $2.50\sigma$ CL, whereas for $\alpha>0$, the exclusion significance decreases to $\sim 2\sigma$ CL, indicating a moderate preference for these models over $\Lambda$CDM.}  

\begin{document}
\maketitle
\flushbottom

\section{Introduction}
\label{sec:intro}

Current cosmological data from the cosmic microwave background (CMB), baryon acoustic oscillations (BAO), and Type Ia supernovae (SNIa) favor the presence of an evolving effective dark energy (DE) component in the Universe, exhibiting a crossing of the phantom divide at $z \sim 0.5$ and with the DE density growing with the expansion prior to that redshift \cite{DESI:2025zgx}. These hints have been corroborated in both model-dependent analyses \cite{Chakraborty:2024xas,Gomez-Valent:2024tdb,Benisty:2024lmj,Poulin:2024ken,Ye:2024ywg,Wolf:2024eph,Wolf:2024stt,Park:2024vrw,Gomez-Valent:2024ejh,Odintsov:2024woi,Tiwari:2024gzo,Giare:2025pzu,Keeley:2025stf,Khoury:2025txd,Pan:2025psn,Lu:2025gki,Wolf:2025jed,Chaussidon:2025npr,Chakraborty:2025syu,Yang:2025mws,Chen:2025wwn,Giani:2025hhs,Cai:2025mas,Braglia:2025gdo,Ozulker:2025ehg,Poulin:2025nfb,Camarena:2025upt,Gomez-Valent:2025mfl,Wang:2025znm,Yang:2025uyv,Yao:2025wlx,Nojiri:2025low,Artola:2025zzb,Mishra:2025goj,Goh:2025upc,Tsujikawa:2025wca,Wolf:2025acj,Adi:2025hyj,Alestas:2025syk,Sharma:2025iux,Efstratiou:2025iqi,Cheng:2025yue,Ghedini:2025epp,Hossain:2025gpr,deCruzPerez:2025dni,Li:2026xaz,Ibarra-Uriondo:2026zbp,Akarsu:2026anp,Park:2026iqa,Jhaveri:2026bla,Wang:2026wrk,Gomez-Valent:2026ept,Toda:2026yum,Giare:2026oti,SolaPeracaula:2026lyk} and in more model-agnostic studies, where the background expansion history and the DE properties are directly reconstructed from the data \cite{DESI:2024aqx,Jiang:2024xnu,DESI:2025fii,Berti:2025phi,Li:2025ops,Gonzalez-Fuentes:2025lei,Gonzalez-Fuentes:2026rgu,Paliathanasis:2026dqk,Wang:2026ske,Schoneberg:2026buf}. The $\Lambda$CDM model is excluded at $\sim 3\sigma$ CL, and the probability of phantom-divide crossing is about $96.7\%$--$98.5\%$ \cite{Gonzalez-Fuentes:2026rgu}. See also Ref.~\cite{Keeley:2025rlg}.

From a theoretical perspective, such a crossing might require the existence of interactions in the dark sector or may even be a smoking gun of modified gravity. In this work, we concentrate on the second possibility, focusing on a subclass of Horndeski scalar-tensor theories \cite{Horndeski:1974wa} containing a non-minimal coupling to gravity $F(\varphi)$ and a potential $V(\varphi)$ for the dark energy scalar field $\varphi$ that drives the late-time cosmic acceleration. We consider the action\footnote{In this paper, we work in natural units by setting $c=1$ and $G_N=1/m_{\rm Pl}^2$, where $m_{\rm Pl}=1.22\times10^{19}\,\mathrm{GeV}$ denotes the Planck mass. Throughout, we use the metric signature $(-,+,+,+)$. We define the curvature tensors according to
$R^{\lambda}{}_{\mu\nu\sigma}=\partial_{\nu}\Gamma^{\lambda}_{\mu\sigma}+\Gamma^{\rho}_{\mu\sigma}\Gamma^{\lambda}_{\rho\nu}-(\nu\leftrightarrow\sigma)$ and
$R_{\mu\nu}=R^{\lambda}{}_{\mu\lambda\nu}$, and the Ricci scalar
$R=g^{\mu\nu}R_{\mu\nu}$. These definitions coincide with the $(+,+,+)$ convention in the sign-classification scheme of \cite{Misner:1974qy}.}
\begin{equation}\label{eq:action}
S = \int d^4x \sqrt{-g} \left[\frac{F(\varphi)}{2\kappa^2} R - \frac{1}{2} \partial_\mu \varphi \, \partial^\mu \varphi - V(\varphi) + \mathcal{L}_M \right] \,,
\end{equation}
with $\mathcal{L}_M$ the Lagrangian density for the matter sector, which includes the contribution of the standard model of particle physics and the required extensions to account for massive neutrinos and cold dark matter (CDM). 

Although the late-time forms of $F(\varphi)$ and $V(\varphi)$ can, in principle, be reconstructed from constraints on specific DE parametrizations \cite{Boisseau:2000pr,Gannouji:2006jm} (see also the more recent works \cite{Efstratiou:2025iqi,Pan:2025psn}, and \cite{Ye:2026yqk} for a reconstruction of non-minimally coupled gravity within the effective field theory of DE), in this work we consider the simple ansatz
\begin{equation}\label{eq:main_functions}
F(\varphi) = 1 + \alpha \varphi^2 \quad \text{and} \quad V(\varphi) = V_0 + \beta \varphi \,,
\end{equation}
assuming that Taylor expansions are valid due to the small field excursion throughout cosmic history, i.e., we neglect higher-order terms in $\varphi$, which could, in general, be present in the effective field theory. We do not include a linear term in $F(\varphi)$, since it can always be removed by a trivial constant shift in $\varphi$, and we set $\kappa = (8\pi G_N)^{1/2}$, where $G_N$ is Newton's constant\footnote{The parameter $\kappa$ could equally well be treated as a free parameter of the theory and included in the Monte Carlo analysis. For simplicity, however, we fix it to the value mentioned in the main text.}. For $\alpha=0$ (or, equivalently, $F=1$) one retrieves the minimally coupled scenario.

A non-minimal coupling to gravity of the form adopted in Eq.~\eqref{eq:main_functions} is required for a consistent renormalization of a scalar field theory in curved spacetime \cite{BirrellDavies1982}, so it is natural to analyze its cosmological implications in the context of DE. The action in Eq.~\eqref{eq:action}, with this particular choice of $F(\varphi)$, has been extensively studied since the discovery of cosmic acceleration (see, e.g., \cite{Chiba:1999wt,Perrotta:1999am,Riazuelo:2001mg,Perrotta:2002sw}), and it has regained interest in light of the latest observations. In \cite{Ye:2024ywg,Wolf:2024stt}, it was shown that this model can produce a crossing of the phantom divide such as the one preferred by the data when considering negative values of the non-minimal coupling (i.e., $\alpha < 0$) not only in the metric formalism, but also in the Palatini one \cite{SanchezLopez:2025uzw}. For sufficiently large absolute values of $\alpha$, reasonable initial conditions for the scalar field and regardless of the sign of the slope of the potential, this configuration gives rise to the so-called thawing gravity behavior, in which gravity remains close to standard during a significant fraction of cosmic history, until the moment when the potential starts to compete with non-relativistic matter at low redshifts. This enables a transition from an effective phantom phase to quintessence, in very good agreement with CMB+BAO+SNIa data.

However, as pointed out in \cite{Wolf:2025jed}, thawing gravity inevitably enhances the growth of large-scale structure below $z \sim 1.5$ for parameter values that produce the required crossing, leading to an effective gravitational coupling at cosmological scales that is $\sim 40$--$100\%$ larger than in General Relativity (GR) at present, in tension with low-redshift data from redshift-space distortions; see, e.g., \cite{Toda:2024fgv,Toda:2026yum}. Moreover, in the absence of a screening mechanism -- which is not present in the action \eqref{eq:action} -- the region of parameter space with $\alpha < 0$ that produces the crossing completely spoils the agreement with local gravity constraints. Similar problems are also encountered in scalar-tensor theories with a minimal coupling to gravity, a non-canonical kinetic term including higher-derivative contributions, and a broken shift symmetry through a field-dependent potential, despite their ability to produce the crossing \cite{Tsujikawa:2025wca,Wolf:2025acj} (see also \cite{Calderon:2026hbr})\footnote{If, instead, the shift symmetry is broken by means of field-dependent terms in the kinetic sector \cite{Naidoo:2026umv,Hallam:2026qsk} or if a pure momentum
transfer interaction between the scalar and CDM is also included \cite{Pookkillath:2026wyg}, structure growth at linear scales remains close to that predicted in the standard model.}. This raises concerns about the viability of these theories or, at the very least, points to the need to include additional terms in the action to keep these effects under control. 

In this paper, we demonstrate that the action~\eqref{eq:action}, with the coupling and potential given in Eq.~\eqref{eq:main_functions}, can not only give rise to a crossing of the phantom divide but also fully satisfy stringent local constraints in the region of parameter space with $\alpha > 0$. This possibility, which requires some degree of fine-tuning, was already pointed out in \cite{Adam:2025kve} and was recently further discussed in \cite{Adam:2026ajg}\footnote{This work was posted on arXiv during the last stages of the preparation of our manuscript.}. However, those authors neither performed a comprehensive Monte Carlo exploration of this region of parameter space, nor considered the effect of the variation of $G$ on SNIa intrinsic luminosities \cite{Amendola:1999vu,Garcia-Berro:1999cwy,Wright:2017rsu,Desmond:2019ygn,Efstratiou:2025iqi}, and they based their analysis on significantly weaker local constraints than those adopted here. Furthermore, \cite{Adam:2025kve} argued that the potential in Eq.~\eqref{eq:main_functions} should be modified to ensure that DE remains subdominant at early times to not spoil the correct description of key processes such as recombination or phenomena occurring at even higher redshifts. As we show here, however, this modification is not needed as long as the decaying mode for $\varphi$ in the radiation-dominated era is already negligible, which might be a natural assumption. Meanwhile, in \cite{Adam:2026ajg}, the authors modified the non-minimal coupling $F(\varphi)$ to freeze the scalar field at the minimum of the effective potential during matter domination, thereby preventing the development of background instabilities. By contrast, we show that these instabilities are harmless provided the scalar field remains sufficiently close to the origin after matter--radiation equality, a scenario that may arise in the context of Higgs-type symmetry-breaking mechanisms.

In addition to analyzing the positive-$\alpha$ scenario, we update previous constraints on the model in the $\alpha<0$ region using the re-calibrated SNIa sample from the Dark Energy Survey (DES), known as the DES-Dovekie sample \cite{DES:2025sig}. This updated sample addresses important issues affecting the previous DES-Y5 sample \cite{DES:2024hip,DES:2024jxu}, which was used in earlier analyses \cite{Wolf:2025jed,SanchezLopez:2025uzw}. We compare the model dynamics in both regions of parameter space and assess their respective fitting performance using the most recent cosmological data. Specifically, we combine the DES-Dovekie SNIa sample with CMB data from Planck PR4 \cite{Planck:2018vyg,Efstathiou:2019mdh,Rosenberg:2022sdy} and BAO measurements from the second data release of the Dark Energy Spectroscopic Instrument (DESI) \cite{DESI:2025zgx}. 

Before closing the Introduction, we note that the authors of \cite{Garcia-Garcia:2026nzy} recently presented an exhaustive study comparing the non-minimal coupling in the $\alpha<0$ scenario with other single scalar field models of DE (namely, standard quintessence and massive Galileon). Although their work updated the constraints using DES-Dovekie, as we do here, we note that their setup is slightly different from ours. On the one hand, they include a mass term to the scalar field potential, which we discard as the main desired features of this model are induced by the linear term in the potential, $\beta$. In fact, their contours show that $m^2=0$ at $1-2\sigma$ CL (cf. Fig. 19 of \cite{Garcia-Garcia:2026nzy}). Moreover, our CMB likelihood choice is slightly different, since they use \texttt{plik} {\it Planck} PR3 CMB power
spectra and {\it Planck+ACT} CMB lensing as their baseline. Nevertheless, apart from the aforementioned changes to the data set, we focus on the $\alpha>0$ case and compare it with the $\alpha<0$ case for completeness.

This paper is organized as follows. In Sec.~\ref{sec:model}, we first present the field equations of the model~\eqref{eq:action} and discuss the existing local constraints, together with the theoretical expressions required to apply them. We then examine the cosmological background evolution and derive approximate analytical solutions for positive and negative values of $\alpha$. These solutions help justify our choice of initial conditions and clarify the conditions under which phantom crossing can occur. We consider that some of these aspects have not been explained in sufficient detail in previous works. We also present the basic equations governing the evolution of matter perturbations. In Sec.~\ref{sec:method}, we describe the methodology used to solve the Einstein--Boltzmann system of equations and the strategy adopted to scan and constrain the parameter space of the theory. We also summarize the data sets employed in the analysis. Sec.~\ref{sec:results} is devoted to the discussion of the results, while Sec.~\ref{sec:conclusions} presents our conclusions. Two appendices complement the information contained in the main text.


\section{The model}\label{sec:model}

\subsection{Field equations}

The modified Einstein equations are obtained from the variation of the action \eqref{eq:action} with respect to the metric tensor. They read:
\begin{equation}\label{eq:EinsEq}
F(\varphi)G_{\mu\nu}-\nabla_\nu\nabla_\mu F(\varphi)+ g_{\mu\nu}\Box F(\varphi)=\kappa^2 (T_{\mu\nu}^{(M)}+T^{(\phi)}_{\mu\nu})\,,
\end{equation}
with $G_{\mu\nu}=R_{\mu\nu}-g_{\mu\nu}R/2$ the Einstein tensor, $T_{\mu\nu}^{(M)}$ the energy-momentum tensor of matter and 

\begin{equation}
T^{(\phi)}_{\mu\nu}=\partial_\mu\varphi\partial_\nu\varphi-g_{\mu\nu}\left(\frac{1}{2}\partial_\theta\varphi\partial^\theta\varphi+V(\varphi)\right)
\end{equation}
the energy-momentum tensor of the scalar field, which we deliberately write in the same form as in uncoupled scalar field models in order to leave the new terms arising from the non-minimal coupling on the left-hand side of Eq. \eqref{eq:EinsEq}, treating them as being of pure gravitational nature.  

If we vary, instead, the action \eqref{eq:action} with respect to the scalar field, we are led to the modified Klein-Gordon (KG) equation:

\begin{equation}\label{eq:KG}
\Box\varphi-V^{\prime}(\varphi)+\frac{F^\prime(\varphi)}{2\kappa^2}R=0\,,
\end{equation}
with the prime denoting a derivative with respect to $\varphi$. The dynamics of the scalar field is controlled not only by its potential but also by the curvature of spacetime through the non-minimal coupling. It is therefore convenient to define the effective potential\footnote{We emphasize that this effective potential is introduced only at the level of the KG equation. Field-independent shifts in the numerator of the second term leave the scalar-field dynamics unchanged. We choose a shift such that $F(\varphi) R/(2\kappa^2)-V(\varphi)=R/(2\kappa^2)-V_{\rm eff}(\varphi,R)$ in the action \eqref{eq:action}.}

\begin{equation}\label{eq:eff_pot}
V_{\rm eff}(\varphi,R)\equiv V(\varphi)-\frac{(F(\varphi)-1)}{2\kappa^2}R\,,
\end{equation}
such that Eq. \eqref{eq:KG} can be written in a form closer to the  KG equation found in uncoupled scenarios:
\begin{equation}\label{eq:KGcompact}
\Box\varphi-V_{\rm eff}^{\prime}(\varphi,R)=0\,.
\end{equation}
The non-minimal coupling can modify the geometry of spacetime relative to GR both by affecting the dynamics of the scalar field and by shifting the value of $F(\varphi)$ away from unity. As we will see below, such shifts may exist even in regimes where the scalar-field dynamics is strongly suppressed. In these cases, the gravitational coupling entering the modified Einstein equations is simply shifted from $\kappa^2/(8\pi)$ to $\kappa^2/(8\pi F)$. When curvature induces dynamics to $\varphi$, additional changes appear due to the non-trivial scalar-field kinetic structure of Eq. \eqref{eq:EinsEq}.

Finally, before closing this section, we briefly discuss the connection between the model~\eqref{eq:action} and the original Brans--Dicke (BD) theory \cite{BransDicke1961,Brans1962,Dicke1962},
\begin{equation}
S_{\rm BD}=\int d^4x\sqrt{-g}\left[\frac{1}{16\pi}\left(R\psi-\frac{\omega_{\rm BD}}{\psi}\partial^\mu\psi\partial_\mu\psi\right)+\mathcal{L}_m\right]\,,
\end{equation}
which features a non-canonical kinetic term. The Brans--Dicke field $\psi$ has dimensions of mass squared in natural units and $\omega_{\rm BD}$ is the (dimensionless) Brans--Dicke parameter. The BD action can be rewritten in terms of a canonical kinetic term -- as in Eq.~\eqref{eq:action} -- by means of the field redefinition
\begin{equation}
\psi = \frac{2\pi}{\omega_{\rm BD}}\varphi^2\,.
\end{equation}
The non-minimal coupling in the transformed Brans--Dicke action then takes the form $\tilde{F}(\varphi)=\kappa^2\varphi^2/(4\omega_{\rm BD})$. Therefore, the model~\eqref{eq:action} is equivalent to Brans--Dicke theory only in the limit $|\alpha|\varphi^2\gg1$, with $\alpha/\kappa^2=1/(4\omega_{\rm BD})$. An analogous result is obtained by instead performing a field transformation such that the non-minimal coupling in Eq.~\eqref{eq:main_functions} takes the Brans--Dicke form. This transformation is
\begin{equation}
\psi=\frac{8\pi}{\kappa^2}\left(1+\alpha\varphi^2\right)\,,
\end{equation}
which leads to the following Brans--Dicke function in the transformed action~\eqref{eq:action}:

\begin{equation}
\omega_{\rm BD}(\psi)=\frac{\kappa^4\psi}{4\alpha(\kappa^2\psi-8\pi)}\,.
\end{equation}
Again, the two models become equivalent in the limit $|\alpha|\varphi^2\gg1$, with $\alpha/\kappa^2=1/(4\omega_{\rm BD})$. However, this limit is incompatible with current observational constraints, implying that the two theories coincide only in a phenomenologically excluded region of parameter space. In the region where the model~\eqref{eq:action} provides a viable description of the data, it is not equivalent to the original Brans--Dicke theory and therefore warrants an independent study, even in the absence of a scalar potential. For recent in-depth cosmological analyses of Brans--Dicke theory with a cosmological constant, we refer the reader to Refs.~\cite{Avilez:2013dxa,SolaPeracaula:2019zsl,SolaPeracaula:2020vpg,Joudaki:2020shz}. See also \cite{Euclid:2025vml} for forecasts based on Euclid data.


\subsection{Effective gravitational coupling and local constraints}\label{sec:Geff}

The model \eqref{eq:action} does not include higher-order derivative terms to screen the modified gravity effects at local scales as is done in Vainshtein \cite{Vainshtein:1972sx} or K-mouflage mechanisms \cite{Babichev:2009ee} through the introduction of $\Box\varphi\partial_\mu\varphi\partial^\mu\varphi$ or $(\partial_\mu\varphi\partial^\mu\varphi)^2$ terms in the action, respectively. Therefore, the dynamics of the scalar field affect physics on both cosmological and local scales. The field evolves throughout cosmic history, modifying the gravitational strength that governs the background expansion (see Sec. \ref{sec:background} for details), while also inducing a distinction between the latter and the coupling experienced by two test masses, $G_{\rm eff}$. It is therefore of utmost importance to have a theoretical control of $G_{\rm eff}$, since it is subject to very tight local constraints, which translate into very strong bounds on the model parameters. 

In this section, we compute the effective gravitational coupling that is measured locally on Earth and in the Solar System. Although this is a very well-known result, we deem it useful to provide the details of the calculation here for completeness and also to emphasize the presence of terms that have been incorrectly neglected in recent works, as in \cite{Adam:2025kve,Adam:2026ajg}.

In these environments, the gravitational potential is sufficiently small that deviations from the Minkowski metric can be treated as small perturbations, i.e.,
\begin{equation}\label{eq:pert}
g_{\mu\nu}=\eta_{\mu\nu}+h_{\mu\nu}\,, \qquad |h_{\mu\nu}|\ll1.
\end{equation}
Our first step is to solve the coupled system of the modified Einstein and KG equations, Eqs.~\eqref{eq:EinsEq} and \eqref{eq:KG}, in vacuum around a spherical, massive, pressureless object, as the Earth is to first approximation. We do so considering a static source and in the weak-field limit. At sufficiently large distances from the source, the scalar field takes its cosmological value, $\varphi_c$, which varies with the cosmic expansion -- we compute it in Sec. \ref{sec:background}. Due to the presence of the source, though, the scalar field receives some correction  around the massive object,  $\delta\varphi$, \begin{equation}
\varphi=\varphi_c+\delta\varphi\,,\qquad |\delta\varphi|\ll|\varphi_c|\,,
\end{equation}
which is also small due to the weak-field assumption. 

Using the first-order expansions of the geometrical quantities in $h_{\mu\nu}$, given in Appendix~\ref{sec:geom}, Eqs.~\eqref{eq:EinsEq} and \eqref{eq:KG} can be written, respectively, in the perturbed form,

\begin{equation}\label{eq:mix1}
(1+\alpha\varphi_c^2)G_{\mu\nu}(h)+2\alpha\varphi_c\left[\eta_{\mu\nu}\partial^{\alpha}\partial_\alpha\delta\varphi-\partial_\mu\partial_\nu\delta\varphi\right]=\kappa^2 T_{\mu\nu}^{(M)}\,,
\end{equation}

\begin{equation}\label{eq:mix2}
\partial^\mu\partial_\mu\delta\varphi+\frac{\alpha\varphi_c}{\kappa^2}\left(\partial^\mu\partial^\nu h_{\mu\nu}-\partial^\mu\partial_\mu h\right)=0\,.
\end{equation}
Note that we have neglected the scalar field potential, which is responsible for cosmic acceleration in the late Universe. At local scales, its effects are negligible, in complete analogy with the cosmological constant in GR. We also neglect the cosmic time variation of $\varphi_c$, so by symmetry $h_{\mu\nu}$ and $\delta\varphi$ can only depend on the radial coordinate.

Eqs.~\eqref{eq:mix1}--\eqref{eq:mix2} are coupled in the variables $h_{\mu\nu}$ and $\delta\varphi$. Fortunately, the field redefinition 
\begin{equation}
h_{\mu\nu}=H_{\mu\nu}-\frac{2\alpha\varphi_c\,\delta\varphi}{1+\alpha\varphi_c^2}\,\eta_{\mu\nu}
\end{equation}
removes $\delta\varphi$ from the Einstein equations, leaving

\begin{equation}\label{eq:uncoupledEins}
G_{\mu\nu}(H)=\frac{\kappa^2T_{\mu\nu}^{(M)}}{1+\alpha\varphi_c^2}\,,
\end{equation}
with $T_{\mu\nu}^{(M)}=\rho u_\mu u_\nu$, $u^\mu=(1,\vec{0})$ the 4-velocity of the (static) source and $\rho$ its energy density. Therefore, we can solve for $H_{\mu\nu}$ without knowing $\delta\varphi$. The trace and 00 component of Eq. \eqref{eq:uncoupledEins} read, respectively,

\begin{equation}
\delta^{ij}\nabla^2H_{ij}-\nabla^2H_{00}-\partial_i\partial_j H_{ij}=-\frac{\kappa^2\rho}{1+\alpha\varphi_c^2}\,,
\end{equation}

\begin{equation}
\partial_i\partial_jH_{ij}-\delta^{ij}\nabla^2 H_{ij}=\frac{2\kappa^2\rho}{1+\alpha\varphi_c^2}\,,
\end{equation}
where $\nabla^2=\delta^{ij}\partial_i\partial_j$. Summing the two equations yields
\begin{equation}
-\nabla^2 H_{00}=\frac{\kappa^2\rho}{1+\alpha\varphi_c^2}
\quad\Longrightarrow\quad
H_{00}(r)=\frac{\kappa^2}{1+\alpha\varphi_c^2}\frac{M}{4\pi r}\,.
\end{equation}
On the other hand, the $0i$ and $ij$ components of Eq. \eqref{eq:uncoupledEins} lead to $H_{0i}=0$ and $H_{ij}=H_{00}\delta_{ij}$. The resulting metric can be plugged into the scalar field equation \eqref{eq:mix2}, whose solution reads

\begin{equation}
\delta\varphi(r) = \frac{\frac{\alpha\varphi_c}{\kappa^2}H_{00}(r)}{1+\frac{6\alpha^2\varphi_c^2}{\kappa^2}\frac{1}{1+\alpha\varphi_c^2}}=\frac{\alpha\varphi_c}{1+\alpha\varphi_c^2+\frac{6\alpha^2\varphi_c^2}{\kappa^2}}\,\,\frac{M}{4\pi r}\,.
\end{equation}
Hence, coming back to $h_{\mu\nu}$ we find 
\begin{equation}\label{eq:h00}
h_{00}(r) =  \frac{\kappa^2}{1+\alpha\varphi_c^2}\frac{M}{4\pi r}\left[\frac{\kappa^2(1+\alpha\varphi_c^2)+8\alpha^2\varphi_c^2}{\kappa^2(1+\alpha\varphi_c^2)+6\alpha^2\varphi_c^2}\right]\,,
\end{equation}
\begin{equation}\label{eq:hij}
h_{ij}(r) = \frac{\kappa^2}{1+\alpha\varphi_c^2}\frac{M}{4\pi r}\left[\frac{\kappa^2(1+\alpha\varphi_c^2)+4\alpha^2\varphi_c^2}{\kappa^2(1+\alpha\varphi_c^2)+6\alpha^2\varphi_c^2}\right]\delta_{ij}\,.
\end{equation}
Now, we are ready to compute the effective gravitational coupling, $G_{\rm eff}$. We study the motion of a free-falling particle, which is subject to the geodesic equation. In the non-relativistic limit, we find 

\begin{equation}
\frac{d^2x^i}{dt^2}\approx\frac{h_{00,i}}{2} = -\frac{M}{8\pi r^2}\frac{\kappa^2 }{1+\bar{\alpha}\bar{\varphi}_c^2}\left[\frac{\kappa^2(1+\alpha\varphi_c^2)+8\alpha^2\varphi_c^2}{\kappa^2(1+\alpha\varphi_c^2)+6\alpha^2\varphi_c^2}\right]\hat{r}^i\equiv -\frac{G_{\rm eff}M}{r^2}\hat{r}^i\,.
\end{equation}
Hence, in terms of the dimensionless coupling constant and scalar field, defined as 

\begin{equation}\label{eq:dimensionless}
\bar{\alpha}\equiv \alpha/\kappa^2 \quad {\rm and}\quad  \bar{\varphi}\equiv \kappa\varphi\,,
\end{equation}
we have

\begin{equation}
\label{eq:Geff}
G_{\rm eff}(z)= \frac{\kappa^2 }{8\pi }\frac{1}{1+\bar{\alpha}\bar{\varphi}_c^2}{}\left[\frac{1+\bar{\alpha}\bar{\varphi}_c^2(1+8\bar{\alpha})}{1+\bar{\alpha}\bar{\varphi}_c^2(1+6\bar{\alpha})}\right]\,,
\end{equation}
where the redshift dependence is inherited from the cosmic evolution of $\bar{\varphi}_c$. Three observations are in order. First, Newton's gravitational constant, $G_N=\kappa^2/(8\pi)$, is recovered not only when the non-minimal coupling vanishes, i.e., in the GR limit, but also when $\bar{\varphi}_c=0$. This means that, even in the presence of a non-zero non-minimal coupling and in the absence of a screening mechanism, there can be stages in the cosmic evolution at which $G_{\rm eff}(z)=G_N$. Second, for $|\bar{\alpha}|\bar{\varphi}_c^2\ll 1$, Eq. \eqref{eq:Geff} can be approximated by

\begin{equation}\label{eq:GeffTaylor}
G_{\rm eff}(z)=1+\bar{\alpha}\bar{\varphi}_c^2(2\bar{\alpha}-1)+\mathcal{O}[(\bar{\alpha}\bar{\varphi}_c^2)^2]\,.
\end{equation}
This expression tells us that for $\bar{\alpha}=1/2$ the effective gravitational coupling only receives corrections of higher order and, therefore, deviations from $G_N$ can be kept small for larger values of $\bar{\varphi}_c$ than those required if $\bar{\alpha}\ne 1/2$. Third, $G_{\rm eff}(z)=\kappa^2/(8\pi F(\varphi_c))$ only if $|\bar{\alpha}|\ll 1$. For $|\bar{\alpha}|=\mathcal{O}(1)$, as preferred by current cosmological data, it receives corrections that must be duly taken into account, even if the product $\bar{\alpha}\bar{\varphi}_c^2$ is small in absolute value. According to Eq. \eqref{eq:GeffTaylor}, neglecting them would result in an error of $200\bar{\alpha}\times\bar{\alpha}\bar{\varphi}_c^2\,\%$ in the estimation of $G_{\rm eff}(z)$, which can be non-negligible.

Clearly, any modified gravity model must satisfy very stringent local constraints. We need to recover the gravitational strength measured on Earth and in the Solar System at present, i.e., $G_{\rm eff}(z=0)=G_N$, and fulfill the constraints on the parametrized post-Newtonian (PPN) parameters $\gamma^{\rm PPN}$ and $\beta^{\rm PPN}$ \cite{Will1993}  (also known as Eddington's parameters \cite{FujiiMaeda2003}), which were measured with exquisite precision by the Cassini \cite{Bertotti:2003rm} and MESSENGER \cite{Park:2017zgd} spacecraft, respectively. These constraints read\footnote{$\gamma^{\rm PN}$ can be trivially obtained from Eq. \eqref{eq:hij} upon imposing $G_{\rm eff}(z=0)=G_N$ and using the definition $g_{ij}=\left[1+\gamma^{\rm PPN}r_s/r+\mathcal{O}(r_s^2/r^2)\right]\delta_{ij}$ with $r_s=2GM$ the Schwarzschild radius. $\beta^{\rm PPN}$ can be obtained in an analogous way from Eq. \eqref{eq:h00}, but now using $g_{00}=-1+r_s/r-\beta^{\rm PPN}r_s^2/(2r^2)+\mathcal{O}(r_s^3/r^3)$. However, this would require going to second order in perturbations, which we avoid here. For more details, we refer the reader to \cite{Will1993,FujiiMaeda2003}. See also \cite{Karam:2026sqg}.} 

\begin{equation}\label{eq:gammaPN}
\gamma^{\rm PPN}-1 = \frac{-4\bar{\alpha}^2\bar{\varphi}_{c,0}^2}{1+\bar{\alpha}\bar{\varphi}_{c,0}^2+8\bar{\alpha}^2\bar{\varphi}_{c,0}^2}=(2.1\pm 2.3)\cdot 10^{-5}\,\, (68\%\,{\rm CL})\,,
\end{equation}

\begin{equation}\label{eq:betaPN}
\beta^{\rm PPN}-1 =\frac{1}{4}\frac{\bar{\alpha}\bar{\varphi}_{c,0}(1+\bar{\alpha}\bar{\varphi}_{c,0}^2)}{1+\bar{\alpha}\bar{\varphi}_{c,0}^2+6\bar{\alpha}^2\bar{\varphi}_{c,0}^2}\frac{d\gamma^{\rm PPN}}{d\bar{\varphi}_{c,0}}=(-2.7\pm 3.9)\cdot 10^{-5}\,\, (68\%\,{\rm CL})\,,
\end{equation}
with $\bar{\varphi}_{c,0}\equiv \bar{\varphi}(z=0)$. Since GR is recovered when $\gamma^{\rm PPN}=\beta^{\rm PPN}=1$, departures of these parameters from unity signal deviations from GR. Lastly, the relative time variation of the gravitational constant, which can be inferred from Eq. \eqref{eq:Geff},

\begin{equation}
\frac{\dot{G}_{\text{eff}}}{G_{\text{eff}}} =-2\bar{\alpha}\bar{\varphi}_c\dot{\bar{\varphi}}_c\left[\frac{1}{1+\bar{\alpha}\bar{\varphi}_c^2}-\frac{1+8\bar{\alpha}}{1+\left(1+8\bar{\alpha}\right)\bar{\alpha}\bar{\varphi}_c^2}+\frac{1+6\bar{\alpha}}{1+\left(1+6\bar{\alpha}\right)\bar{\alpha}\bar{\varphi}_c^2}\right]\,,
\end{equation}
is constrained by Lunar Laser Ranging (LLR) measurements \cite{Biskupek:2020fem},

\begin{equation}\label{eq:Gdot}
\left. \frac{\dot{G}_{\rm{eff}}}{G_{\rm{eff}}} \right|_{z = 0} = (-5.0\pm9.6)\cdot 10^{-15}\; \mathrm{yr}^{-1} \, (68\%\,{\rm CL})\,.
\end{equation}
This is the most stringent measurement available to date. For other (weaker) constraints on $\dot{G}_{\rm eff}/G_{\rm eff}$, we refer the reader to Table 20 of the review \cite{Uzan:2024ded}. 

Since the non-minimally coupled model is recovered for $\bar{\alpha}=0$, in which no phantom-divide crossing can occur, such a crossing is only possible if this parameter departs from zero and, indeed, takes values of order unity, as will be shown in Sec.~\ref{sec:background}. We simply note this here to illustrate that, in order to satisfy the constraints~\eqref{eq:gammaPN}, \eqref{eq:betaPN}, and~\eqref{eq:Gdot} while allowing an effective phantom-divide crossing, $\bar{\varphi}_{c,0}$ must be sufficiently close to zero, typically below $10^{-3}$.  As will be shown below, this is only possible for $\bar{\alpha}>0$, since negative values of $\bar{\alpha}$ necessarily lead to a violation of the local constraints. Achieving the desired phenomenology requires therefore positive values of the non-minimal coupling and the fine-tuning of the initial value of the scalar field, so that $\bar{\varphi}(z=0)\approx0$. Note, however, that the degree of fine-tuning is alleviated when $\bar\alpha \approx 1/2$. Not only does this value of $\bar{\alpha}$ suppress deviations of $G_{\rm eff}$ from $G_N$ in the $\alpha\varphi^2\ll1$ limit, but also that of $\dot{G}_{\rm eff}/G_{\rm eff}$. Nevertheless, no suppression is observed in the deviations of the PPN parameters from GR for $\bar\alpha \approx 1/2$. Thus, even if $\alpha$ is fixed to this value in the model, one cannot completely avoid some degree of fine-tuning if both cosmological and local constraints are to be satisfied without a screening mechanism. Interestingly, our posterior mean value of $\bar{\alpha}$ lies close to $1/2$. Further discussion of this point can be found in Sec. \ref{sec:results}.


\subsection{Background dynamics in the expanding Universe}\label{sec:background}

We want to study the cosmological implications of the non-minimal coupling. As usual, we assume the symmetries embedded in the Cosmological Principle, so we use the Friedmann-Lema\^itre-Robertson-Walker (FLRW) metric, considering flat spatial hypersurfaces (i.e., no spatial curvature). The line element written in spherical coordinates is

\begin{equation}\label{eq:line_element}
ds^2=-dt^2+a^2(t)\left(dr^2+r^2d\Omega^2\right)\,.
\end{equation}
The modified Friedmann and pressure equations are derived from the Einstein field equations \eqref{eq:EinsEq} upon substitution of the metric \eqref{eq:line_element} and the non-minimal coupling specified in Eq. \eqref{eq:main_functions}. They read\footnote{From this point onward, we omit the subscript $c$ from the scalar field, which was introduced in Sec.~\ref{sec:Geff} to distinguish the cosmological scalar field from its local value.}, respectively,

\begin{equation}\label{eq:Friedmann_eq}
3H^2=\frac{\kappa^2(\rho+\frac{\dot{\varphi}^2}{2}+V)-6\alpha H\varphi\dot{\varphi}}{1+\alpha\varphi^2}\,,
\end{equation}

\begin{equation}\label{eq:pressure_eq}
-(2\dot{H}+3H^2)=\frac{\kappa^2(p+\frac{\dot{\varphi}^2}{2}-V)+2\alpha[\dot{\varphi}^2+\varphi(\ddot{\varphi}+2H\dot{\varphi})]}{1+\alpha\varphi^2}\,,
\end{equation}
where the dots denote derivatives with respect to the cosmic time $t$. Here $p$ and $\rho$ refer to the energy density and pressure of all the species included in the energy-momentum tensor $T_{\mu\nu}^{(M)}$, which can be treated as perfect fluids at the background level. They are covariantly self-conserved, since $\varphi$ does not couple directly to the matter sector, but only through gravity. Hence, 
\begin{equation}\label{eq:cons}
\nabla^\mu T_{\mu\nu}^{(M)}=0\quad\Longrightarrow\quad\dot{\rho}+3H(p+\rho)=0\,,
\end{equation}
and the evolution of the energy density $\rho_i$ and pressure $p_i$ of each cosmic fluid component $i$ is governed by the same dependence on the scale factor (or, equivalently, the redshift) as in the standard cosmological model. This also applies to neutrinos, for which we assume the baseline normal-ordering mass hierarchy, with a single massive neutrino of mass 0.06 eV. For species that remain either relativistic or non-relativistic throughout the cosmic eras considered here, the energy density scales as $\rho_i\propto a^{-3(1+w_i)}$, where $w_i=p_i/\rho_i$ is the corresponding constant equation-of-state (EoS) parameter. In particular, for baryons and CDM, $w_b=w_{\rm cdm}=w_m=0$, where the subscript $m$ denotes the combination of the former, while for photons $w_\gamma=1/3$.

\begin{figure}[t!]
    \centering
\begin{subfigure}{0.45\textwidth}
    \centering
    \includegraphics[width=\textwidth]{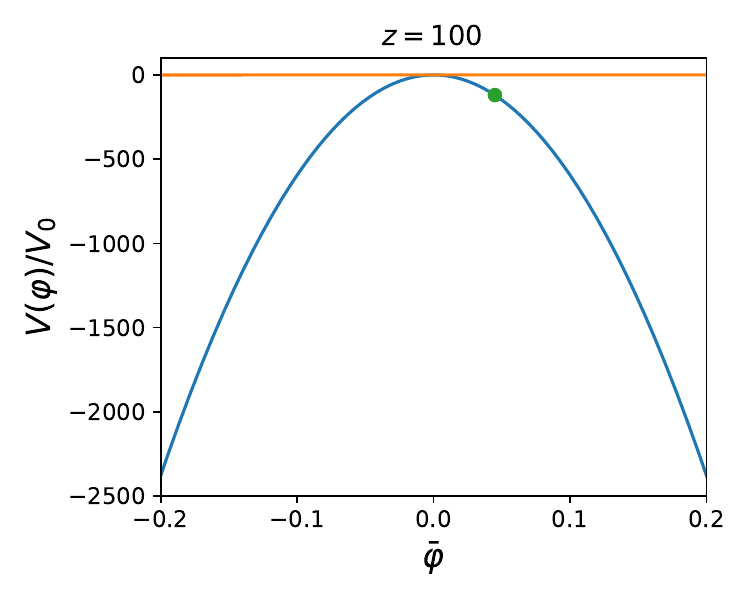}
\end{subfigure}\hfill
\begin{subfigure}{0.45\textwidth}
    \centering
    \includegraphics[width=\textwidth]{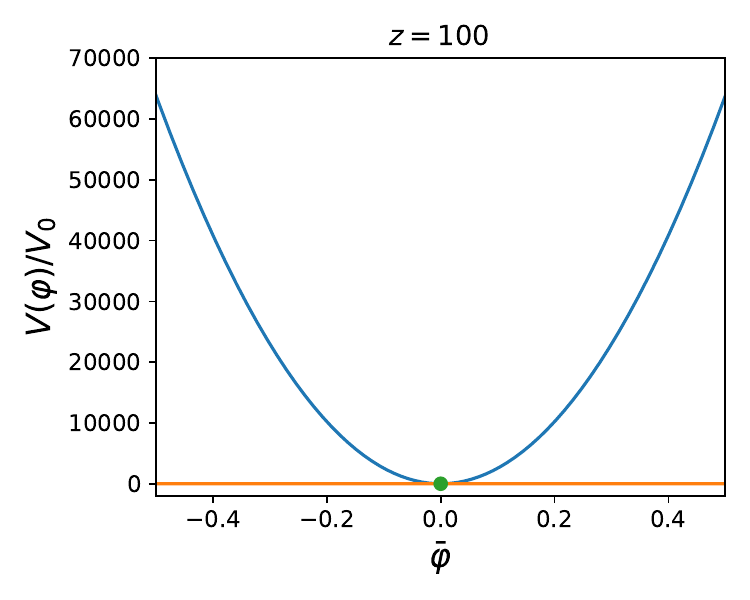}
\end{subfigure}
\begin{subfigure}{0.45\textwidth}
    \centering
    \includegraphics[width=\textwidth]{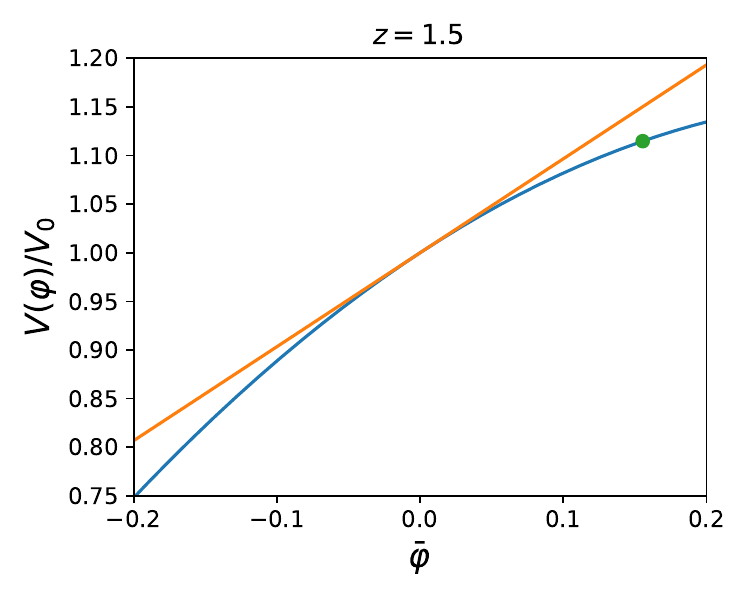}
\end{subfigure}\hfill
\begin{subfigure}{0.45\textwidth}
    \centering
    \includegraphics[width=\textwidth]{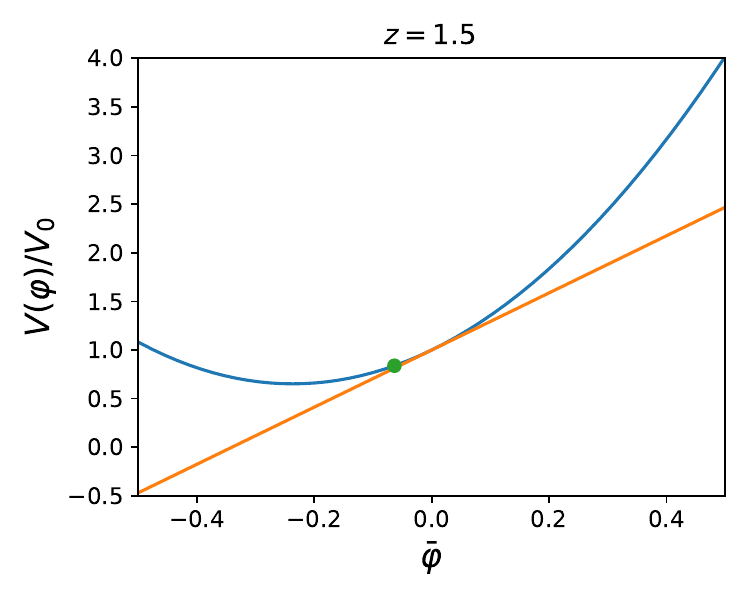}
\end{subfigure}
\begin{subfigure}{0.45\textwidth}
    \centering
    \includegraphics[width=\textwidth]{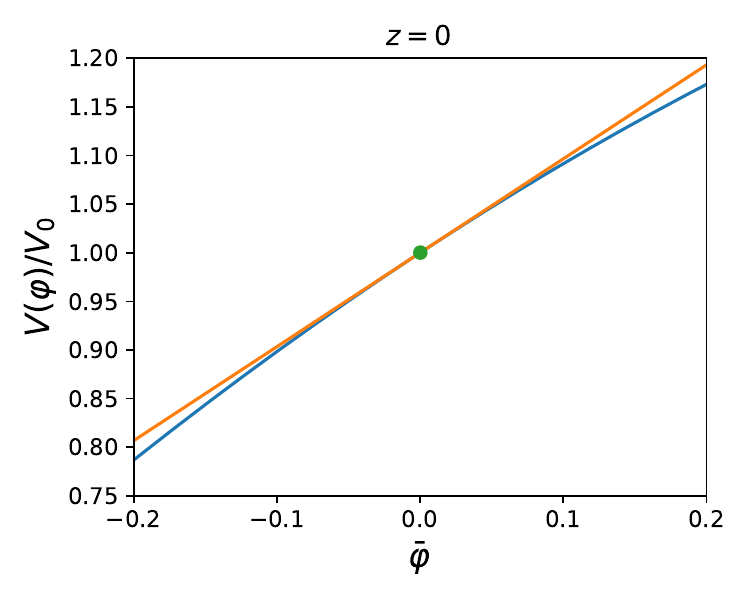}
\end{subfigure}\hfill
\begin{subfigure}{0.45\textwidth}
    \centering
    \includegraphics[width=\textwidth]{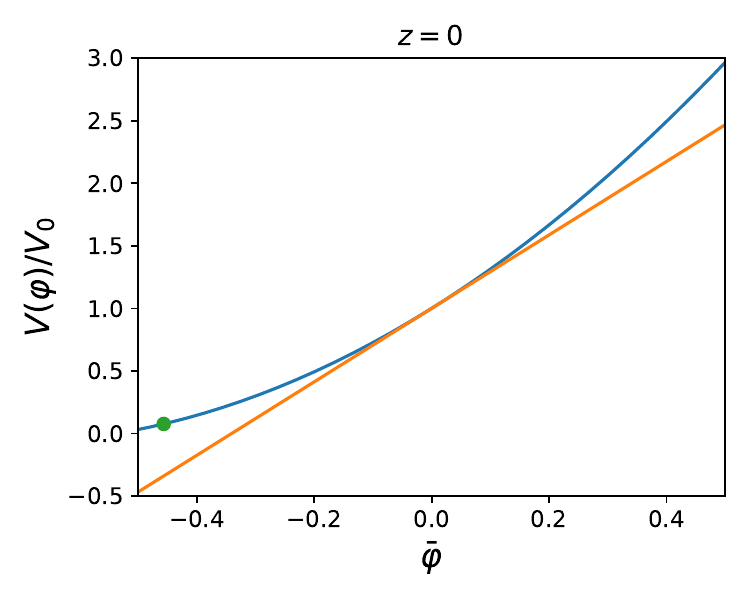}
\end{subfigure}
    \caption{Bare potential $V(\varphi)$ (orange line), effective potential $V_{\rm eff}(\varphi)$ (blue curve) and the value of the scalar field (green dot) at different redshifts for the best-fit parameters, cf. Table \ref{tab:results}. The left panels correspond to $\alpha>0$, while the right panels correspond to $\alpha<0$.}
    \label{fig:potential}
\end{figure}

The KG equation \eqref{eq:KGcompact} in a FLRW background takes the following form: 

\begin{equation}\label{eq:KG_FLRW}
    \ddot{\varphi}+3H\dot{\varphi}+\frac{\partial V}{\partial\varphi}-\frac{6\alpha\varphi}{\kappa^2}(\dot{H}+2H^2)=0\,.
\end{equation}
We will focus on parameter values for which the linear potential \eqref{eq:main_functions} becomes relevant only at late times. Consequently, at high redshifts (i.e., during the matter- and radiation-dominated epochs, hereafter referred to as MDE and RDE, respectively), the third term in the KG equation is negligible compared to the friction term. Hence, if the scalar field evolves during these epochs, its dynamics can only be driven by the non-minimal coupling. In Fig. \ref{fig:potential} we show the effective potential and the value of the scalar field at different redshifts for the best-fit values inferred from our fitting analyses (cf. Sec. \ref{sec:results}).

Let us begin by analyzing the dynamics of the model during the RDE. Since $R\propto\dot{H}+2H^2\sim \kappa^2\rho_m$ (up to subdominant corrections of order $\alpha$), very deep in the RDE the source term in Eq. \eqref{eq:KG_FLRW} can be neglected, as it is suppressed relative to the friction term, since the latter is proportional to the radiation energy density.  Therefore, the derivative of the scalar field evolves as 
\begin{equation}\label{eq:decay_mode}
\dot{\varphi}(a)= Ca^{-3}
\end{equation}
and its kinetic energy scales as $\dot{\varphi}^{2}\sim a^{-6}$. Consequently, the decaying mode -- which undergoes a kination phase -- eventually becomes negligible, irrespective of the sign of $\alpha$, leaving the scalar field frozen at a constant value with vanishing kinetic energy. This fully justifies our choice of initial conditions $\varphi_{\rm ini}={\rm const.}$ and $\dot{\varphi}_{\rm ini}=0$, as long as $z_{\rm ini}$ is sufficiently large -- in our implementation to the Einstein-Boltzmann solver we set it to $z_{\rm ini}=10^{11}$ (cf. the details in Sec. \ref{sec:method}).

However, this situation eventually comes to an end, as non-relativistic matter becomes increasingly important, making the source term in the KG equation more relevant. As a result, the scalar field may become dynamical even during the RDE, depending on the value of the coupling parameter $\alpha$. It is therefore useful to estimate analytically when this occurs. Let us assume
\begin{equation}\label{eq:condition}
|\bar{\alpha}|\bar{\varphi}^2\ll 1\,.
\end{equation}
This is a very natural condition, since the coupling function  should not depart excessively from its canonical value at the Big Bang Nucleosynthesis (BBN) epoch, so we expect $F(\varphi)\simeq 1$ deep in the RDE. Under condition \eqref{eq:condition}, we find the following solutions of the KG equation:

\begin{figure}[t!]
    \centering
\includegraphics[width=0.45\linewidth]{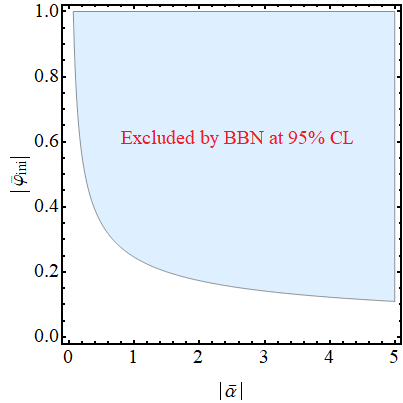}
    \caption{Exclusion plot in the $(|\bar{\alpha}|,|\bar{\varphi}_{\rm ini}|)$ plane obtained by imposing the BBN bound on the gravitational strength from \cite{Alvey:2019ctk}. The shaded area (in blue) corresponds to the excluded region of parameter space. See the main text for further details.}
    \label{fig:exclusion_BBN}
\end{figure}

\begin{equation}\label{eq:RDE_pos}
\varphi(a)\Big|_{\rm RDE} = \frac{\varphi_{\rm ini}}{\sqrt{a}}\sqrt{\frac{\Omega_r^0}{3\bar{\alpha}\,\Omega_m^0}} \,\,I_1\left(2\sqrt{a}\sqrt{\frac{3\bar{\alpha}\Omega_m^0}{\Omega_r^0}}\right)\qquad {\rm for}\,\alpha>0\,,
\end{equation}
\begin{equation}\label{eq:RDE_neg}
\varphi(a)\Big|_{\rm RDE} = \frac{\varphi_{\rm ini}}{\sqrt{a}}\sqrt{\frac{\Omega_r^0}{3|\bar{\alpha}|\,\Omega_m^0}} \,\,J_1\left(2\sqrt{a}\sqrt{\frac{3|\bar{\alpha}|\Omega_m^0}{\Omega_r^0}}\right)\qquad {\rm for}\,\alpha<0\,,
\end{equation}
with $\Omega_r^0=\kappa^2\rho_r^0/(3H_0^2)$ the radiation density parameter and $\rho_r^0$ the current radiation energy density, computed as if all the neutrino species were still massless today. All the details of the calculation of Eqs. \eqref{eq:RDE_pos} and \eqref{eq:RDE_neg} are provided in Appendix \ref{sec:background_sol}. The functions $I_1$ and $J_1$ are the first order modified and standard Bessel functions of the first kind. Here we have already neglected the decaying modes. Note that both functions can be approximated by $J_1(x),I_1(x)\sim x/2+\mathcal{O}(x^2)$ for $x\ll 1$, so we recover the constant field solution $\varphi(a)\to\varphi_{\rm ini}$ when $a\to 0$, as expected. Moreover, we find that the departure from the $\varphi\simeq {\rm const.}$ behavior happens when $x=\mathcal{O}(1)$, so at the approximate redshift

\begin{equation}\label{eq:z_nc}
z_{nc}\simeq \frac{12|\bar{\alpha}|\Omega_m^0}{\Omega_r^0}-1\simeq 4.5\cdot 10^4|\bar{\alpha}|-1\,.
\end{equation}
For values of $|\bar{\alpha}|=\mathcal{O}(1)$, as preferred by current data (cf. Sec. \ref{sec:results}), and typical values of $\Omega_m^0\simeq 0.3$ and $\Omega_r^0\simeq 8\cdot 10^{-5}$, we find $z_{nc}\sim 4.5\times10^4$, five orders of magnitude smaller than the BBN redshift ($z_{\rm BBN}\sim 10^9$). Therefore, $\varphi_{\rm BBN}\simeq\varphi_{\rm ini}$, so if $\varphi_{\rm ini}$ yields a value of $F$ consistent with the BBN bounds, the model will automatically satisfy them at the BBN epoch as well, where $|F-1|_{\rm BBN}$ must be smaller than $\sim 5-6\%$ at 95\% CL \cite{Alvey:2019ctk} to avoid spoiling the primordial abundances of light elements (see also \cite{Uzan:2024ded}). This constraint can be cast as an upper bound on $|\bar{\alpha}|\bar{\varphi}_{\rm ini}^2$ leading to an excluded region in the
$(|\bar{\alpha}|,|\bar{\varphi}_{\rm ini}|)$ plane, as shown in
Fig.~\ref{fig:exclusion_BBN}. We also note that, according to Eq. \eqref{eq:z_nc}, the field thaws before the end of the RDE only if $|\alpha|\gtrsim 0.08$.

During the MDE, the approximate solution for the scalar field (valid when  $|\bar{\alpha}|\bar{\varphi}^2\ll 1$) depends again on the sign of $\bar{\alpha}$. If $\bar{\alpha}>0$, we find 

\begin{equation}\label{eq:pos_MDE}
    \varphi(a)\Big|_{\rm MDE}=\varphi(a_*)\left(\frac{a}{a_*}\right)^{\frac{3}{4}\left[-1+\sqrt{1+\frac{16}{3}\bar{\alpha}}\right]    }\quad {\rm for}\,\,\bar{\alpha}>0
\end{equation}
deep in the MDE (with $a_*$ some pivot scale factor in that era), whereas for negative $\bar{\alpha}$ the result depends on whether it is smaller or larger than $-3/16$: 

\begin{figure}[t!]
    \centering
\begin{subfigure}{0.45\textwidth}
    \centering
    \includegraphics[width=\textwidth]{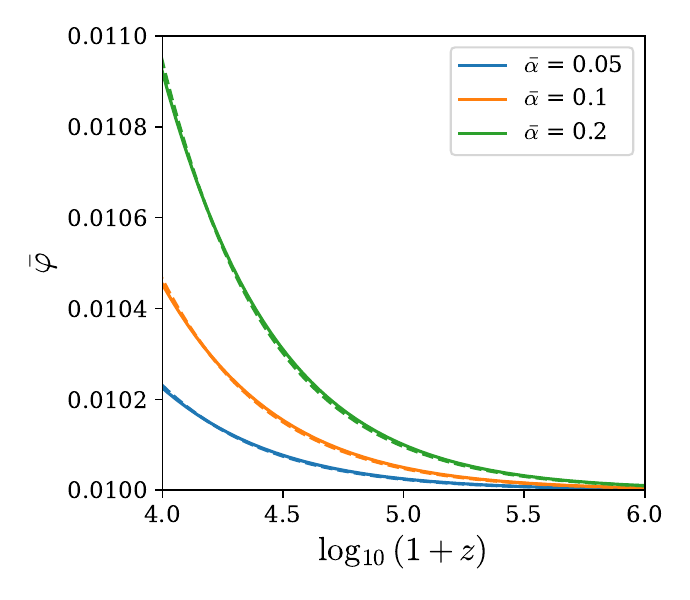}
\end{subfigure}\hfill
\begin{subfigure}{0.45\textwidth}
    \centering
    \includegraphics[width=\textwidth]{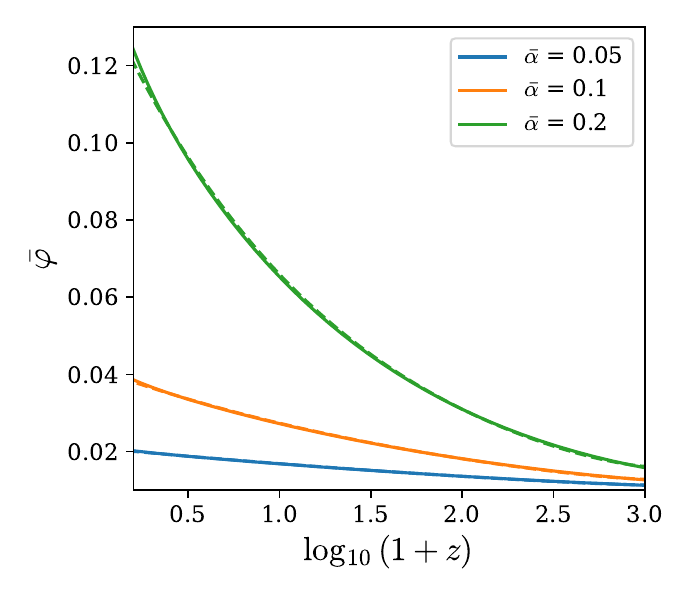}
\end{subfigure}
\begin{subfigure}{0.45\textwidth}
    \centering
    \includegraphics[width=\textwidth]{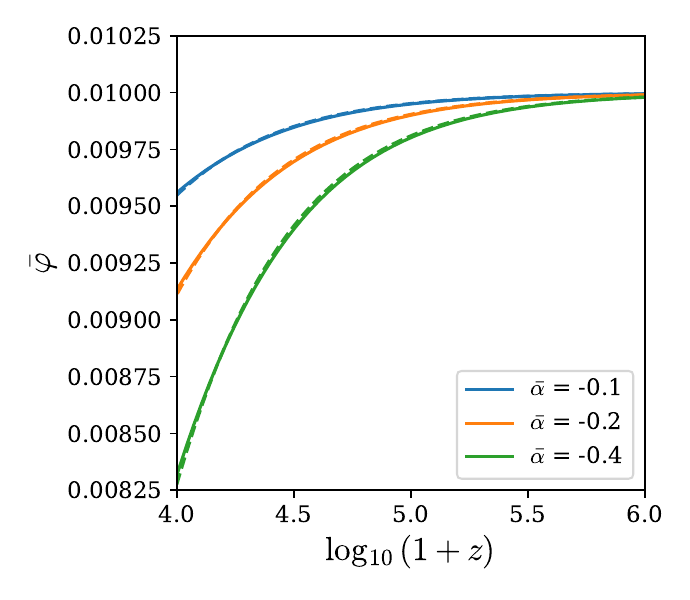}
\end{subfigure}\hfill
\begin{subfigure}{0.45\textwidth}
    \centering
    \includegraphics[width=\textwidth]{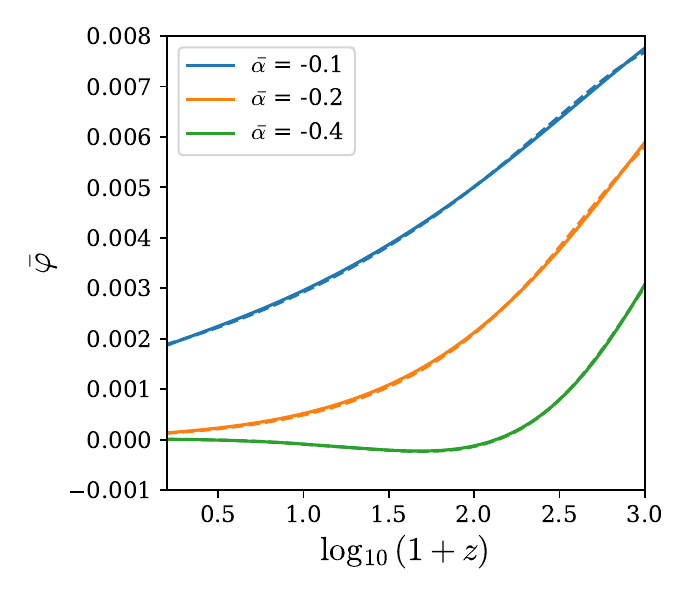}
\end{subfigure}
    \caption{Evolution of the scalar field in the RDE (left column) and the MDE (right column), both for positive (upper row) and negative (lower row) values of $\alpha$. Dashed lines correspond to the approximate analytical solutions.}
    \label{fig:scalarfield}
\end{figure}

\begin{equation}
  \varphi(a)\Big|_{\rm MDE}=a^{-3/4}\left[d_0\cosh\left(\frac{3\xi}{4}\,\ln(a)\right)+d_1\sinh\left(\frac{3\xi}{4}\,\ln(a)\right)\right] \quad {\rm if}\,\,-\frac{3}{16}<\bar{\alpha}<0\,,  
\end{equation}

\begin{equation}\label{eq:MDE_osc}
  \varphi(a)\Big|_{\rm MDE}=a^{-3/4}\left[d_0\cos\left(\frac{3\xi}{4}\,\ln(a)\right)+d_1\sin\left(\frac{3\xi}{4}\,\ln(a)\right)\right] \quad {\rm if}\,\,\bar{\alpha}<-\frac{3}{16}\,, 
\end{equation}
with $\xi(\bar{\alpha})\equiv\sqrt{|1+\frac{16}{3}\bar{\alpha}|}$ and $\{d_0,d_1\}$ integration constants. All the details of these calculations are also provided in Appendix \ref{sec:background_sol}. In Fig.~\ref{fig:scalarfield}, we compare the exact (numerical) solutions of the KG equation~\eqref{eq:KG_FLRW} with the corresponding analytical approximations discussed above for several values of the parameter $\bar{\alpha}$, including the decaying modes whenever they are present. The agreement is apparent.

We can now estimate the relative growth (decrease) of the scalar field with respect to its initial value for positive (negative) values of $\bar{\alpha}$ by the end of the MDE, which we take to occur at $z\simeq 1$. This is a useful exercise, as it allows us to determine the typical values of $\varphi_{\rm ini}$ required to obtain a realistic cosmology. To do so, we match the solutions obtained during the radiation- and matter-dominated epochs by imposing continuity at the radiation-matter equality time (i.e., at $z_{\rm eq}=\Omega_m^0/\Omega_r^0-1\simeq 3750$). Although this is only a rough approximation, it is sufficiently accurate for our purposes. We will obtain an analytical expression for the ratio $\varphi_{\rm ini}/\varphi(z=1)$.  

For $\bar{\alpha}>0$ we arrive at the following expression:

\begin{equation}\label{eq:ratio_pos}
\frac{\varphi_{\rm ini}}{\varphi(z=1)}\approx\frac{\sqrt{3\bar{\alpha}}}{I_1(2\sqrt{3\bar{\alpha}})}\left(\frac{2}{3750}\right)^{\frac{3}{4}\left[-1+\sqrt{1+\frac{16\bar{\alpha}}{3}}\right]} \quad {\rm for}\,\,\bar{\alpha}>0\,.
\end{equation}
For $\bar{\alpha}<-3/16$, instead, we find 

\begin{figure}[t!]
    \centering
\includegraphics[width=1\linewidth]{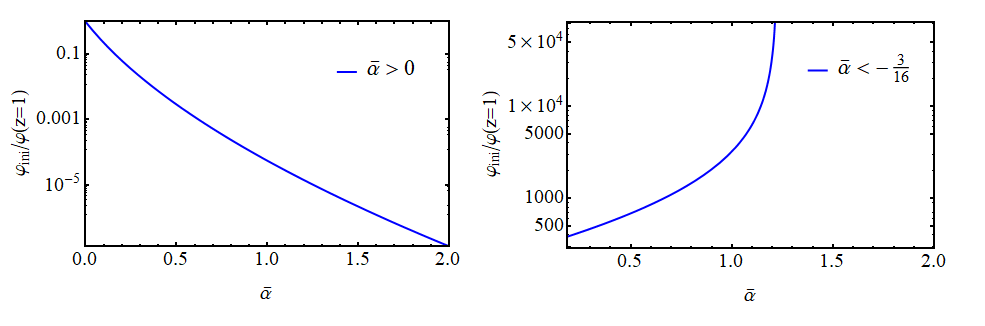}
    \caption{{\it Left plot:} Typical enhancement of the scalar field relative to its initial value during the late MDE, as a function of $\bar{\alpha}$, for $\bar{\alpha}>0$. The initial conditions are imposed deep within the RDE. {\it Right plot:} Suppression of $\varphi(z=1)$ relative to $\varphi_{\rm ini}$ for the model with $\bar{\alpha}<-3/16$.}
    \label{fig:ratios}
\end{figure}

\begin{equation}\label{eq:ratio_neg}
\frac{\varphi_{\rm ini}}{\varphi(z=1)}\approx\frac{\sqrt{3|\bar{\alpha}|}}{J_1(2\sqrt{3|\bar{\alpha}|})}\left(\frac{2}{3750}\right)^{-3/4} \quad {\rm for}\,\,\bar{\alpha}<-3/16\,.
\end{equation}
where we have neglected in both cases the decaying modes and the modulating effect of the oscillations in Eq. \eqref{eq:MDE_osc}. We plot Eqs. \eqref{eq:ratio_pos} and \eqref{eq:ratio_neg} in Fig. \ref{fig:ratios} to illustrate that (i) for values of $\varphi_{\rm ini}$ obeying the condition $|\bar{\alpha}|\bar{\varphi}_{\rm ini}^2\ll 1$, the scalar field is very suppressed in the last stages of the MDE if $\bar{\alpha}<-3/16$, as preferred by the data. In fact, under this condition, the data cannot distinguish between the cases $\varphi_{\rm ini}=0$ and $\varphi_{\rm ini}\neq0$, since the cosmic evolution during the radiation- and matter-dominated epochs is very similar in both scenarios. Notice that, for these values of $\bar{\alpha}$, the decay towards $\bar{\varphi}\to 0$ already occurs during the RDE. This motivates the choice of the initial conditions $\varphi_{\rm ini}=0$ and $\dot{\varphi}_{\rm ini}=0$ for negative $\alpha$, thereby reducing the number of free parameters in the Monte Carlo analysis. This corresponds in very good approximation to set the scalar field at the minimum of the effective potential \eqref{eq:eff_pot}, yielding $V_{\rm eff}=V_0$  during the RDE and MDE; and (ii) if $\bar{\alpha}$ is positive and of order one, due to the fact that the effective potential is in this case convex, the scalar field is expelled towards large values; it can increase by several orders of magnitude\footnote{We can always restrict ourselves to $\varphi_{\rm ini}\geq0$, since flipping the sign of the scalar field is equivalent to flipping the sign of $\beta$ -- all other terms in the action \eqref{eq:action} are invariant under $\varphi\to-\varphi$.}. This will force us to consider values of $\varphi_{\rm ini}$ extremely close to zero to obey the local constraints and keep the model dynamics stable. This is somewhat unnatural and demands additional structure beyond that in the action \eqref{eq:action} (or at least, more complicated shapes of the functions \eqref{eq:main_functions}), e.g., some Higgs-type symmetry breaking mechanism that confines the field close to zero before the moment of imposing our initial conditions\footnote{See \cite{Adam:2026ajg} for a modification of the non-minimal coupling $F(\varphi)$ that makes the field freeze at the minimum of the modified effective potential during the RDE and MDE epochs, thereby rendering the model stable.}. We will discuss further details about the initial conditions in Sec. \ref{sec:method}.

We note that the analytical solutions obtained for $\varphi(a)$ under the assumption $|\bar{\alpha}|\bar{\varphi}^2\ll1$ in the RDE and MDE can also be readily used to derive analytical expressions for the gravitational strength experienced by test masses through direct substitution into Eq. \eqref{eq:Geff}, as well as for the corrections to the Universe's expansion rate, for both positive and negative values of $\alpha$. We leave this exercise for the interested reader.

At sufficiently low redshifts in the late Universe, the scalar field potential $V(\varphi)$ \eqref{eq:main_functions} becomes important and significantly modifies the shape of the effective potential \eqref{eq:eff_pot} in the region where the scalar field resides (see again Fig. \ref{fig:potential}). It is interesting to analyze the evolution of the effective DE EoS parameter during this stage of cosmic expansion to determine the conditions under which crossing the phantom divide is possible. We obtain the expression of the aforementioned EoS parameter by establishing a correspondence between the modified Friedmann and pressure equations, Eqs. \eqref{eq:Friedmann_eq} and \eqref{eq:pressure_eq}, respectively, and their counterparts in GR with a DE component, given by
\begin{equation} \label{eq:effFried}
3H^2=\kappa^2(\rho+\rho_{\rm DE})\qquad;\qquad -(2\dot{H}+3H^2)=\kappa^2(p+p_{\rm DE})\,.
\end{equation}
The effective DE density and pressure, $\rho_{\rm DE}$ and $p_{\rm DE}$, encapsulate the modified gravity effects that characterize the non-minimally coupled model under study. From Eqs. \eqref{eq:effFried}, it is clear that the effective DE EoS parameter can be constructed as follows:

\begin{equation}\label{eq:EoS}
w_{\rm DE}=\frac{p_{\rm DE}}{\rho_{\rm DE}}=\frac{-1+\frac{1}{V}\left[\frac{\dot{\varphi}^2}{2}-\alpha\varphi^2 p+\frac{2\alpha}{\kappa^2}\left\{\dot{\varphi}^2+\varphi\left(\ddot{\varphi}+2H\dot{\varphi}\right)\right\}\right]}{1+\frac{1}{V} \left[\frac{\dot{\varphi}^2}{2}-\alpha\varphi^2\rho-6\frac{\alpha}{\kappa^2}H\varphi\dot{\varphi}\right]}\,.
\end{equation}
Note that if $\alpha \neq 0$ DE receives contributions from matter and radiation. That is, if we map our theory onto GR, modified gravity effects are interpreted as part of the effective DE. For $\alpha=0$ we retrieve the EoS of a minimally coupled scalar field, i.e., 

\begin{equation}
w=\frac{\frac{\dot{\varphi}^2}{2}-V(\varphi)}{\frac{\dot{\varphi}^2}{2}+V(\varphi)}\,,
\end{equation}
from which it follows that $-1\leq w\leq 1$. Hence, standard quintessence cannot cross the phantom divide\footnote{In an uncoupled scenario, one needs at least two uncoupled scalar fields to explain the crossing,  one of them with the opposite sign in the kinetic term, see, e.g., \cite{Gomez-Valent:2025mfl,Goh:2025upc}.}. For $\alpha\neq 0$, however, \eqref{eq:EoS} deviates from this result and, as we will see below, can allow for scenarios in which the phantom divide is crossed.

Under the initial conditions that we are considering for the scalar field, in the last stages of the MDE the potential is much larger than the kinetic energy of the scalar field, regardless of the sign of $\alpha$. Therefore, all the terms in \eqref{eq:EoS} involving derivatives of the scalar field are suppressed in front of $V$. In addition,  $|\alpha|\varphi^2\ll 1$, so we are allowed to Taylor-expand $w_{\rm DE}$, treating the terms in square brackets of the numerator and the denominator as perturbations, i.e., 

\begin{equation}
    w_{\rm DE} \equiv \frac{-1 + \varepsilon_1}{1+ \varepsilon_2}\,,\, {\rm with}\,\,\varepsilon_1,\varepsilon_2\ll 1\,.
\end{equation}
We are led to the following expression:

\begin{equation}\label{eq:EoS2}
w_{\rm DE}= -1+\frac{1}{V(\varphi)}\left\{\dot{\varphi}^2-\alpha\varphi^2\rho_m+\frac{2\alpha}{\kappa^2}\left(\dot{\varphi}^2+\varphi\ddot{\varphi}-H\varphi\dot{\varphi}\right)\right\}+\mathcal{O}(\varepsilon_i^2)
\end{equation}
to first order in the perturbed quantities. We have also neglected radiation, as its contribution is negligible during the epoch of interest. Let us now explore the possibility of achieving a crossing of the phantom divide in different scenarios.


%

\subsection*{Case 1: \textmd{$\alpha>0$, $\beta=0$}}

If $\beta=0$, upon substituting the solution \eqref{eq:pos_MDE} in terms of the cosmic time in Eq. \eqref{eq:EoS2}, we find for $\alpha>0$ during the last stages of the MDE:

\begin{equation}\label{eq:approx}
w_{\rm DE}(t) \Big|_{\rm{end\;MDE}}\approx -1+\frac{C^2 t^{4\zeta/3-2}}{3V_0}\left[\frac{4}{3}\zeta^2(1+4\bar{\alpha})-4\bar{\alpha}-\frac{20}{3}\bar{\alpha}\zeta\right]\,,
\end{equation}
with  $\zeta(\bar{\alpha})\equiv 3(-1+\sqrt{1+16\bar{\alpha}/3})/4>0$ and $C$ a constant\footnote{$C$ is proportional to $\varphi_{\rm ini}$; we consider values of $C$ small enough to keep Eq. \eqref{eq:approx} a good approximation.}. The term in square brackets is positive if $\bar{\alpha}>3/2$ and negative if $\bar{\alpha}<3/2$, causing the effective DE to behave as quintessence or phantom DE, respectively. Only the latter case can account for the phantom phase preceding the crossing favored by current observations. But what happens in the late Universe? In the late Universe $H\approx \rm constant$ as a first approximation in a realistic scenario, since the expansion should be dominated by the cosmological constant, $V_0$. The solution to the KG equation \eqref{eq:KG_FLRW} in that setup for positive $\alpha$ and $\beta = 0$ reads

\begin{equation}
    \varphi(t) = Ae^{\omega_+t},\quad \omega_+ = \frac{H_0}{2}(-3+\sqrt{9+48\bar{\alpha}}),
\end{equation}
where we have neglected the decaying mode $\omega_-$. Substituting into Eq. \eqref{eq:EoS2} yields

\begin{equation}
    w_{\rm DE}(t)\Big|_{\rm late\;Universe} \approx -1 + \frac{A^2\omega_+e^{2\omega_+t}}{V_0}\left[(1+4\bar{\alpha})\omega_+ -2\bar{\alpha}H_0\right].
\end{equation}
It can be shown that the term inside square brackets remains positive for all (positive) values of $\alpha$. Thus, in this scenario the effective DE is quintessence-like at low redshift, implying that a crossing is indeed expected for $0<\bar{\alpha}<3/2$ and $\beta=0$. Nevertheless, we find that $1+w_{\rm DE}$ in the late Universe is proportional to $\bar{\alpha}\bar{\varphi}^2$, so it is not possible to simultaneously satisfy the local constraints and produce a sufficiently large deviation from $w_{\rm DE}=-1$   to reproduce the quintessence-like phase preferred by current observations.

\subsection*{Case 2: \textmd{$\alpha>0$, $\beta>0$}}
If, instead, we consider $\beta>0$, both features are possible. Under some conditions, the scalar field reaches a maximum at the end of the MDE (cf. Appendix \ref{sec:background_sol} for details). Near it, we of course have $\varphi>0$, $\dot{\varphi}\approx 0$ and $\ddot{\varphi}<0$, so Eq. \eqref{eq:EoS2} reduces to

\begin{equation}\label{eq:endMDE_pos}
w_{\rm DE}\Big|_{\rm max}\approx -1+\frac{1}{V(\varphi)}\left(-\alpha\varphi^2\rho_m+2\bar{\alpha}\varphi\ddot{\varphi}\right)\Big|_{\rm max}<-1\,.
\end{equation}
After that, the scalar field evolves towards the origin, so $\dot{\varphi}<0$. If the initial conditions are set such that $\varphi(z=0)\approx 0$ in order to satisfy the local constraints described in Sec. \ref{sec:Geff}, we find
\begin{equation}\label{eq:w0}
w_{\rm DE}(z=0)\approx -1+\frac{\dot{\varphi}^2}{V_0}(1+2\bar{\alpha})>-1\,.
\end{equation}
Therefore, if $\beta>0$ it is possible to produce the desired crossing, while also fulfilling the local constraints.

We now  determine the order of magnitude of the dimensionless parameters $\bar{\alpha}$ and

\begin{equation}
\bar{\beta}\equiv \frac{\kappa}{H_0^2}\beta \,.
\end{equation}
required to reproduce the observations. In the late Universe, where $H\approx H_0$ is approximately constant and $\varphi\approx 0$, the KG equation \eqref{eq:KG_FLRW} reduces to 

\begin{equation}
\ddot{\varphi}+3H_0\dot{\varphi}+\beta =0\,.
\end{equation}
Neglecting the rapidly decaying mode, $\varphi(t)$ depends linearly on cosmic time, with
$\dot{\varphi}\approx-\beta/(3H_0)$. Substituting this result into Eq.~\eqref{eq:w0}, we obtain

\begin{equation}
1+w_{\rm DE}(z=0)\approx \frac{\bar{\beta}^2}{27\Omega_{V_0}^0}(1+2\bar{\alpha})\,,
\end{equation}
where $\Omega_{V_0}^0=\kappa^2V_0/(3H_0^2)$ is the density parameter associated with the constant term in the potential. For the right-hand side to take values of order $0.1$--$0.2$, as preferred by current observations, we require $\bar{\beta}=\mathcal{O}(1)$.

As for $\bar{\alpha}$, in Appendix \ref{sec:background_sol} we show that $\bar{\alpha}<9/2$ for the scalar field to have a maximum. This is a necessary condition to drive the scalar field towards the origin at low redshift, and hence satisfy the local constraints. Using Eq. \eqref{eq:pos_late_mde}, one can easily see that the value of the scalar field and its second derivative with respect to the cosmic time at the aforesaid maximum take the following form: 

\begin{equation}
\label{eq:phi_m}
\bar{\varphi}\Big|_{\rm max}\approx \frac{4\bar{\beta}a_{\rm m}^3}{\Omega_m^0(4\bar{\alpha}-18)}\left(\frac{\zeta(\bar{\alpha})-3}{3\zeta(\bar{\alpha})}\right)\,,
\end{equation}
\begin{equation}
\label{eq:ddphi_m}
\ddot{\bar{\varphi}}\Big|_{\rm max}=\frac{4H_0^2\bar{\beta}[3-\zeta(\bar{\alpha})]}{4\bar{\alpha}-18}\,,
\end{equation}
with $a_m$ the scale factor at the maximum, which is also controlled by the initial value of the scalar field.  Substituting these expressions into Eq. \eqref{eq:endMDE_pos} yields 

\begin{figure}[t!]
    \centering
\includegraphics[width=0.6\linewidth]{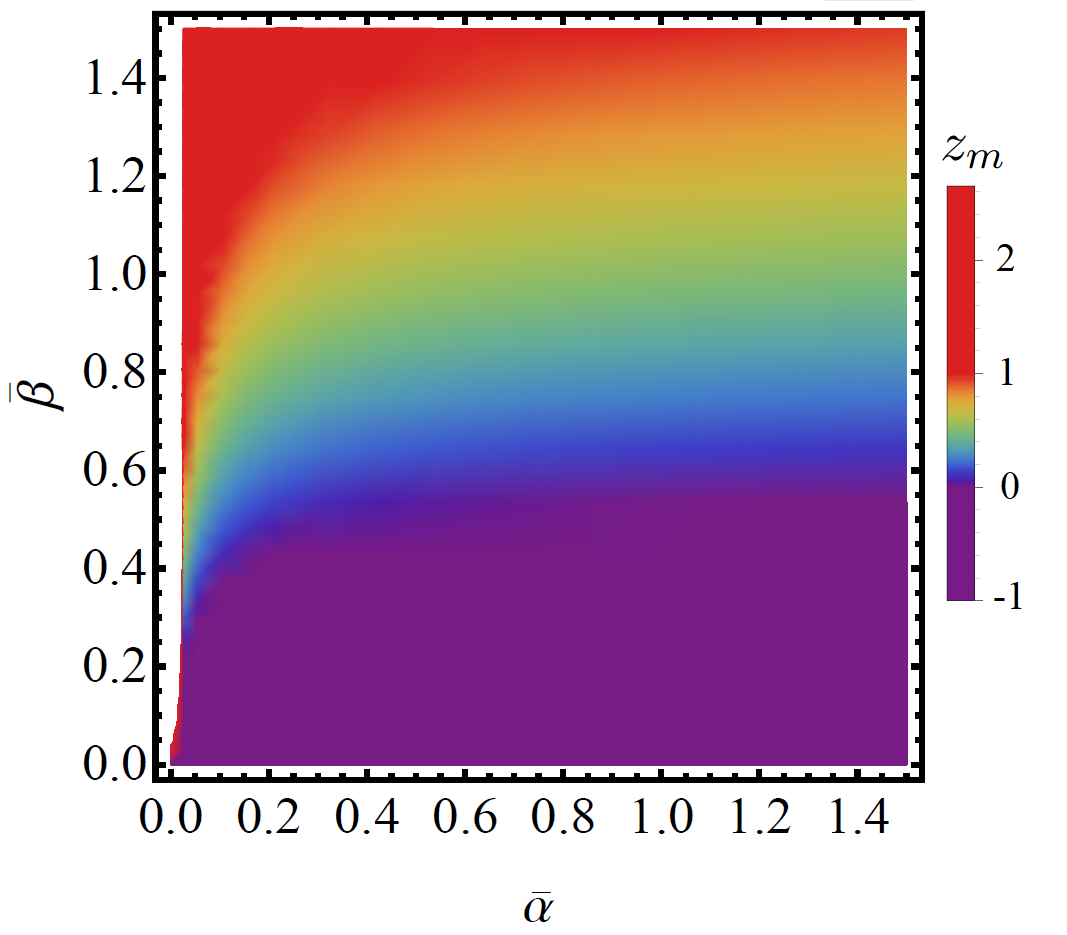}
    \caption{Color map of the redshift corresponding to the maximum of the scalar field,
$z_m(\bar{\alpha},\bar{\beta})$, in the $\bar{\alpha},\bar{\beta}>0$
scenario (case 2 in the main text), where phantom behavior occurs. The map is obtained from
Eq.~\eqref{eq:wmax} by setting $w_{\rm DE}=-1.1$, with
$\Omega_{V_0}^0=0.7$ and $\Omega_m^0=0.3$.}
    \label{fig:map}
\end{figure}

\begin{equation}\label{eq:wmax}
1+w_{\rm DE}\Big|_{\rm max}\approx \frac{-16\bar{\alpha}\bar{\beta}^2[3-\zeta(\bar{\alpha})]^2[2\zeta(\bar{\alpha})+1] a_m^{3}}{\zeta(\bar{\alpha})(4\bar{\alpha}-18)\left[9\zeta(\bar{\alpha})\Omega_{V_0}^0\Omega_m^0(4\bar{\alpha}-18)+4\bar{\beta}^2[\zeta(\bar{\alpha})-3]a_m^3\right]}\,.
\end{equation}
By fixing the density parameters $\Omega_{V_0}^0$ and $\Omega_m^0$ to representative values, together with $1+w_{\rm DE}\big|_{\rm max}= -\mathcal{O}(10^{-1})$, as preferred by current observations, one can use the previous expression to determine $a_m(\bar{\alpha},\bar{\beta})$ or, equivalently, the corresponding redshift $z_m(\bar{\alpha},\bar{\beta})$. Fig. \ref{fig:map} is generated in this way, adopting $1+w_{\rm DE}\big|_{\rm max}= -0.1$. We find that, for the values $\bar{\beta}\sim 1$ required for a late-time quintessence phase, the scalar field reaches its maximum at a redshift close to the value preferred by current observations over a wide range of $\bar{\alpha}$, approximately $0.2 \lesssim \bar{\alpha} \lesssim 1.2$.

It is worth mentioning that in this scenario the effective DE density $\rho_{\rm DE}$ becomes negative at high redshift, making $w_{\rm DE}$ diverge when $\rho_{\rm DE}=0$.  This is not problematic, however, since $\rho_{\rm DE}$ is strictly an effective quantity and the total, physical density $\rho_{\rm tot}$ remains positive at all times. Indeed, for the best-fit values (see Sec. \ref{sec:results}), $\rho_{\rm DE}$ is negative at $z > 6.5$. Since at these epochs the energy budget of the Universe is overwhelmingly dominated by matter and radiation, $\rho_{\rm tot}>0$ and the expansion history remains physically viable.

\subsection*{Case 3: \textmd{$\alpha<0$, $\beta=0$}}

We can repeat the exercise considering negative values of $\bar{\alpha}$. First, we can convince ourselves that the only viable possibility to generate a crossing is to have $\beta\ne 0$. If $\beta=0$ and $\bar{\alpha}<0$ with $|\bar{\alpha}|\ll 1$, the scalar field evolves as indicated in Eq. \eqref{eq:log}. By neglecting the terms quadratic in the coupling parameter in Eq. \eqref{eq:EoS2}, we find 

\begin{equation}
w_{\rm DE}\Big|_{\rm end\, MDE}\approx -1-\bar{\alpha}\bar{\varphi}^2\frac{\rho_m}{V_0}>-1\,.
\end{equation}
Therefore, in this case the effective dark energy behaves as quintessence during the final stages of the MDE. By contrast, if $\bar{\beta}=0$ and $\bar{\alpha}<0$ with $|\bar{\alpha}|=\mathcal{O}(1)$, the field is already located at the minimum of the effective potential, namely at the origin. As a result, $w_{\rm DE}\Big|_{\rm end\,MDE}\simeq -1$, and there is no significant departure from the behavior of a cosmological constant.

\subsection*{Case 4: \textmd{$\alpha<0$, $\beta>0$}}

The situation is much more favorable if $\bar{\alpha}$ is negative and of order unity, and $\beta>0$, since in this case the field evolves at the end of the MDE as
\begin{equation}
\bar{\varphi}(t)\Big|_{\rm end\,MDE}\approx \frac{3\beta\,t^2}{4\bar{\alpha}-18}\,,
\end{equation}
where we have used Eq.~\eqref{eq:pos_late_mde}, retained only the second term, and expressed the result in terms of cosmic time. Substituting this expression into Eq. \eqref{eq:EoS2} we find 

\begin{equation}
\label{eq:w_MDE_neg}
w_{\rm DE}\Big|_{\rm end\,MDE}\approx -1 +\frac{4t^2}{V(\varphi)}\left(\frac{3\beta}{4\bar{\alpha}-18}\right)^2(1+2\bar{\alpha})\,,
\end{equation}
which leads to phantom behavior if $\bar{\alpha}<-1/2$, as desired. In the late Universe, $H\approx$ constant, and, neglecting decaying modes,

\begin{equation}
\varphi(t)\Big|_{\rm late\,Universe}\approx \frac{\beta}{12\bar{\alpha}H_0^2}\,,
\end{equation}
so
\begin{equation}
\label{eq:w_LU_neg}
w_{\rm DE}\Big|_{\rm late\,Universe}\approx -1-\frac{\bar{\beta}^2}{144\bar{\alpha}}\frac{\rho_m}{V(\varphi)}>-1\,.
\end{equation}
This is, of course, only a rough estimate, but it is sufficient to show that, if we aim to explain the crossing of the phantom divide preferred by current observations with negative values of $\bar{\alpha}$, then $\bar{\alpha}$ must be of order one and accompanied by positive values of $\bar{\beta}$, also of order one or larger.

To summarize, a crossing of the phantom divide as preferred by cosmological data requires $\beta>0$ and of order unity and can be achieved with both negative and positive values of $\alpha$ satisfying the bounds $\bar{\alpha}<-1/2$ and $\bar{\alpha}<9/2$, respectively. The former case leads to large (negative) values of the scalar field at $z = 0$, in tension with the local gravitational constraints discussed in Sec. \ref{sec:Geff}. Remarkably, the latter case allows for the fulfillment of the local constraints by choosing $\varphi_{\rm ini}$ such that $\varphi(z=0) \approx0$.


\subsection{Linear perturbations and effective gravitational strength at cosmological scales}

In this section, we compute the linear perturbation equations corresponding to our model, following the formalism of Ma and Bertschinger \cite{Ma:1995ey}. We employ the synchronous gauge, work in Fourier space and consider only scalar modes. The perturbed metric reads
\begin{equation}
    ds^2 = a^2(\tau)\left[-d\tau^2+(\delta_{ij}+h_{ij})dx^idx^j\right]\,,
\end{equation}
with
\begin{equation}
    h_{ij}(\vec{x},\tau) = \int d^3k \: e^{i\vec{k}\cdot \vec{x}}\left[\hat{k}_i\hat{k}_j h(\vec{k},\tau)+\left(\hat{k}_i\hat{k}_j-\frac{1}{3}\delta_{ij}\right)6\eta(\vec{k},\tau)\right],
\end{equation}
and $\tau$ the conformal time. Scalar perturbations in Fourier space are encoded in the two functions  $h$ (the trace) and $\eta$.  We also expand the scalar field around the background\footnote{In this subsection $\bar{\varphi}$ represents the background value of the dimensionful scalar field, rather than the dimensionless one introduced in Eq. \eqref{eq:dimensionless}.}:
\begin{equation}
    \varphi = \bar{\varphi}(\tau) + \delta\varphi(\vec{x},\tau).
\end{equation}
The perturbed cosmological equations read:
\newline\newline
\noindent\textit{Scalar field equation ---}
\begin{equation}
\label{phi_per}
   \delta\varphi''+2\mathcal{H}\delta\varphi'- \left[6\frac{\alpha}{\kappa^2}(\mathcal{H}'+\mathcal{H}^2)-k^2\right]\delta\varphi -\frac{\alpha}{\kappa^2}\bar{\varphi}\left(h''+3\mathcal{H}h'-4k^2\eta\right) +\frac{h'}{2}\bar{\varphi}' = 0\,,
\end{equation}
\textit{Time-time equation ---}
\begin{equation}
\label{tt_per}
\begin{split}
    -2\frac{\alpha}{\kappa^2}\left[3\mathcal{H}^2+k^2\right]\bar{\varphi}\delta\varphi -6 \frac{\alpha}{\kappa^2}\mathcal{H}(\bar{\varphi}\delta\varphi'+\bar{\varphi}'\delta\varphi) + \bar{\varphi}'\delta\varphi' +a^2\beta\delta\varphi \\
    - \frac{1}{\kappa^2}\left[(1+\alpha\bar{\varphi}^2)\mathcal{H}+\alpha\bar{\varphi}\bar{\varphi}'\right]h' + \frac{2k^2\eta}{\kappa^2}\left(1+\alpha\bar{\varphi}^2\right) = a^2\delta {T^0}_0\,,
\end{split}
\end{equation}
\textit{Longitudinal time-space equation ---}
\begin{equation}
    \label{0i_eq}
    k^2\left[2\eta'\frac{1+\alpha\bar{\varphi}^2}{\kappa^2}-2\frac{\alpha}{\kappa^2}\left(\bar{\varphi}'\delta\varphi+\bar{\varphi}\delta\varphi'-\mathcal{H}\bar{\varphi}\delta\varphi\right)-\bar{\varphi}'\delta\varphi\right] = a^2(\bar{\rho}+\bar{P})\theta\,,
\end{equation}
\textit{Trace space-space equation ---}
\begin{equation}
\label{trace_eq}
\begin{split}
    \left(-2\mathcal{H}h'-h''+2k^2\eta \right)\frac{1+\alpha\bar{\varphi}^2}{\kappa^2}-2\frac{\alpha}{\kappa^2}\Biggl\{3\left(2\mathcal{H}'+\mathcal{H}^2\right)\bar{\varphi}\delta\varphi+3\mathcal{H}(\bar{\varphi}'\delta\varphi+\bar{\varphi}\delta\varphi') + \\
     +3(\bar{\varphi}''\delta\varphi+2\bar{\varphi}'\delta\varphi'+\bar{\varphi}\delta\varphi'')+2k^2\bar{\varphi}\delta\varphi+h'\bar{\varphi}\bar{\varphi}' \Biggr\} -3\bar{\varphi}'\delta\varphi' +3a^2\beta\delta\varphi = a^2 \delta {T^i}_i\,,
\end{split}
\end{equation}
\textit{Longitudinal traceless space-space equation ---}
\begin{equation}
    \label{ij_eq}
    (1+\alpha\bar{\varphi}^2)\left[2\mathcal{H}(h'+6\eta')+h''+6\eta''-2k^2\eta \right] + 4k^2\alpha\bar{\varphi}\delta\varphi + 2\alpha(h'+6\eta')\bar{\varphi}\bar{\varphi}' = -3\kappa^2a^2(\bar{\rho}+\bar{P})\sigma\,.
\end{equation}
In this section, a prime denotes a derivative with respect to conformal time, and $\mathcal{H}=a'/a$. The variables $\bar{\rho}$ and $\bar{P}$ represent the total background energy density and pressure, respectively, while the total velocity divergence $\theta$ and anisotropic stress $\sigma$ are defined by
\begin{equation}
    (\bar{\rho}+\bar{P})\theta \equiv ik^i\delta {T^0}_i, \quad(\bar{\rho}+\bar{P})\sigma \equiv -\left(\hat{k}_i\hat{k}_j -\frac{1}{3}\delta_{ij}\right){\Sigma^i}_j,\quad {\Sigma^i}_j \equiv {T^i}_j-\frac{1}{3}\delta^i_j T.
\end{equation}
Before proceeding, two points are worth stressing. Firstly, in the GR limit ($\alpha\to 0$) and for a constant scalar-field potential ($\beta\to 0$) the background scalar field has no dynamics and the equations successfully reduce to those of $\Lambda$CDM \cite{Ma:1995ey}. Secondly, the standard conservation law, Eq. \eqref{eq:cons}, holds at all orders in perturbation theory, since the action \eqref{eq:action} does not have any term that couples matter and radiation to the scalar field. Therefore, the right-hand side of the first-order Einstein equations (i.e., the energy-momentum tensor perturbations, which are derived from the Boltzmann equation) are unmodified with respect to GR. 

To conclude the analysis of linear perturbations, let us derive the linear growth equation for the matter density contrast, which is defined as
\begin{equation}
    \delta_m \equiv \frac{\delta\rho_m}{\bar{\rho}_m}.
\end{equation}
We will restrict ourselves to deep sub-horizon scales, in the so-called quasi-static approximation:
\begin{equation}
\label{QSA}
    k \gg \mathcal{H} , \quad X' \sim \mathcal{H}X,\; X'' \sim \mathcal{H}^2X,
\end{equation}
where $X$ is a generic function (e.g., the scalar field). That is, all quantities are assumed to evolve on Hubble timescales, so that time derivatives can be neglected in favor of spatial gradients (the only exception is $h'$, which, as we will see, is proportional to $\delta_m'$). Under these approximations, the perturbed time-time, traceless and scalar field equations can be combined to obtain
\begin{equation}
\label{h_growth}
    h'' + \frac{a'}{a}h' = -\frac{\kappa^2}{1+\alpha\bar{\varphi}^2}\frac{\kappa^2(1+\alpha\bar{\varphi}^2) + 8\alpha^2\bar{\varphi}^2}{\kappa^2(1+\alpha\bar{\varphi}^2) + 6\alpha^2\bar{\varphi}^2}a^2\bar{\rho}_m\delta_m.
\end{equation}
 Now, in order to relate the metric perturbation $h$ to the density contrast $\delta$, we use the perturbed conservation equations $\nabla_\mu {T^\mu}_\nu = 0$, which for non-relativistic matter reduce to
\begin{equation}
\label{continuity_pert}
    \delta_m' + \frac{h'}{2} = 0.
\end{equation}
Substituting this relation into Eq. \eqref{h_growth} finally yields 
\begin{equation}
\label{eq:delta_growth}
    \delta_m'' + \frac{a'}{a}\delta_m'  -4\pi \left[\frac{\kappa^2}{8\pi}\frac{1}{1+\alpha\bar{\varphi}^2}\frac{\kappa^2(1+\alpha\bar{\varphi}^2) + 8\alpha^2\bar{\varphi}^2}{\kappa^2(1+\alpha\bar{\varphi}^2) + 6\alpha^2\bar{\varphi}^2}\right]a^2\bar{\rho}_m\delta_m = 0.
\end{equation}
The term inside square brackets corresponds to the effective gravitational constant $G_{\text{eff}}$ computed in Sec. \ref{sec:Geff}. That is, the cosmological effective gravitational constant governing linear structure growth within the horizon coincides with that associated with a spherically symmetric source in the weak-field limit, as expected in the absence of a screening mechanism.

\begin{figure}[t]
\centering
\begin{subfigure}{0.49\textwidth}
    \centering
    \includegraphics[width=\textwidth]{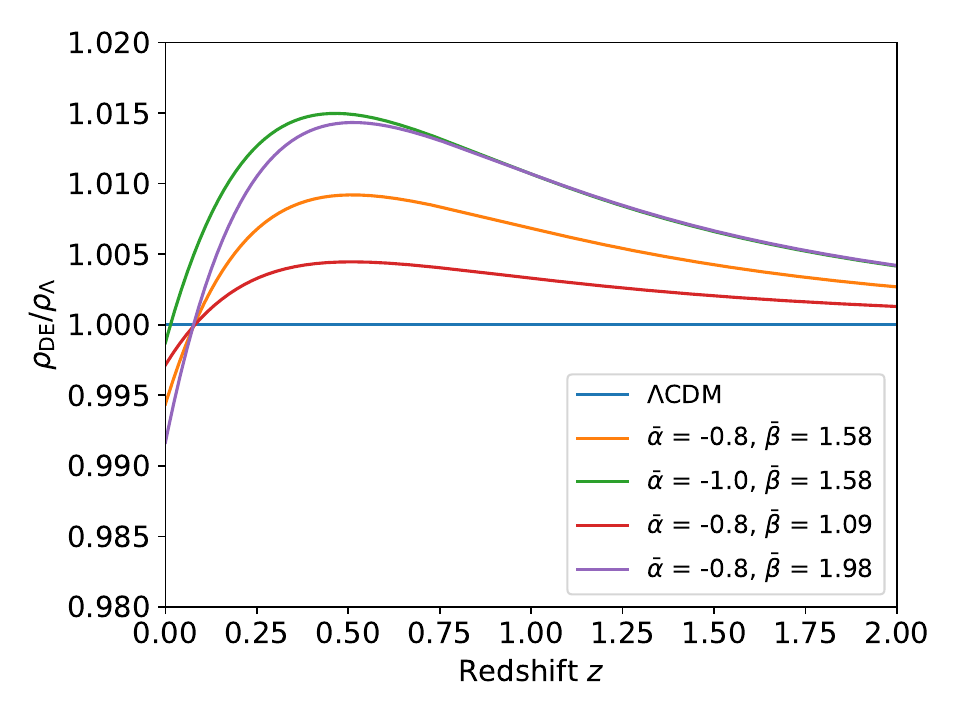}
\end{subfigure}\hfill
\begin{subfigure}{0.49\textwidth}
    \centering
    \includegraphics[width=\textwidth]{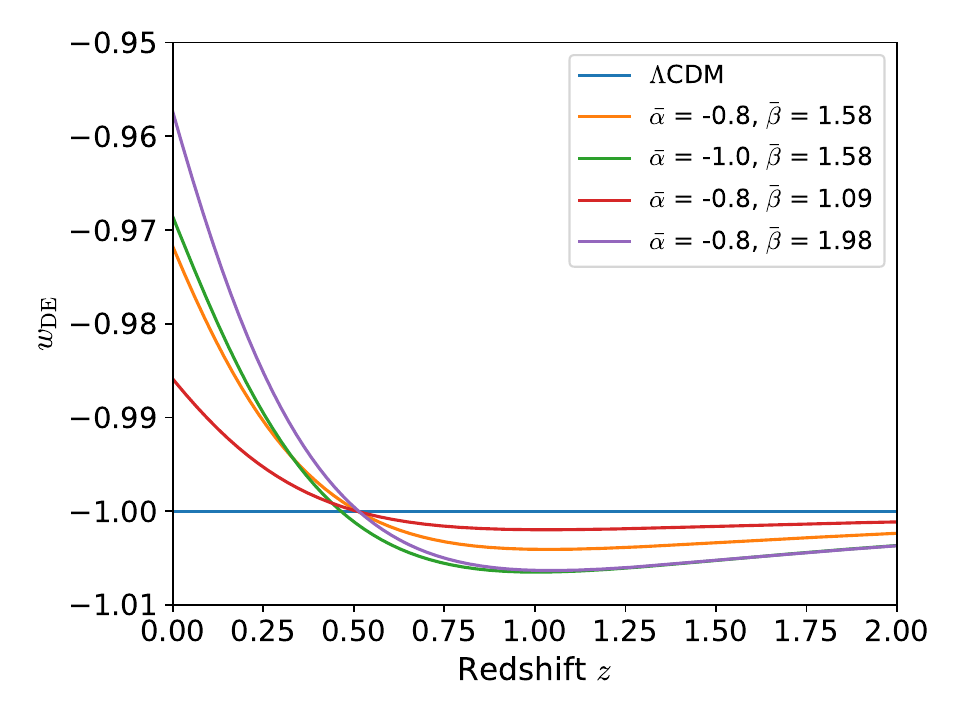}
\end{subfigure}
\begin{subfigure}{0.49\textwidth}
    \centering
    \includegraphics[width=\textwidth]{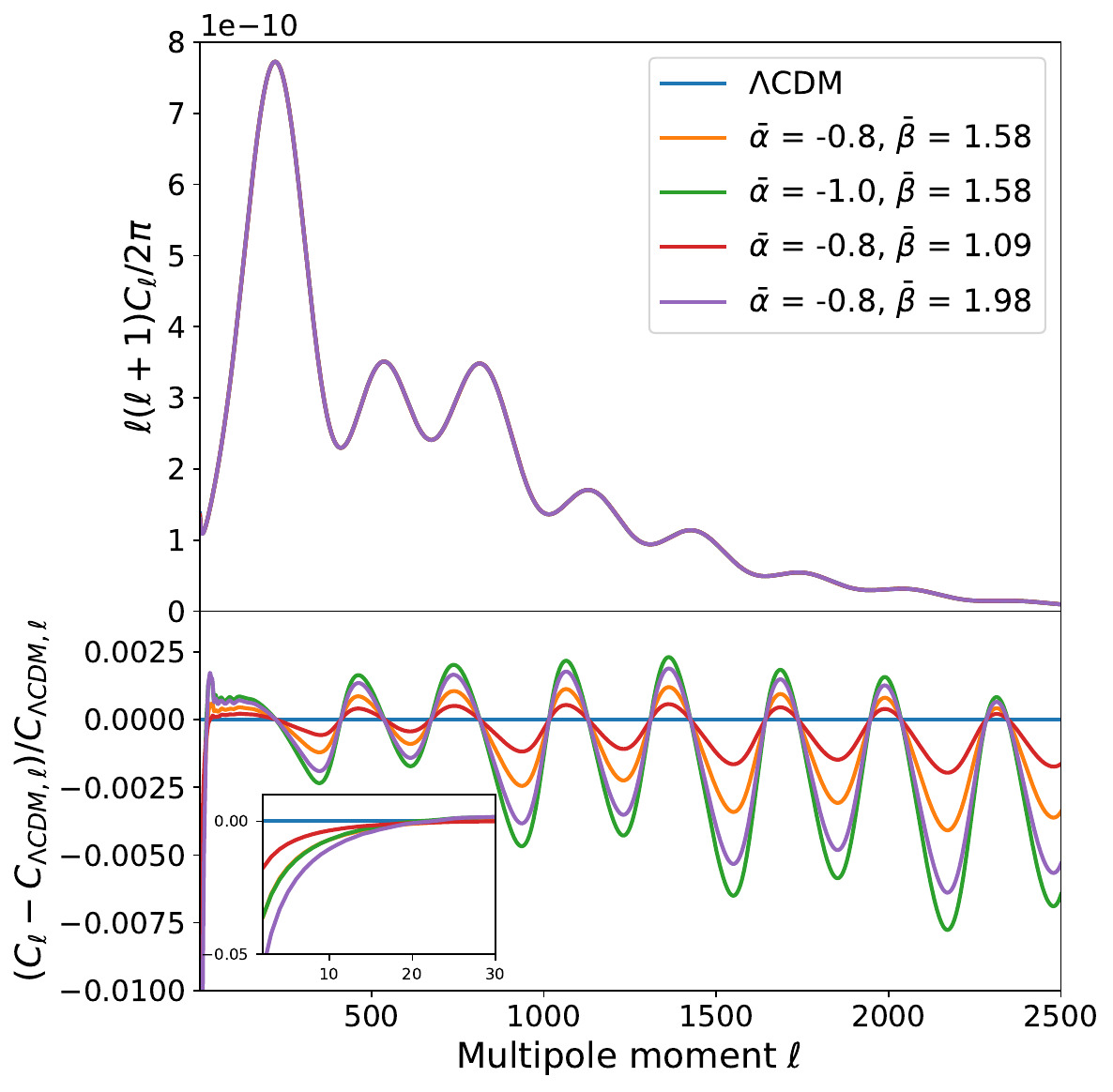}
\end{subfigure}\hfill
\begin{subfigure}{0.49\textwidth}
    \centering
    \includegraphics[width=\textwidth]{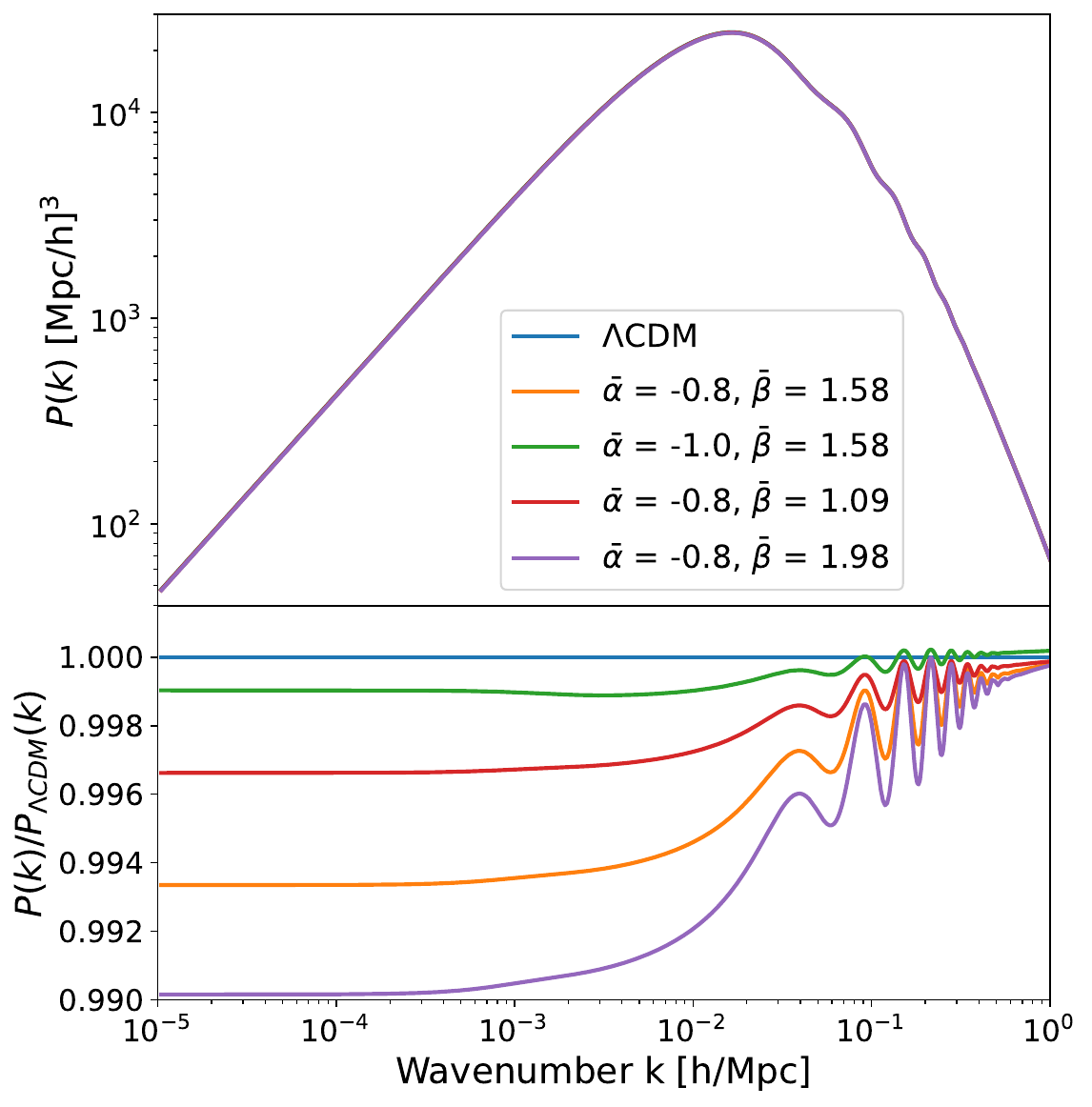}
\end{subfigure}
\caption{Effective DE density and EoS parameter (upper row), together with the corresponding CMB temperature and linear matter power spectra (lower row) for different values of $\bar\alpha<0$ and $\bar\beta$. All remaining parameters are fixed to the same fiducial values across all models. The bottom panels of each spectrum shows the relative deviation from $\Lambda$CDM. For the CMB temperature power spectrum, we  provide an inner plot showing the low-multipole region to facilitate visualization of the differences induced by the ISW effect. }
\label{fig:spectra_neg}
\end{figure}

\begin{figure}[t]
\centering
\begin{subfigure}{0.49\textwidth}
    \centering
    \includegraphics[width=\textwidth]{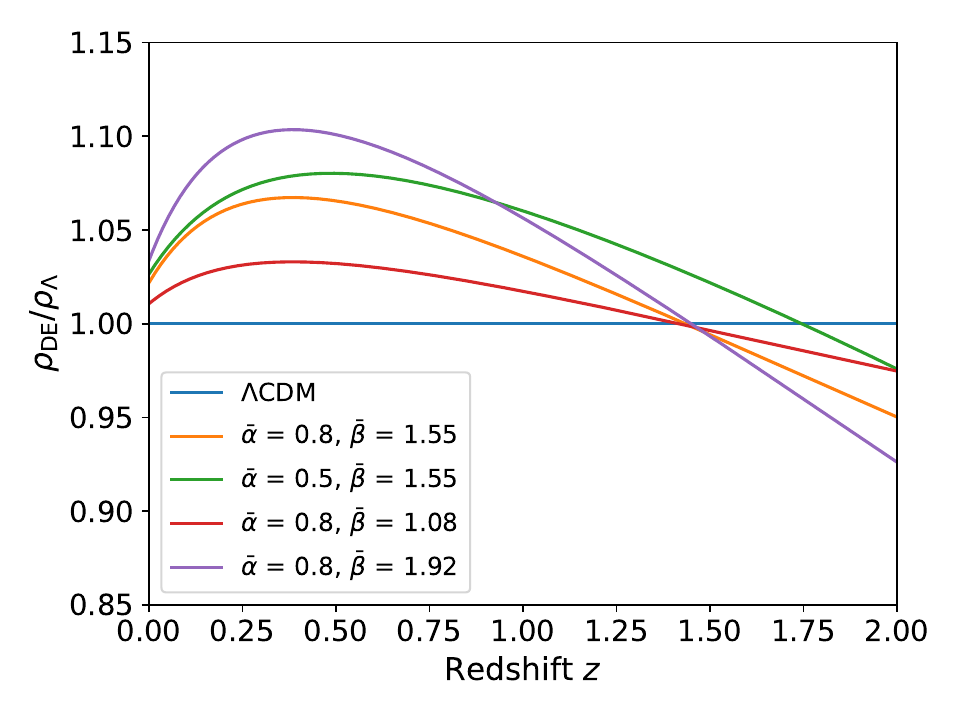}
\end{subfigure}\hfill
\begin{subfigure}{0.49\textwidth}
    \centering
    \includegraphics[width=\textwidth]{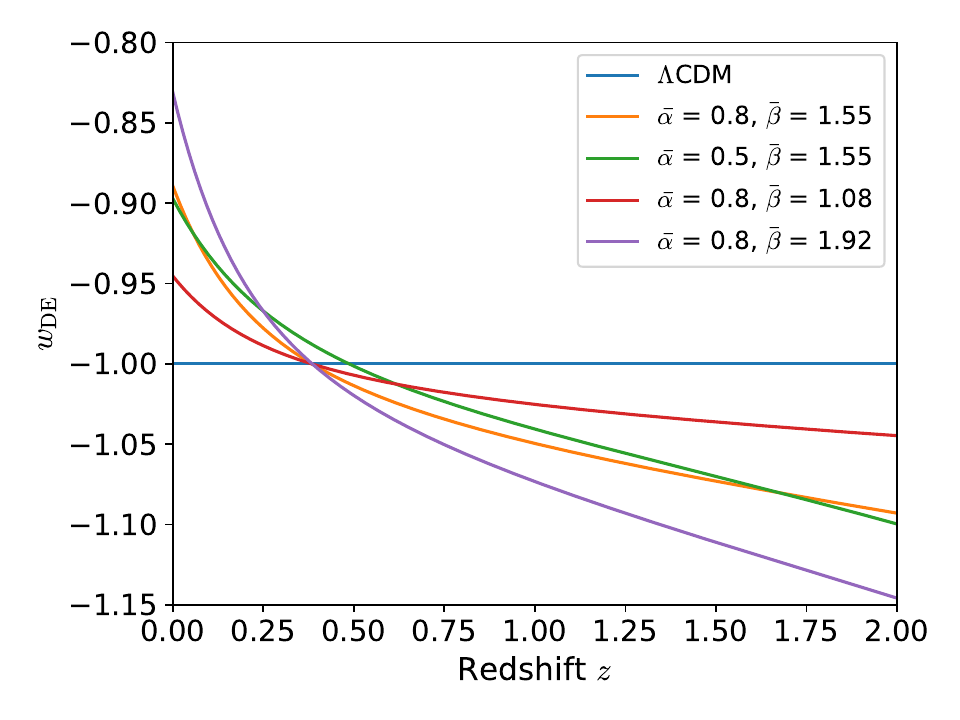}
\end{subfigure}
\begin{subfigure}{0.49\textwidth}
    \centering
    \includegraphics[width=\textwidth]{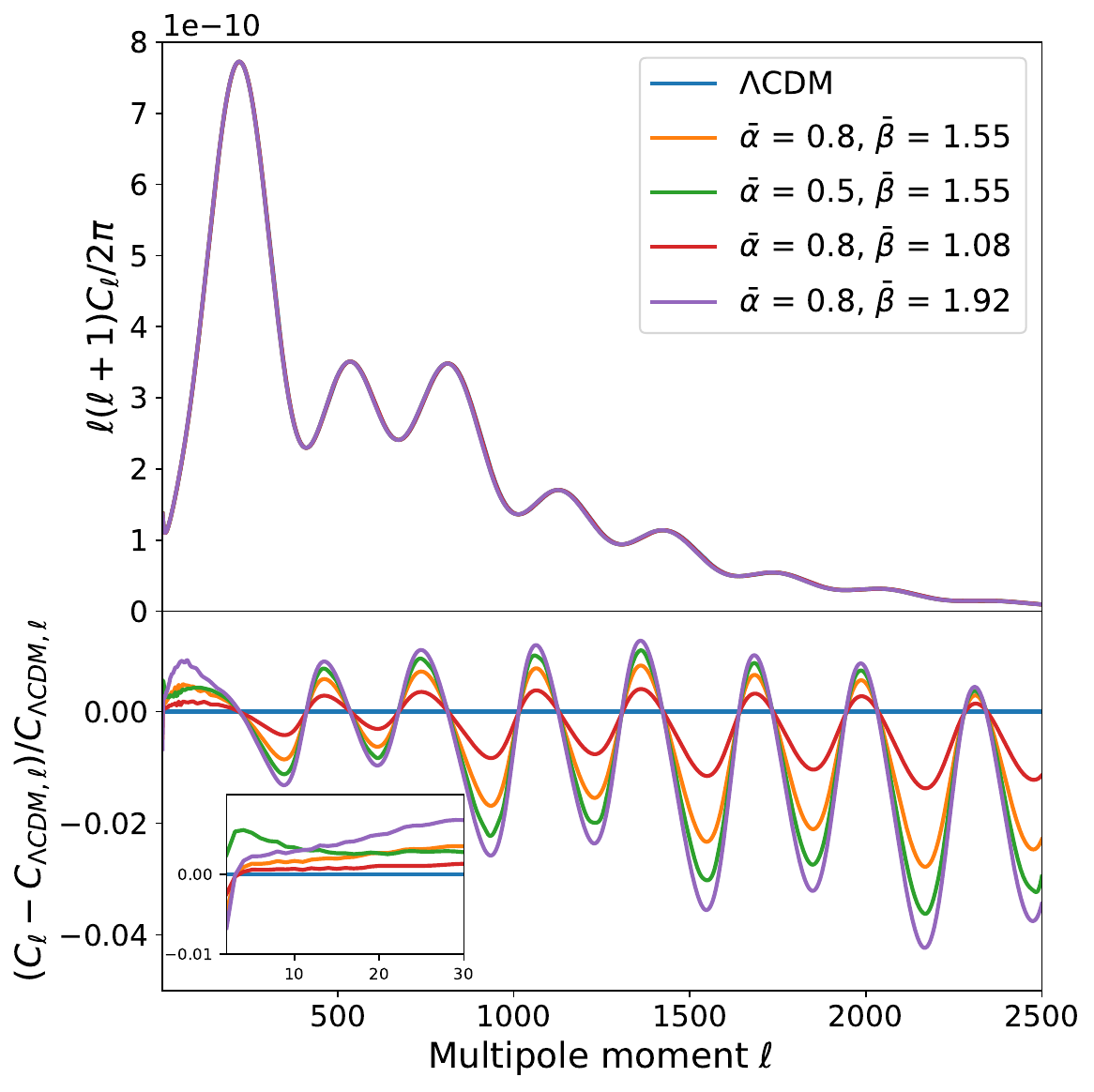}
\end{subfigure}\hfill
\begin{subfigure}{0.49\textwidth}
    \centering
    \includegraphics[width=\textwidth]{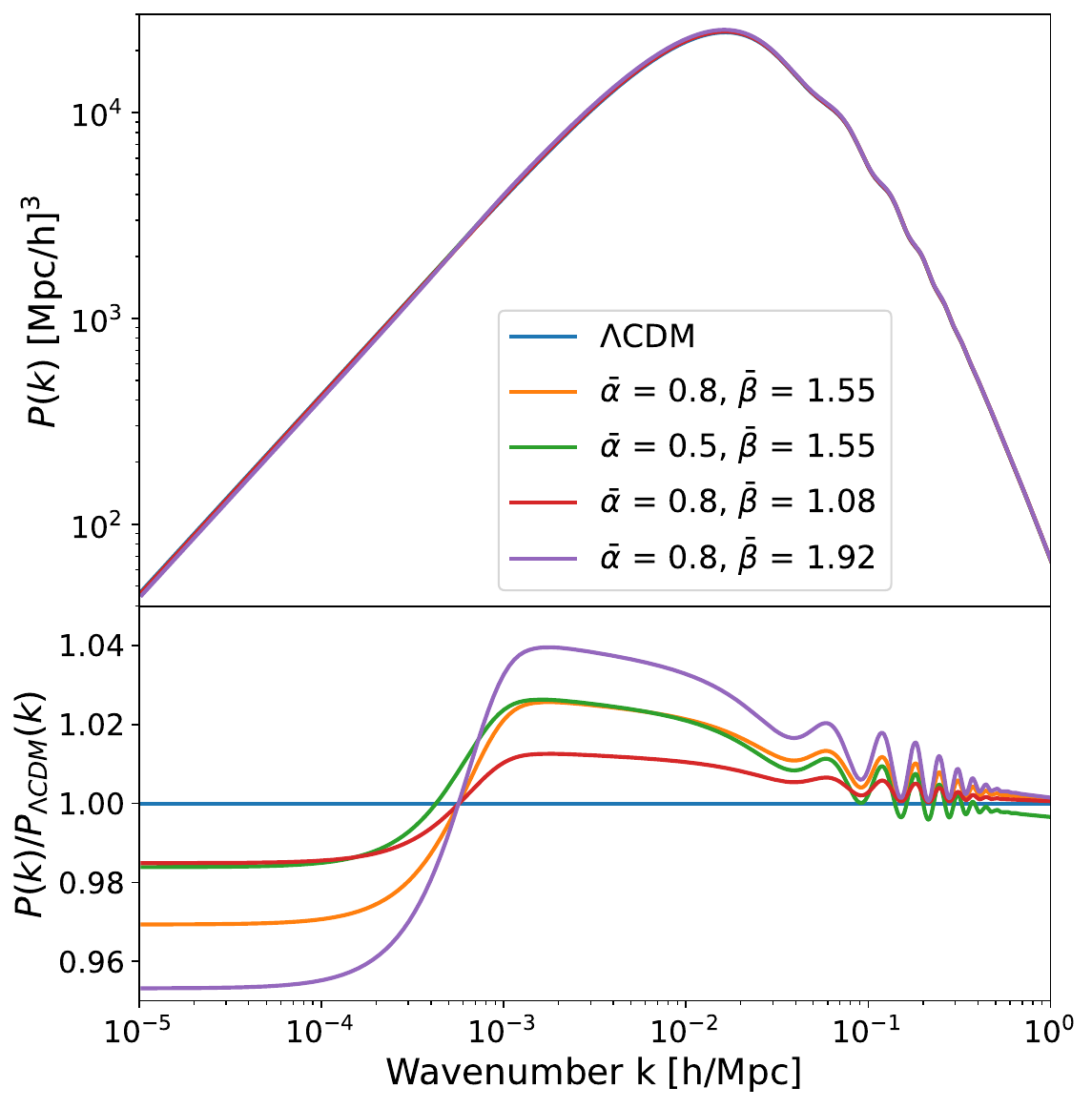}
\end{subfigure}
\caption{The same as in Fig. \ref{fig:spectra_neg}, but for $\bar\alpha>0$. The parameter $\varphi_{\rm ini}$ is chosen such that $\varphi(z=0) = 0$ -- see Sec. \ref{sec:method} for a detailed description of the scalar field's initial conditions.}
\label{fig:spectra_pos}
\end{figure}

In Figs. \ref{fig:spectra_neg} and \ref{fig:spectra_pos}, we show the CMB temperature and linear matter power spectra, together with the effective DE density and EoS parameter, for different values of $\bar\alpha$ and $\bar\beta$ in the $\alpha<0$ and $\alpha>0$ models, respectively. The latter background plots are aimed at easing the interpretation of the spectra. We fix the other parameters to the same fiducial values, but in the case of $\alpha>0$ we tune $\varphi_{\rm ini}$ to automatically satisfy the local constraints (cf. Sec. \ref{sec:method} for details). We find that values of $\bar{\alpha}$ and $\bar{\beta}$ of order unity, as preferred by current data, can lead to physically viable power spectra, with deviations from $\Lambda$CDM reaching the subpercent level. This is true for both positive and negative values of $\alpha$.  

The CMB temperature fluctuations depend on DE through the integrated Sachs-Wolfe (ISW) effect \cite{SachsWolfe1967}, gravitational lensing \cite{Lewis:2006fu} (e.g., by altering the energy fraction contained in pressureless matter, which has an impact on the clustering), and the distance to the last scattering surface (LSS). They are all sensitive to the late-time cosmic history, with the kernels of the ISW effect and CMB lensing peaking at $z\lesssim 1$ and $z\sim 2-3$, respectively. The former affect the low multipoles of the CMB spectrum, while the latter modifies the large-multipole range. 

The amplitude of these oscillations decreases with $\alpha$ and increases with $\beta$, both in the positive and negative $\alpha$ scenarios. The reason is that both larger values of $\beta$ and lower (or more negative) values of $\alpha$ lead to more DE density at low $z$, diluting matter away and thus reducing the lensing effect, which results in sharper peaks/troughs. Furthermore, the acoustic peaks are shifted towards the left. This is because a larger DE density leads to a larger Hubble rate at late times, which results in a smaller distance to the LSS. This increases the angular size of the peaks, with the corresponding multipole moment becoming smaller. We remark that, at the background level, the differences among the various curves only reside in the effective DE density, since the matter energy density, optical depth and the parameters characterizing the shape of the primordial power spectrum are fixed to the same values in all models.

As for the matter power spectrum, we observe that matter density perturbations at sub-horizon scales are enhanced with respect to super-horizon scales. This behavior is expected from an attractive  fifth force mediated by the scalar field between matter particles, which can only contribute to matter clustering at causally connected scales -- that is, inside the horizon. In the super-horizon limit ($k\ll \mathcal{H}$), $G_{\rm eff}\to G_N$, see Ref. \cite{DeFelice:2011hq}. Nevertheless, we note that the datasets we employ (see Sec. \ref{sec:method}) are largely insensitive to these scales. The main exception is the CMB low-multipole region, which exhibits large uncertainties due to cosmic variance (see, however, \cite{Toda:2026yum}). 

We focus on the effect of $\bar{\alpha}$ and $\bar{\beta}$ at subhorizon scales. The first parameter affects matter clustering both through the background expansion and the effective gravitational constant, cf. Eq. \eqref{eq:delta_growth}, whereas $\bar{\beta}$ only has an impact through the background functions. Using Eq. \eqref{eq:GeffTaylor}, which is valid in the limit of interest $|\alpha|\varphi^2\ll1$, one can see that, for $|\bar{\alpha}|$ not exceedingly large, $G_{\rm eff}$ grows with $\bar{\alpha}$ if $\bar{\alpha}>1/4$. For the values considered in Figs. \ref{fig:spectra_neg} and \ref{fig:spectra_pos}, $|\bar{\alpha}|>1/4$ and of order one, so $G_{\rm eff}$ grows in all cases with the absolute value of $\bar{\alpha}$\footnote{One can see that $G_{\rm eff}$ eventually decreases  again with $\bar{\alpha}$ for sufficiently large values of this parameter, $\bar{\alpha}\gg1$.}.

Let us start by analyzing the effect of these two parameters for the case $\bar{\alpha}<0$, focusing on the $P(k)$ spectra shown in Fig. \ref{fig:spectra_neg}. All models produce a larger DE density than that encountered in $\Lambda$CDM, since both larger values of $\bar{\beta}$ and more negative values of $\bar{\alpha}$ lead to a more pronounced phantom behavior in the past, following the thawing of the scalar field (see Eq. \eqref{eq:w_MDE_neg})\footnote{We note that for all these models $\rho_{\rm DE}\to V_0$ at sufficiently high redshift, with the value of $V_0$ taken to be the same throughout this discussion.}. This pushes all the corresponding curves of the ratio $P/P_{\Lambda{\rm CDM}}$ below unity. However, there is a competing effect, namely the enhancement of $P(k)$ due to the increase in $G_{\rm eff}$ induced by more negative values of $\bar{\alpha}$. This can partially counteract the aforementioned effect arising purely from the background evolution. In fact, the curve with $\bar\alpha = -1$ exhibits larger values of $P(k)$ than that corresponding to $\bar\alpha=-0.8$ -- despite having more DE -- due to the enhancement of $G_{\rm eff}$.

Now, we turn our attention to the model with $\alpha>0$, which we analyze by adjusting $\varphi_{\rm ini}$ such that $\varphi(z=0)=0$, thereby ensuring that the local constraints are satisfied. These results are displayed in Fig. \ref{fig:spectra_pos}. In this case, it is much more difficult to disentangle the effects of $\bar{\alpha}$ and $\bar{\beta}$ on the shape of $P(k)$, mainly because the value of $\varphi_{\rm ini}$ plays an important role in the phenomenology of the model and is adjusted in a non-trivial way -- for instance, larger values of $\bar{\alpha}$ are accompanied by lower values of $\varphi_{\rm ini}$ (see Fig. \ref{fig:phi_ini_color}), which clearly introduces a compensating effect at the level of $G_{\rm eff}$. The effective DE density curves of the various models cross, such that models with a smaller DE density in the past have a larger DE density at low redshifts. This likewise gives rise to compensating effects, but they are clearly subdominant. We therefore limit ourselves to presenting the plot, which reflects a complex and non-intuitive phenomenology. For instance, the matter power spectrum at sub-horizon scales appears largely insensitive to $\alpha$, whereas at super-horizon scales lower values of $\alpha$ lead to more structure --- despite the increase in DE density. Another non-trivial feature is the opposite effect $\beta$ has on large scales (where it suppresses structure formation) and small scales (where it enhances it).

Lastly, we briefly comment on the behavior of vector and tensor perturbations in this non-minimally coupled scalar field model. As in $\Lambda$CDM, vector perturbations decay with cosmic expansion and are therefore neglected in our analysis. On the other hand, although the running of the gravitational coupling affects the friction term in the equation governing the propagation of gravitational waves and, consequently, their amplitude \cite{Schutz:1986gp,Saltas:2014dha,Lombriser:2015sxa,Nishizawa:2017nef,Belgacem:2017ihm,Belgacem:2018lbp}, their propagation speed, $c_{gw}$, coincides with the speed of light \cite{Creminelli:2017sry,Ezquiaga:2017ekz}. Therefore, the model satisfies the very stringent constraint on $c_{gw}$, $|c_{gw}/c-1|<5\cdot 10^{-16}$ at $z\sim 0$, reported by the LIGO-VIRGO Collaboration in \cite{LIGOScientific:2017vwq}. Recently, measurements of the gravitational wave luminosity distances from the LIGO--Virgo--KAGRA Collaboration, together with the corresponding redshift inferred using two alternative indirect techniques (based on  features in the mass spectrum and statistical host galaxy association)
have already been employed to constrain modified gravity \cite{LIGOScientific:2026uyd}. However, the resulting constraints are still quite loose compared to those derived from the cosmological probes used in this paper (see, e.g., Fig. 5 in \cite{Lagos:2026pbr} for a comparison of the constraints obtained for the model with $\alpha<0$). Future measurements with the gravitational-wave interferometer LISA might provide direct constraints on the evolution of the effective gravitational coupling out to redshifts $z\sim\mathcal{O}(1\text{--}10)$ relative to its present-day value thanks to the detection of the signals with electromagnetic counterpart emitted by supermassive black hole mergers. These observations are expected to constrain the ratio $G_{\rm eff}(z)/G_N$ at the few-percent level \cite{Tamanini:2016zlh,LISACosmologyWorkingGroup:2019mwx}, providing further constraining power for the model under discussion.


\section{Methodology and data}\label{sec:method}

We have implemented the model in \texttt{hi\_class} \cite{Zumalacarregui:2016pph, Bellini:2019syt, Blas:2011rf} to solve the Einstein–Boltzmann equations governing the background and linear perturbations. This allows us to compute the theoretical quantities needed to establish the connection with observations. The correct functioning of our implementation in \texttt{hi\_class} has been duly tested by comparing the background output with that obtained with our Python solver\footnote{The code is available at \url{https://github.com/Joel-Argudo/Modified-Gravity-Background-Solver}.} as well as with the background and linear perturbation solutions obtained from our own implementation of the model in \texttt{CLASS}\footnote{The code is available at \url{https://github.com/Joel-Argudo/class_public}.}. For a thorough analysis of the consistency of \texttt{hi\_class} with other Einstein-Boltzmann solvers for modified theories of gravity, we refer the reader to \cite{Bellini:2017avd}.

We employ the following state-of-the-art cosmological data:
\begin{itemize}
    \item {\it Cosmic Microwave Background (CMB):} We use \textit{Planck} temperature (T), polarization (E) and cross (TE) power spectra \cite{Planck:2018vyg}. We use the \texttt{Commander} and \texttt{simall} likelihoods for low-$\ell$ and  \texttt{NPIPE (PR4) CamSpec} high-$\ell$ likelihood \cite{Efstathiou:2019mdh, Rosenberg:2022sdy}, as well as PR4 lensing \cite{Carron:2022eyg}.
    \item {\it Baryon Acoustic Oscillations (BAO):} We use measurements of the BAO scale, i.e. transverse angular comoving $D_M$, line-of-sight $D_H$ and angle-averaged $D_V$ distances relative to the sound horizon $r_d$ at redshifts $0.295\leq z \leq 2.33$ and their correlations from DESI DR2 as described in Table IV of \cite{DESI:2025zgx}.
    \item {\it Type Ia Supernovae (SNIa):} We include distance modulus measurements from the recalibrated DES SNIa compilation Dovekie \cite{DES:2025sig}, which supersedes the previous DES-Y5 \cite{DES:2024hip, DES:2024jxu}.
\end{itemize}
As discussed in Sec. \ref{sec:Geff}, the cosmological evolution of the scalar field induces a time variation of $G_{\rm eff}$ on local scales in the absence of a screening mechanism, causing the gravitational strength experienced by SNIa to differ from $G_N$. This, of course, affects the standardization process of these objects by shifting their absolute luminosities $L$ and, hence, their absolute magnitudes, $M$. In order to account for the impact of the modification of $G_{\rm eff}$, we follow \cite{Desmond:2019ygn} and study the influence of an unscreened fifth-force by modeling the luminosity as $L\propto G^C$ \footnote{See \cite{Amendola:1999vu,Garcia-Berro:1999cwy} for older works that, instead of simulating the effect of a varying Chandrasekhar mass $M_{\rm Ch}\sim G^{-3/2}$ on the supernovae light curves to infer the impact of such a variation on the relation $L(G)$, directly took for granted a concrete relation of the type $L\sim G^{-\gamma}$ with $\gamma>0$ of order unity.}, which leads to the following correction in the absolute magnitude,

\begin{equation}\label{eq:Mcorr}
M(G_{\rm eff}) = M(G_N) -2.5C\log_{10}\left(\frac{G_{\rm eff}}{G_N}\right)\,.
\end{equation}
We take $C=1.46$, as obtained in \cite{Desmond:2019ygn} from a linear fit to the relation $L(G)$ derived in \cite{Wright:2017rsu} (see the left panel of their Fig.~7). For $\alpha<0$, it makes little sense to consider this correction, since, as discussed above, the region of parameter space preferred by the cosmological data already violates the local constraints displayed in Sec.~\ref{sec:Geff}. Consequently, the viability of this model would require a screening mechanism, which is left unspecified in our study but is assumed to exist. In this setup, we just assume that $G_{\rm eff}=G_N$ in the SNIa environment. For $\alpha>0$, instead, it is possible to satisfy the local constraints without relying on a screening mechanism. Nevertheless, there can still be deviations of $G_{\rm eff}$ from $G_N$ in the local environments of SNIa at past redshifts. Thus, the shift in Eq.~\eqref{eq:Mcorr} could in principle play a role. However, we find that it is negligible, since in this setup $|G_{\rm eff}/G_N-1|\ll 1$ for all $z$. For the best-fit model, we find $|G_{\rm eff}/G_N-1|\sim 0.2\,\%$ at most and, therefore, the induced shift in the absolute magnitude is much smaller than the uncertainties of the SNIa apparent magnitudes, as we will explicitly show in Sec. \ref{sec:results}. Thus, also in this case, we can safely neglect this correction. 

\begin{figure}[t]
    \centering
    \includegraphics[width=0.55\linewidth]{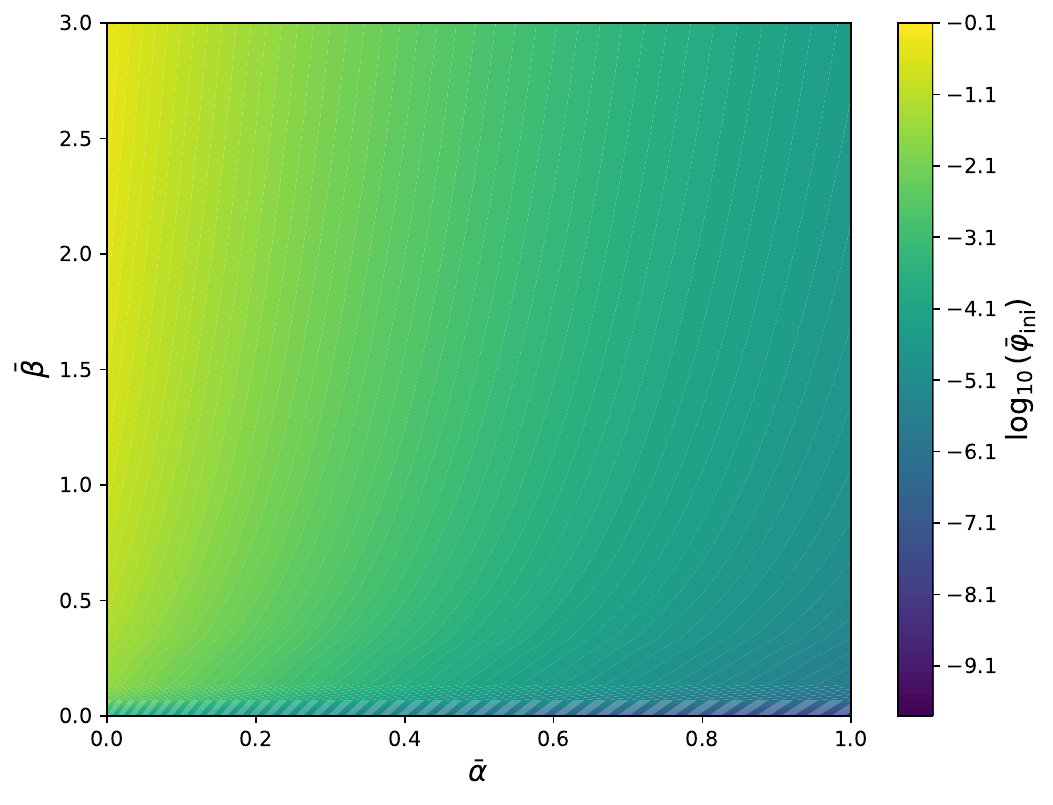}
    \caption{Color plot of the initial value of the scalar field $\bar{\varphi}_{\rm ini}$ required to fulfill the condition $\bar{\varphi}(z=0) \approx 0$, as a function of $\bar{\alpha}$ and $\bar{\beta}$, for the model $\alpha>0\,+\,$\discretionary{}{}{}LC. It is obtained from the grid interpolation described in Sec. \ref{sec:method}. The remaining parameters are fixed to the mean values of Table \ref{tab:results}.}
    \label{fig:phi_ini_color}
\end{figure}

We perform the Monte Carlo Markov Chain (MCMC) analysis using the standard Metropolis-Hastings \cite{Metropolis:1953am, Hastings:1970aa} algorithm fixing Gelman-Rubin \cite{Gelman:1992zz} stopping criterion $R-1=0.03$ with \texttt{Cobaya} \cite{Torrado:2020dgo} in two distinct scenarios:

\begin{enumerate}
    \item[a)] {{\it Non-minimal coupling with $\alpha<0$}: Initial conditions are set to $\varphi_{\rm ini} =0$ and $ \varphi_{\rm ini}^\prime = 0$. As discussed in Sec. \ref{sec:background}, considering $\varphi_{\rm ini}\ne 0$ does not lead to any appreciable differences for $|\bar{\alpha}|>0.08$, which is the region of parameter space preferred by the data. This is because $\bar{\varphi}_{\rm ini}\to 0$ before matter-radiation equality. Consequently, for these values of the coupling and in the realistic regime $|\bar{\alpha}|\bar{\varphi}_{\rm ini}^2\ll 1$, no significant features are imprinted on the observables used in this work to constrain the model. A more general approach would leave $\bar{\varphi}_{\rm ini}$ free in the MCMC analysis. However, this procedure would not introduce any significant difference in our results and, in addition, would complicate the comparison with previous works in the literature, which made use of initial conditions closer to those employed in this work by us.}
    \item[b)] {\it Non minimal coupling with $\alpha>0$ satisfying local constraints}, which we dub ``$\bm{\alpha>0}\,\bm+\,$\discretionary{}{}{}\textbf{LC}'': Initial conditions are set to $\varphi_{\rm ini}^\prime = 0$ and $\varphi_{\rm ini}$ tuned to satisfy the boundary condition $\varphi(z=0) \sim 0$, which ensures all local gravitational constraints are fulfilled (see Sec. \ref{sec:Geff}). However, instead of performing a shooting algorithm at each step of the Monte Carlo run, which would slow down the sampling, $\varphi_{\rm ini}$ and $H_0$ are estimated using a cubic interpolation over a grid in the parameter space spanned by $\omega_{\rm b}+\omega_{\rm cdm}$, $\alpha$, $\beta$, $V_0$, and the corresponding values of $\varphi_{\rm ini}$ and $H_0$\footnote{\texttt{hi\_class}, for the scalar-field potential and minimal coupling considered in this paper, takes as input the parameters $(\tau,n_s,A_s,\omega_{\rm b},\omega_{\rm cdm},H_0,\alpha,\beta,\varphi_{\rm ini})$ and uses a shooting algorithm to compute the value of $V_0$ required to reach the specified value of $H_0$. For the MCMC of $\alpha>0+$LC, we vary $(\tau,n_s,A_s,\omega_{\rm b},\omega_{\rm cdm},V_0,\alpha,\beta)$, interpolate $\varphi_{\rm ini}$ (so that $\varphi(z=0)\sim 0$) and $H_0$, and feed these values to \texttt{hi\_class}, treating $V_0$ as an initial guess that is tuned to precisely match the value of $H_0$. The use of the grid makes this recursive process extremely fast.}. The grid points used for the interpolation are themselves obtained using a shooting algorithm\footnote{The file with the grid can be found in \url{https://github.com/Joel-Argudo/phi_ini_grid}.}, and we perform the interpolation using the \texttt{RegularGridInterpolator} class from the \texttt{scipy} Python package. Using this procedure, we manage to keep $\bar\varphi(z=0)$ at the $\mathcal{O}(10^{-5})$ level (cf. Table \ref{tab:results}). For illustrative purposes, we show in Fig. \ref{fig:phi_ini_color} the interpolated value of $\bar{\varphi}_{\rm ini}$ as a function of $\bar\alpha$ and $\bar\beta$, obtained by fixing sensible values for the remaining parameters.
\end{enumerate}

We use \texttt{halofit} to include non-linearities \cite{Smith:2002dz,Takahashi:2012em}, assuming mild departures from GR and $\Lambda$CDM, as done in previous works (see, e.g., \cite{Ye:2024ywg,Wolf:2024stt,Wolf:2025jed,Garcia-Garcia:2026nzy}). A more rigorous approach should include modified gravity effects (see e.g. \cite{Yi:2026sqj}). This is beyond the scope of this paper and is left for future work. 

We adopt flat, sufficiently broad priors on the sampled parameters, i.e., we take them to be largely uninformative, much broader than the likelihood. Our Markov chains are processed and analyzed with the aid of \texttt{GetDist} \cite{Lewis:2019xzd}. In addition to fitting the $\Lambda$CDM and non-minimally coupled models in the two configurations described above, we also consider the Chevallier-Polarski-Linder (CPL) parametrization, also known as $w_0w_a$CDM, of the DE EoS parameter,
$w_{\rm DE}(a)=w_0+w_a(1-a)$ \cite{Chevallier:2000qy,Linder:2002et}, which has been widely used to phenomenologically describe possible dynamics of the DE sector (see, e.g., \cite{DESI:2025zgx,Gonzalez-Fuentes:2025lei,Gonzalez-Fuentes:2026rgu}). It can mimic quite well the behavior encountered in freezing and thawing models of DE, while also allowing for a phenomenological description of scenarios involving a crossing of the phantom divide, either from phantom to quintessence -- as preferred by current data in the absence of new physics before recombination, which is required to accommodate the data when the distance-ladder calibration from SH0ES \cite{Riess:2021jrx} is taken into account \cite{Poulin:2024ken,Gonzalez-Fuentes:2026rgu,Schoneberg:2026buf} -- or vice versa. We use the CPL parametrization as a baseline model, as is commonly done in the literature.

We assess the goodness of fit of each model with the minimum chi-square $\chi_{\rm min}^2 = -2\log \mathcal{L_{\rm max}}$, where $\mathcal{L_{\rm max}}$ is the maximum likelihood. We note, however, that the model with non-minimal coupling with $\alpha<0$ has two more parameters than $\Lambda$CDM (namely, $\alpha$ and $\beta$)\footnote{This is also the case of CPL, which includes $w_0$ and $w_a$ on top of the $\Lambda$CDM parameters.}. In the MCMC analysis of the $\alpha>0$+LC model, we vary exactly the same number of parameters as in the $\alpha<0$ scenario. One could therefore conclude that, also in this case, the model has two additional parameters relative to $\Lambda\mathrm{CDM}$. However, this parameter counting is partly a consequence of the sampling strategy adopted to explore the parameter space efficiently. In principle, one could instead allow $\varphi_{\rm ini}$ to vary freely in the MCMC and include the contribution of the local constraints described in Sec. \ref{sec:Geff} directly in the likelihood. Treating $\varphi_{\rm ini}$ as an additional free parameter on top of $\alpha$ and $\beta$ would, however, be computationally prohibitive, since the resulting likelihood would be non-negligible only within an extremely narrow region of parameter space. For this reason, we adopt the grid method described above, which effectively fixes $\varphi_{\rm ini}$ as a function of the other parameters entering the fit. This procedure automatically restricts the sampling to the narrow region of parameter space compatible with the local constraints. Although $\varphi_{\rm ini}$ is not varied freely in the MCMC analysis, it could nevertheless be regarded as an effective degree of freedom that is tuned to satisfy these constraints at each step of the Monte Carlo. From a conservative perspective, we could therefore treat $\varphi_{\rm ini}$ as an additional parameter and assign three extra degrees of freedom to the $\alpha>0$+LC model relative to $\Lambda\mathrm{CDM}$, rather than two. For the sake of generality, for the $\alpha>0$+LC model we provide the results under both parameter-counting prescriptions.

In any case, it is important to bear in mind the presence of additional parameters when performing the model comparison in order to properly penalize overfitting. To compare the fitting performance of models with a different number of parameters $n_p$ we employ two different methods. First, we use the Akaike information criterion (AIC) \cite{Akaike:1100705}, which for the large number of data points considered in this study is simply defined as
\begin{equation}\label{eq:AIC}
    {\rm AIC} =  \chi^2_{\rm min}+2n_p\,.
\end{equation}
More concretely, we define the relative AIC difference
of model $i$ and the $\Lambda$CDM, i.e., $\Delta\rm{AIC} \equiv AIC_{\rm \Lambda CDM}-AIC_{i}$. The $\Lambda$CDM model is naturally taken as a benchmark. A positive difference $\Delta {\rm AIC}$ is obtained if the penalization factor $2n_p$ that accounts for the use of additional parameters in model $i$ is successfully compensated by a lower $\chi^2_{\rm min}$. Hence, such positive differences are encountered when model $i$ performs better than the standard model even after applying Occam's razor. Values $2 \leq \Delta\textrm{AIC} < 6$ are usually considered to show \textit{positive evidence} according to Jeffreys' scale \cite{Jeffreys1961,Trotta:2005ar,Trotta:2008qt}. Values $\Delta\textrm{AIC} < 2$ entail only \textit{weak evidence} for model $i$; values in the range $6 \leq \Delta\textrm{AIC} < 10$ indicate \textit{strong evidence}, whereas values $\Delta\textrm{AIC}>10$ point to a \textit{very strong} statistical evidence for the model $i$ compared to $\Lambda$CDM. Although, strictly speaking, AIC differences cannot be interpreted universally through Jeffreys' scale (see  \cite{Nesseris:2012cq,Gonzalez-Fuentes:2026rgu} for details), it can work in many situations as a good approximation. It can be considered a proxy of two times the natural logarithm of the  Bayes factor, and it is insensitive to the choice of priors, as long as they are much wider than the likelihood\footnote{More rigorous methods, such as the Frequentist-Bayesian approach described in   \cite{Jenkins:2011va,Keeley:2021dmx,Amendola:2024prl} and recently applied in \cite{Sakr:2025chr,Gonzalez-Fuentes:2026rgu}, are also largely insensitive to broad priors and do not rely on the validity of Jeffreys' scale. However, they require simulations and can be computationally very expensive in the presence of strong non-Gaussianities in the posterior distributions or when the full CMB likelihood is considered, as in the present study. This is why we opt to stick to AIC here.}. This is a welcome feature when theoretical considerations or observational data (beyond those considered in our fitting analysis) are vague in setting robust prior boundaries in parameter space, as it might be in the model under study. 

\begin{figure}[t!]
    \centering
    \includegraphics[width=0.6\linewidth]{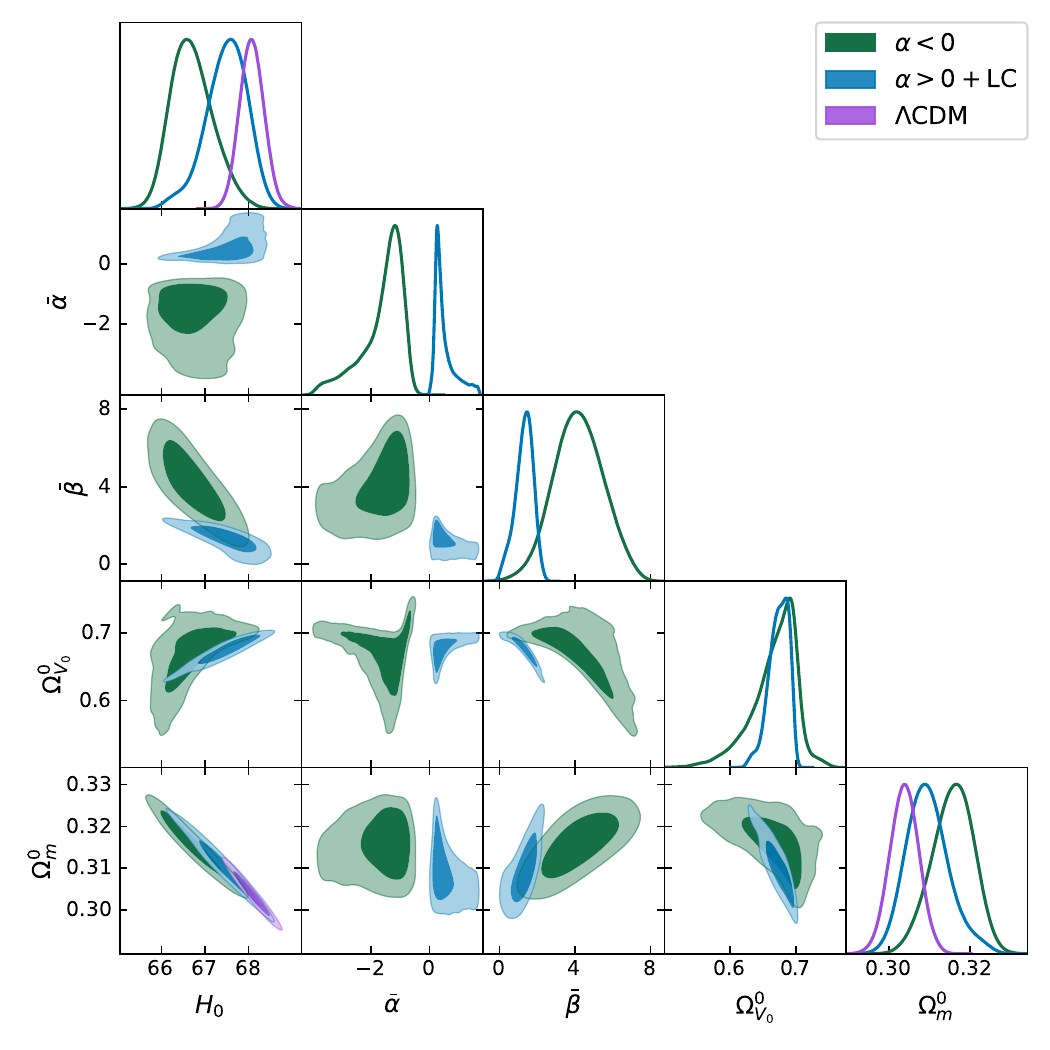}
    \caption{Triangle plot showing the 68\% and 95\% CL marginalized constraints in some of the most relevant planes of both models with non-minimal coupling and $\Lambda$CDM. $H_0$ is given in km/s/Mpc.}
    \label{fig:contours}
\end{figure}

Second, we employ the likelihood-ratio test \cite{NeymanPearson1933} for model comparison, in order to assess whether it leads to the same conclusions as the AIC, thereby providing an independent consistency check. By Wilks' theorem \cite{Wilks1938}, the distribution of the difference between minimum $\chi^2$ values of model $i$ and $\Lambda$CDM is a $\chi^2_{\nu}$ with $\nu=n_{p,i}-n_{p,\Lambda{\rm CDM}}$ degrees of freedom. We first compute the $p$-value associated with the value of $\Delta\chi^2=\chi^2_{{\rm min},\Lambda{\rm CDM}}-\chi^2_{{\rm min},i}$ obtained in our fitting analysis. Then, we determine the confidence level at which we can reject the null hypothesis -- which in our case is the validity of $\Lambda$CDM -- and finally translate these $p$-values into a number of sigmas, $\xi$, by solving the equation 

\begin{equation}\label{eq:pvalue}
p{\rm -value} = 1-\frac{1}{\sqrt{2\pi}}\int_{-\xi}^{\xi}e^{-y^2/2}dy\,.
\end{equation}
This is a fairly standard approach, recently adopted, for instance, by the DESI collaboration in their cosmological analysis \cite{DESI:2025zgx}. We define the exclusion level of the standard model (relative to the considered model $i$)  as $E_{\Lambda{\rm CDM}}\equiv \xi\times\sigma$. As will become apparent in the next section, both the AIC and $E_{\Lambda{\rm CDM}}$ lead to fully consistent interpretations of the results, thereby reinforcing the main conclusions of our work. We remark, however, that Wilks' theorem assumes a well-behaved quadratic expansion in $\ln \mathcal{L}$ (i.e., no significant deviations from Gaussianity) and a parameter space such that it does not reduce to the null hypothesis at the boundary. This is not satisfied in our setup (cf. Fig. \ref{fig:contours}), so the values obtained should be interpreted with caution as estimates of the true value\footnote{Again, a more detailed estimation of $E_{\Lambda{\rm CDM}}$ would require very time-consuming simulations, which we avoid in this work.}.

\section{Results and discussion}\label{sec:results}

\begin{table*}[t!]
    \renewcommand{\arraystretch}{1.1}
   \centering
   \resizebox{\textwidth}{!}{%
   \begin{tabular}{c|cccc}
   \hline
    \textbf{Parameter} & $\mathbf{\Lambda}$\textbf{CDM} & \textbf{CPL} & \textbf{NMC} ($\alpha<0$) & \textbf{NMC} ($\alpha>0\; +\;$\textbf{LC}) \\ 
   \hline \hline
    $H_0\;[\rm km\;s^{-1}\;Mpc^{-1}]$ & $68.07\pm 0.28$ & $67.33\pm 0.54$ & $66.71^{+0.40}_{-0.56}$ & $67.49^{+0.53}_{-0.40}$ \\
    $10^2\omega_{\rm b}$ & $2.230\pm 0.012$ & $2.222\pm 0.012$ & $2.233\pm 0.012$& $ 2.230\pm 0.012$\\
    $10\omega_{\rm cdm}$ & $1.178\pm 0.006$ & $1.190\pm 0.008$ & $1.175\pm 0.006$ & $1.181\pm 0.006$ \\
    $\ln (10^{10} A_s)$ & $3.046^{+0.013}_{-0.015}$ & $3.036\pm 0.014$ & $3.041\pm 0.014$ & $3.046\pm 0.014$ \\
    $n_s$ & $0.967\pm 0.003$ & $0.965\pm 0.004$ & $0.968\pm 0.003$ & $0.969\pm 0.003$ \\
    $\tau_{\rm reio}$ & $0.058^{+0.006}_{-0.007}$ & $0.053\pm 0.007$ & $0.056\pm 0.007$ & $0.059\pm 0.007$ \\
    $w_0$ & $-$ & $-0.811\pm 0.054$ & $-$ & $-$ \\
    $w_a$ & $-$ & $-0.70\pm 0.21$ & $-$ & $-$ \\
    $\bar{\alpha}$ & $-$ & $-$ & $-1.59^{+0.84}_{-0.25}$ & $0.499^{+0.07}_{-0.37}$ \\
    $\bar{\beta}$ & $-$ & $-$ & $4.2\pm 1.3$ & $1.30^{+0.55}_{-0.39}$ \\
    \hline
    $\Omega_{m}^0$ & $ 0.304\pm 0.004$ & $0.313\pm 0.005$ & $0.316^{+0.006}_{-0.005}$& $0.310^{+0.004}_{-0.006}$\\
    $r_{\rm d}\; \left[\mathrm{Mpc} \right]$ & $147.76\pm 0.19$ & $147.54\pm 0.21$ & $147.82\pm 0.19$ & $147.71\pm 0.19$ \\
    $\sigma_{12}(z=0)$ & $0.794\pm 0.006$ & $0.801\pm 0.007$ & $0.802\pm 0.007$ & $0.797\pm 0.006$\\
    $S_8$ & $0.811\pm 0.008$ & $0.824\pm 0.009$ & $0.823\pm 0.010$ & $0.816\pm 0.009$\\
    $\bar{\varphi}_{c} (z=0)$ & $-$ & $-$ & $-0.28^{+0.15}_{-0.12}$ & $\left(\,-2.5^{+5.1}_{-2.5}\,\right)\cdot 10^{-5}$ \\
    $\dot{\bar{\varphi}}_{c} (z=0)\; \left[\mathrm{yr}^{-1}\right]$ & $-$ & $-$ & $\left(\,-8.1^{+6.3}_{-3.3}\,\right)\cdot 10^{-19}$ & $\left(\,-6.2^{+2.2}_{-2.9}\,\right)\cdot 10^{-19}$ \\ 
    $\Omega^0_{V_0}$ & $0.696\pm 0.004$ & $0.683\pm 0.005$ & $0.668^{+0.038}_{-0.020}$ & $0.674^{+0.019}_{-0.011}$ \\
    $\frac{\dot{G}_{\rm{eff}}}{G_{\rm{eff}}} (z=0) \; \left[\mathrm{yr}^{-1}\right]$ & $0$ & $0$ & $\left(\,3.8^{+1.3}_{-2.9}\,\right)\cdot 10^{-11}$ & $\left(\,0.496^{+0.08}_{-0.70}\,\right)\cdot 10^{-15}$\\
    $\frac{G_{\rm{eff}}}{G_{N}} (z=0)$ & $1$ & $1$ &  $1.33^{+0.11}_{-0.20}$& $1.000001\pm 0.000017$ \\
    $\gamma^{\rm PPN} - 1$ & $0$ & $0$ & $-0.274^{+0.055}_{-0.097}$ & $\left(\,-0.4\pm 5.1\,\right)\cdot 10^{-5}$\\
    $\beta^{\rm PPN} - 1$ & $0$ & $0$ & $0.050^{+0.009}_{-0.031}$ & $\left(\,-0.3\pm 3.9\,\right)\cdot 10^{-5}$\\
    
    \hline
    $\chi^2_{\rm min}$ & $12622.30$ & $12611.22$ & $12613.51$ & $12615.45$ \\
    $\Delta{\rm AIC}$ & $-$ & $7.08$ & $4.79$ & $2.85$ ($0.85$) \\
    $p-$value & $-$ & $0.00393$ & $0.0124$ & $0.0326$ ($0.0769$) \\
    $E_{\Lambda \mathrm{CDM}}$ & $-$ & $2.88\sigma$ & $2.50\sigma$ & $2.14\sigma$ ($1.77\sigma$)\\
   \end{tabular}}
   \caption{Mean and $68\%$ CL values for the main cosmological parameters left free in the MCMC runs (upper panel) and the most important derived parameters (middle panel), including those that parameterize the local deviations with respect to GR at $z=0$. The acronym NMC refers to the model with non-minimal coupling studied in this paper. In the lowest panel, we display the values of the statistical quantities employed to assess the relative fitting performance of the models, which we duly discuss in Sec. \ref{sec:results}. For $\alpha>0+$LC, we display $\Delta$AIC, $p$-value and $E_{\Lambda\mathrm{CDM}}$ considering 2 extra parameters, and, in parentheses, considering $\varphi_{\rm ini}$ as an additional degree of freedom. We refer the reader to Sec. \ref{sec:method} for technical details on the definition and use of these indicators.}
   \label{tab:results}
\end{table*}

\begin{figure}[t!]
    \centering
    \includegraphics[width=\linewidth]{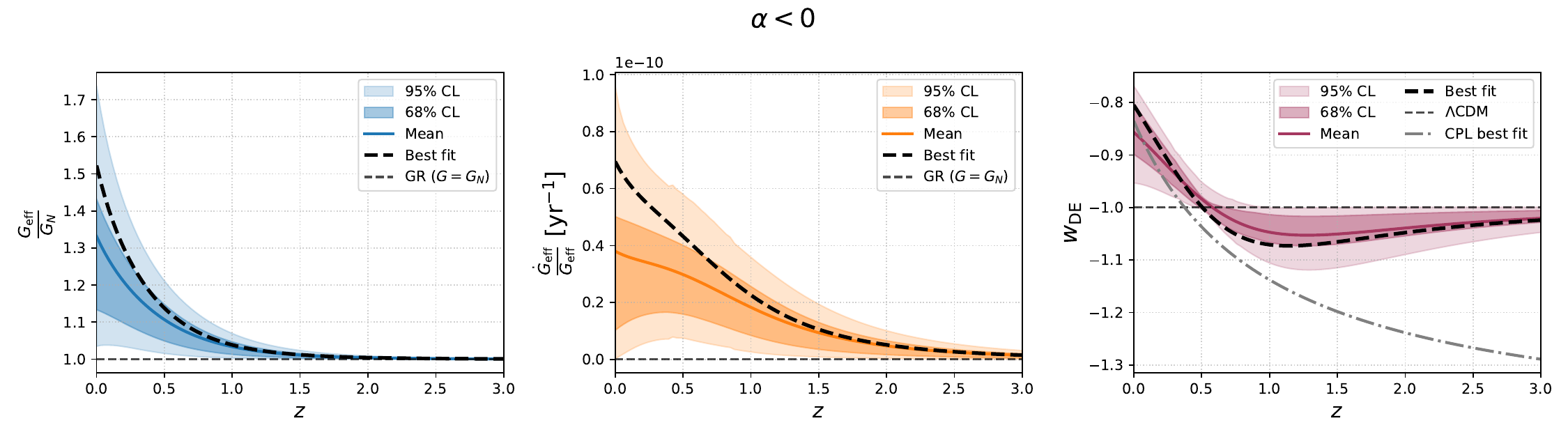}
    \includegraphics[width=\linewidth]{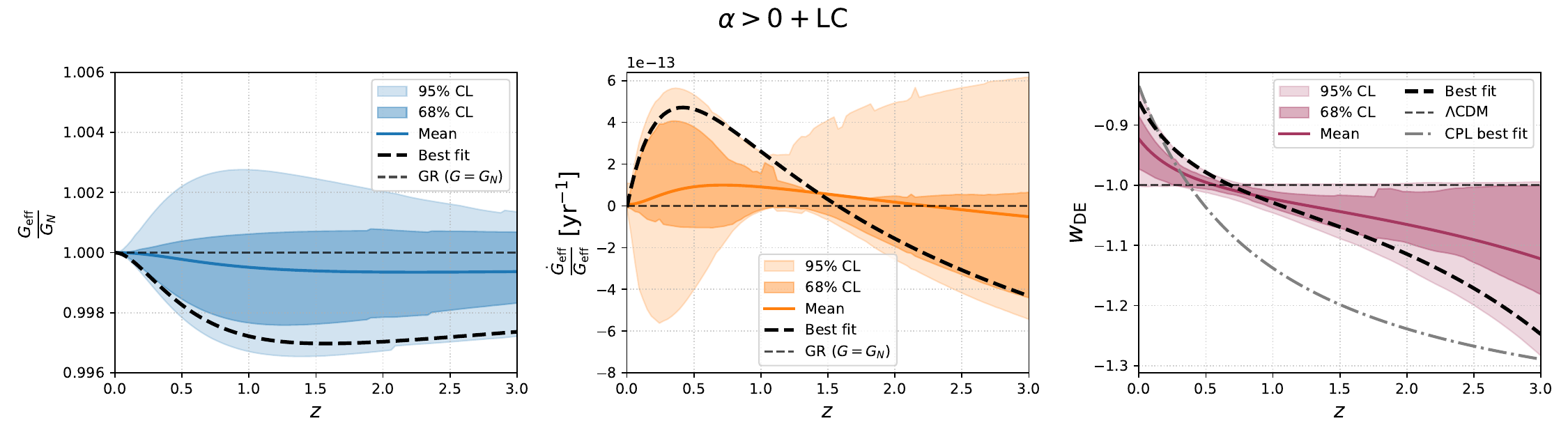}
    \caption{{\it Upper plots:} $68\%$ and $95\%$ CL bands for $G_{\rm eff}(z)/G_N$,  $\dot{G}_{\rm eff}(z)/G_{\rm eff}(z)$ and $w_{\rm DE}(z)$, respectively, obtained with $\alpha<0$. The corresponding best-fit curves are displayed as black dashed lines, while those for $\Lambda$CDM are displayed as gray dashed lines. In the third column, we also include the best-fit curve for CPL. {\it Lower plots:} Same for $\alpha>0+$LC. Note that the scale of $\dot{G}_{\rm eff}/G_{\rm eff}$ (i.e., $10^{-13}$ yr$^{-1}$) is chosen for visualization purposes. It is not indicative of its value at $z=0$, where this ratio is of order $10^{-15}$ yr$^{-1}$, in full agreement with the local constraint \eqref{eq:Gdot}, as displayed in Table \ref{tab:results}.} 
    \label{fig:bands}
\end{figure}

\begin{figure}[t!]
    \centering
    \includegraphics[width=\linewidth]{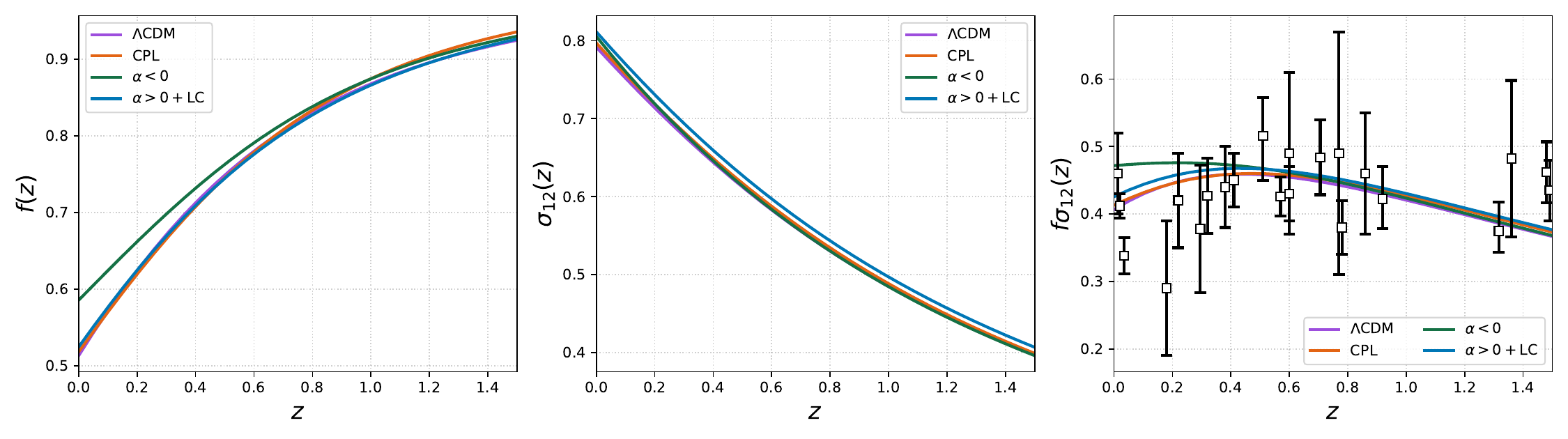}
    \caption{Growth rate $f=d\ln \delta_{\rm m}/d\ln a$, variance of mass fluctuations $\sigma_{12}$ at scale $R_{12}=12$ Mpc and $f \sigma_{12}$ as a function of redshift for the best fits of the four models under study. We have computed the functions at a scale $k=0.05\, \mathrm{Mpc^{-1}}$, which lies well within the linear regime covered by galaxy surveys, and have explicitly checked that the result is stable along several linear scales. In the rightmost plot, we also display the points from ALFALFA \cite{Avila:2021dqv}, 6dFGS+SDSS \cite{Said:2020epb}, GAMA \cite{Simpson:2015yfa, Blake:2013nif}, WiggleZ \cite{Blake_2011}, DR12 BOSS \cite{Gil-Marin:2016wya}, VIPERS \cite{Mohammad:2018mdy}, VVDS \cite{Guzzo:2008ac,Song:2008qt}, FastSound \cite{Okumura:2015lvp}, eBOSS Quasar \cite{eBOSS:2020gbb}, DESI DR1 ShapeFit \cite{DESI:2024jxi} and Stiskalek \cite{Stiskalek:2025oht}. }
    \label{fig:structuregrowth}
\end{figure}

\begin{figure}[t!]
    \centering
    \includegraphics[width=0.4\linewidth]{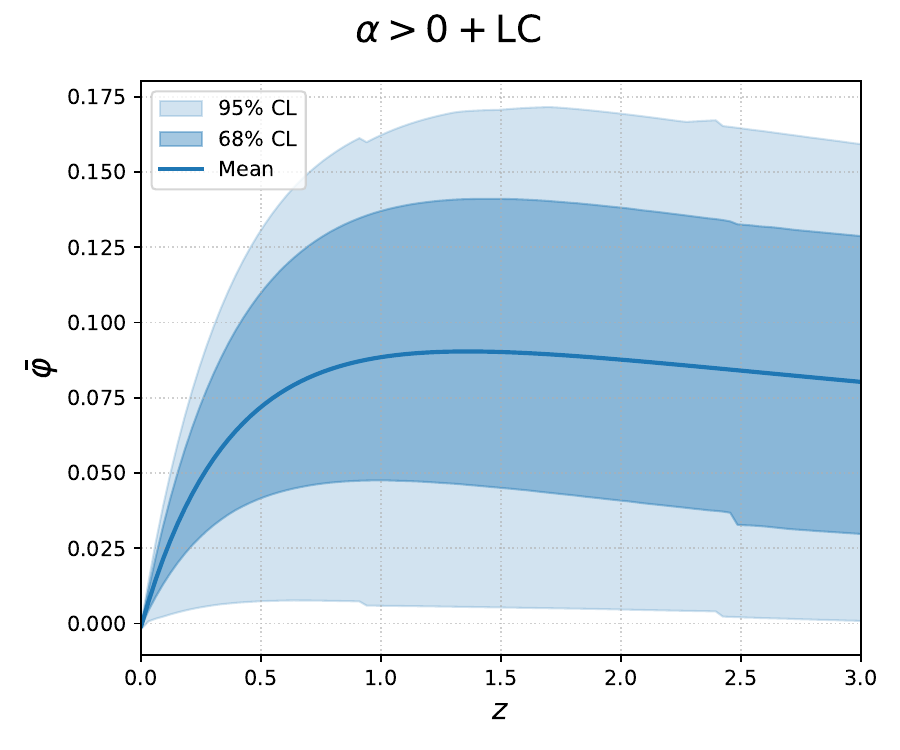}
    \includegraphics[width=0.4\linewidth]{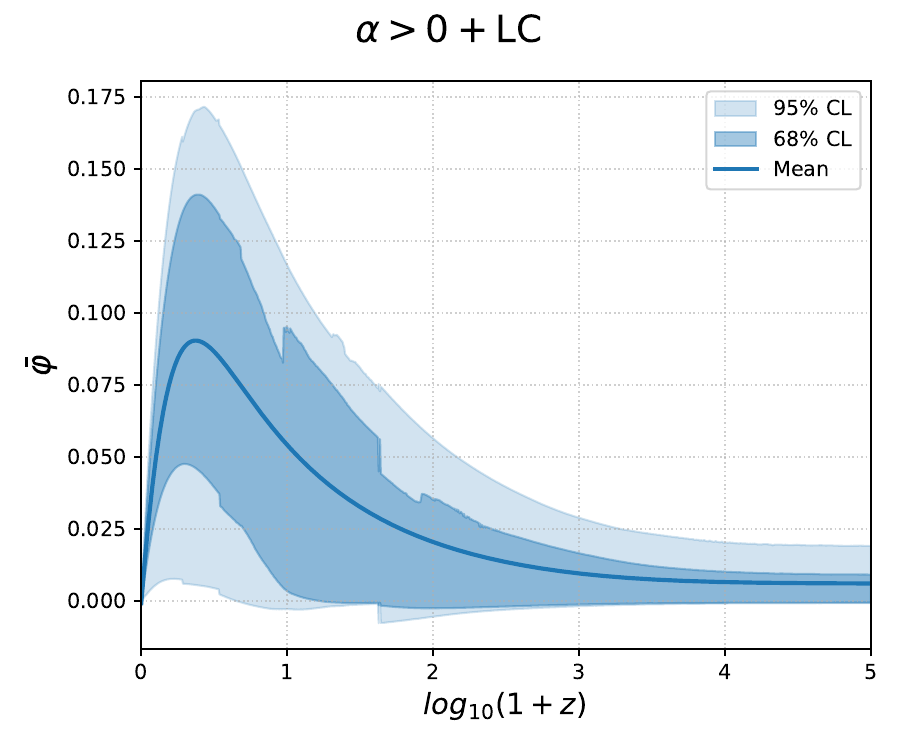}
    \caption{{\it Left plot:} Same as Fig. \ref{fig:bands}, but for $\bar{\varphi}$, in the same redshift range. The redshift $z_m$ at which the scalar field reaches its maximum is consistent with Fig. \ref{fig:map}, given the range of values of $\bar{\alpha}$ and $\bar{\beta}$ preferred by the data (cf. Table \ref{tab:results}); {\it Right plot:} Here, we simply extend the redshift range of the left plot to encompass a substantial fraction of the RDE. }
    \label{fig:bands_pos_phi}
\end{figure}

In Table \ref{tab:results}, we show the parameter constraints obtained for the various models from the corresponding MCMC analyses, following the pipeline described in the previous section. In Fig. \ref{fig:contours}, we display the one- and two-dimensional marginalized posteriors for some of the most relevant parameters of the non-minimally coupled model, whereas in Figs. \ref{fig:bands}-\ref{fig:bands_pos_phi} we reconstruct several cosmological functions of interest, which will help us interpret the results and facilitate the discussion.

Let us start the discussion of the results obtained for the model with non-minimal coupling and $\alpha<0$. We obtain a value of $\chi^2_{\rm min}$ that is roughly 9 units below that found for $\Lambda\mathrm{CDM}$, indicating that the model provides a non-negligible improvement in the description of the data relative to the standard model.

We remark once again that this model is unable to produce a crossing of the phantom divide close to that preferred by the data without strongly violating the local constraints, given in Eqs. \eqref{eq:gammaPN}, \eqref{eq:betaPN}, and \eqref{eq:Gdot}. This is apparent from the posterior constraints on the locally measured value of the gravitational coupling and its time derivative, as well as on the PPN parameters. Hence, this model is viable only in the presence of a screening mechanism capable of rendering deviations from GR sufficiently small locally. The cosmological analysis of the negative-$\alpha$ scenario therefore makes sense only if such a screening mechanism is assumed. In this setup, our constraints on $G_{\rm eff}$ and $\dot{G}_{\rm eff}/G_{\rm eff}$ must be interpreted as being of cosmological nature, affecting the clustering of matter on cosmic linear scales through Eq. \eqref{eq:delta_growth}, while leaving the physics of the Solar System untouched. The upper panel of Fig. \ref{fig:bands} shows the evolution of these quantities as a function of redshift. The modified-gravity effects are switched off at $z\gg 1$ because the scalar field is frozen -- we have thawing gravity. In the late Universe, however, the CMB+BAO+SNIa data set allows for very significant departures from GR at present, with the effective gravitational strength increasing monotonically with the expansion and being allowed to deviate by as much as $\sim$70\% at the $ 95\%$ CL. This has, of course, an impact on the observable $f\sigma_{8}(z)$, which can be measured from peculiar velocities at low redshifts and, over a broader redshift range, through the analysis of redshift-space distortions in galaxy surveys. Here, $f(z)=d\ln\delta_m/d\ln a$ is the growth rate, while $\sigma_{8}$ is the root-mean-square matter fluctuation at the scale $R_{8}=8/h$ Mpc. In the rightmost panel of Fig. \ref{fig:structuregrowth}, we display the best-fit curves of the function $f\sigma_{12}(z)$ for the various models, where the smoothing scale is set to $R_{12}=12$ Mpc, making the quantity insensitive to the value of $h$. This allows us to compare all the models on equal footing, considering their effect on the same scale range; see Refs. \cite{Sanchez:2020vvb,Forconi:2025cwp} for details. It is clear from the plot that the best-fit model with $\alpha<0$ leads to substantially larger values of $f\sigma_{12}(z)$ below $z\sim 0.5$ than the other models, and in particular than $\Lambda\mathrm{CDM}$. This exacerbates the tension with the clustering data already present in the standard model \cite{Macaulay:2013swa,Gomez-Valent:2017idt,Nesseris:2017vor,Gomez-Valent:2018nib,Benisty:2020kdt,Nunes:2021ipq,Nguyen:2023fip,Toda:2024fgv,Toda:2026yum}, as evidenced by the fact that most of the data points in the figure lie below the theoretical curves at those low redshifts, with the only exception being the recent DESI DR1 measurement \cite{DESI:2024jxi}. These findings are fully consistent with the conclusions reached in \cite{Wolf:2025jed,Garcia-Garcia:2026nzy}. However, we identify here that the origin of this enhancement does not lie in a substantially larger amplitude of the matter fluctuations, as the $\sigma_{12}$ curve is very close to those obtained for the other models (see the values in Table \ref{tab:results}), being strongly constrained by the CMB. Rather, it is driven by an increase in the growth rate at low redshifts, which controls the rate of structure formation in the Universe (cf. the leftmost and central plots in Fig. \ref{fig:structuregrowth}). Since the amount of clustering in the past is not significantly enhanced compared to $\Lambda$CDM, the weak-lensing observable $S_8=\sigma_8(z=0)(\Omega_m^0/0.3)^{0.5}$ remains only slightly higher than in the standard model. It is compatible with the measurement from the Kilo Degree Survey (KiDS), $S_8 = 0.815^{+0.016}_{-0.021}$ \cite{Wright:2025xka}, although it is still $\sim 1.8\sigma$ higher than the value measured by DES \cite{DES:2021wwk}, $S_8=0.775^{+0.026}_{-0.024}$, and $\sim 1.5\sigma$ higher than that measured by the Hyper Suprime-Cam \cite{Miyatake:2023njf}, $S_8=0.763^{+0.040}_{-0.036}$. These differences, though, are not statistically significant.

In any case, the fitting performance found for this model is  consistent with the shape of $w_{\rm DE}(z)$ displayed in the rightmost panel of Fig. \ref{fig:bands}. The model is able to reproduce the crossing of the phantom divide, which happens to be in the ballpark of $z=0.5$, as well as the large values of the effective DE EoS parameter in the quintessence region at low redshifts\footnote{The mean value of $|\bar{\alpha}|$ is close to unity, while the mean value of $\bar{\beta}$ is somewhat larger, with $\bar{\beta}\sim 4$ being preferred (cf. Table \ref{tab:results}). This is expected to yield the desired level of quintessence-like behavior, as can be seen from Eq. \eqref{eq:w_LU_neg}.}, as preferred by the CPL parametrization and model-agnostic reconstructions (see, e.g., \cite{Gonzalez-Fuentes:2026rgu}). However, it is somewhat unable to produce sufficiently phantom behavior before the crossing, most probably because we consider a vanishing mass term for the scalar field. This limits the ability of the model to achieve an even stronger exclusion of $\Lambda\mathrm{CDM}$ \cite{Garcia-Garcia:2026nzy}.

We emphasize that our data set differs from that employed in \cite{Garcia-Garcia:2026nzy} in three main respects: (i) we employ different Planck CMB likelihoods; (ii) we adopt a more conservative approach and do not include the ACT DR6 CMB lensing data; and (iii) we consider a massless scalar field, whereas these authors include a mass term in the scalar-field potential. These differences explain why, after properly penalizing the use of additional parameters, we find a somewhat weaker statistical improvement over the standard model than that reported in \cite{Garcia-Garcia:2026nzy}. We find that $\Lambda\mathrm{CDM}$ can be excluded at $2.50\sigma$ CL, with $\Delta\mathrm{AIC}=4.79$, indicating positive evidence in favor of the dynamical dark energy model under consideration. In contrast, the authors of \cite{Garcia-Garcia:2026nzy} found a Bayes ratio that lies slightly within the strong-evidence regime.

We now turn to the results obtained with the $\alpha>0$+LC model, for which a crossing of the phantom divide is also possible and the local constraints can be satisfied without invoking any screening mechanism, at the expense of a considerable degree of fine-tuning of the initial conditions deep in the RDE (see also \cite{Adam:2026ajg}). In this case, $\chi^2_{\rm min}$ decreases by $\sim 7$ units with respect to $\Lambda$CDM, resulting in a milder exclusion of the standard model at the $2.14\sigma$ CL and in a consistently smaller difference in the Akaike information criterion, $\Delta{\rm AIC}=2.85$. This estimate assumes that the model has two additional parameters compared to $\Lambda$CDM. In a more conservative analysis, in which the model is considered to have three additional parameters (see the details in Sec. \ref{sec:method}), $\Delta{\rm AIC}$ decreases to $0.85$, while $E_{\Lambda{\rm CDM}}=1.77\sigma$. The evidence in favor of the $\alpha>0$+LC model can therefore be considered weak, or at most mildly positive in the light of the Occam's razor. We can understand why by looking at the shape of $w_{\rm DE}(z)$ in Fig. \ref{fig:bands}. The model is able to produce a much more pronounced phantom evolution prior to the crossing, with $w_{\rm DE}$ reaching values as low as $-1.2$ or even $-1.3$ in some cases\footnote{The existence of this phantom phase is necessary to reproduce the distance to the LSS measured by Planck in models with standard pre-recombination physics, such as the one studied in this work -- note that the values of the comoving sound horizon at the baryon-drag epoch, $r_{\rm d}$, are fully compatible with those obtained with $\Lambda$CDM, regardless of the sign of $\alpha$ (cf. Table \ref{tab:results}). Without it, this distance would be altered by the excess of DE at low redshift relative to $\Lambda$CDM. As explained in Sec. \ref{sec:background}, in the case of the model with $\alpha>0$+LC, this phase is possible thanks to the existence of a peak in $\bar{\varphi}(z)$. For illustrative purposes, we show its reconstructed shape in Fig. \ref{fig:bands_pos_phi}.}. However, the quintessence phase is considerably more modest than required by the data. In other words, the $\alpha>0$+LC model provides a better description of the phantom phase, but it is accompanied by a poorer description of the quintessence phase. Since the latter occurs over the redshift range where the data are more abundant, this ultimately leads to a less significant improvement in the overall fit compared to the model with $\alpha<0$. 

\begin{figure}[t!]
    \centering
    \includegraphics[width=0.5\linewidth]{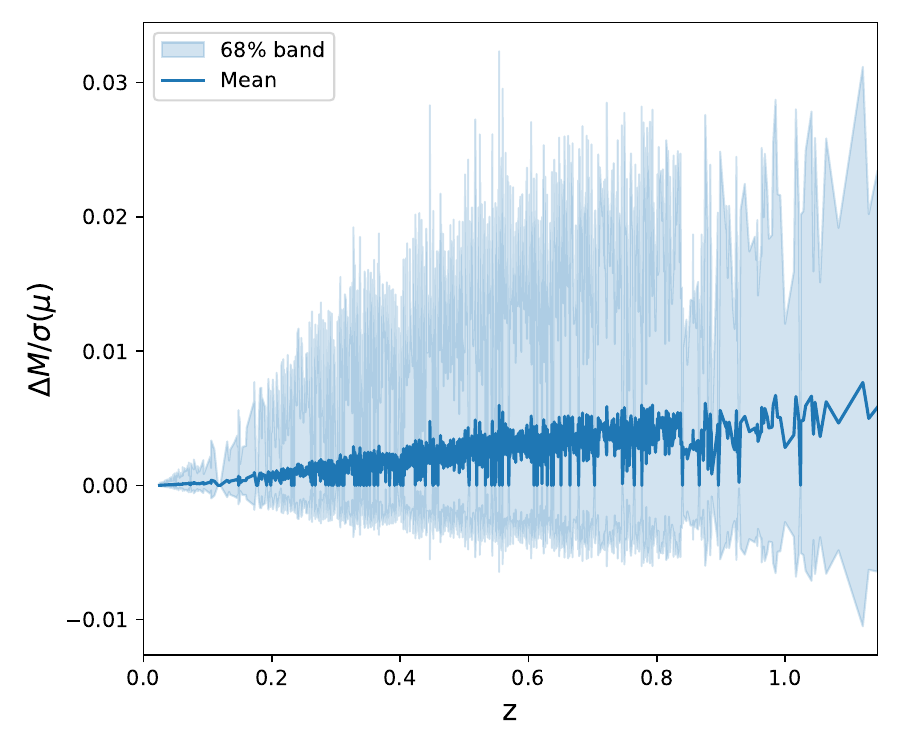}
    \caption{Change in the absolute magnitude induced by $G_{\rm eff}(z)/G_N$ from Eq. \eqref{eq:Mcorr} with respect to the uncertainties from the DES-Dovekie sample. We compute the reconstructed $\Delta M\equiv M(G_{\rm eff})-M(G_N)$ from Eq. \eqref{eq:Mcorr} and evaluate the ratio $\Delta M/\sigma(\mu)$ at the SNIa redshifts, where $\sigma^2(\mu) = C_{ii}$ from the covariance matrix. For ease of visualization, we  plot this function using a linear interpolation between points.}
    \label{fig:DeltaM}
\end{figure}

In the $\alpha>0+$LC scenario, we find $G_{\rm eff}/G_N$ and $\dot{G}_{\rm eff}/G_{\rm eff}$ curves with both concave and convex shapes, which make the uncertainty bands span regions above and below the GR values (see the lower panels of Fig. \ref{fig:bands}). The reason behind this dual behavior is the posterior mean value of $\bar{\alpha}\sim 0.5$ (cf. Table \ref{tab:results}). As discussed in Sec. \ref{sec:Geff}, this is precisely the frontier value that, in the expansion of Eq. \eqref{eq:GeffTaylor}, determines whether $G_{\rm eff}$ is greater or smaller than $G_N$, and also controls the sign of its time derivative. Interestingly, the data require departures from GR to remain small even at high redshifts. Indeed, $G_{\rm eff}$ deviates from $G_N$ by no more than $\sim 0.3\%$, with an extremum occurring around $z\sim 1$. For the best-fit model, we find only a mild increase of $f\sigma_{12}(z)$ compared to the $\Lambda$CDM, which is partially induced by the larger fraction of non-relativistic matter in the universe found within this model. However, this enhancement is much smaller than the one found with $\alpha<0$. Regarding the weak-lensing observable $S_8$, both models yield similar predictions.

Since this model does not assume the existence of a screening mechanism at local scales, it is natural to ask whether these small deviations of $G_{\rm eff}$ from $G_N$ can affect the SNIa likelihood by shifting the effective absolute magnitude of SNIa (see Eq. \eqref{eq:Mcorr} and the related discussion in Sec. \ref{sec:method}). To check whether this is indeed the case, we plot in Fig. \ref{fig:DeltaM} the correction to the absolute magnitude of the SNIa in the DES-Dovekie sample induced by modified-gravity effects, normalized by the corresponding uncertainties in their apparent magnitudes. This function grows with redshift, while its mean remains below $0.5\%$ at all redshifts, with deviations reaching at most $\sim 3\%$ at the $2\sigma$ CL. Hence, for the small deviations of $G_{\rm eff}$ from $G_N$ allowed by the data in the context of the $\alpha>0$+LC model, these corrections are tiny and can be safely neglected to an excellent approximation. We note, however, that these effects could play an important role in other modified-gravity models that predict larger deviations at local scales. A self-consistent analysis of such models using SNIa data must therefore account for these effects.

It is also natural to compare the fitting performance of the two models with non-minimal coupling with that offered by the CPL parametrization. The latter provides the best fit to the data and is the only dynamical DE model studied in this paper that is strongly favored over $\Lambda$CDM according to the AIC, with $\Delta\mathrm{AIC}=7.08$. Therefore, despite the use of the recalibrated DES-Dovekie sample, the signal for dynamical dark energy in the CPL framework remains strong according to Jeffreys' scale, and the standard model is excluded at $\sim 2.9\sigma$ CL. Our results for this model are in agreement with \cite{DES:2025sig,Garcia-Garcia:2026nzy}\footnote{See also \cite{Camilleri:2026rer} for recent results obtained with the Supernovae Unite compilation, which combines in a consistent way the SNIa from the Pantheon+ and DES-SN5YR samples. These data are not yet publicly available.}.  

It is important to bear in mind that CPL is a phenomenological parametrization of the DE EoS parameter, while the other two are derived from an action and offer a microphysical description of DE in terms of a non-minimally coupled scalar field. However, it is also worth noting that the good performance of the model with $\alpha<0$ relies on a screening mechanism, whose existence and plausibility remain to be established. By contrast, the model with $\alpha>0$+LC, despite not requiring any screening, leads to a poorer fit and is subject to a high degree of fine tuning. The initial value of the scalar field must be finely tuned to sufficiently small values to prevent values of $\bar{\alpha}$ of order unity from driving the field to exceedingly large values and thereby inducing an excessively large departure from GR (see the left plot of Fig. \ref{fig:ratios}, and Fig. \ref{fig:phi_ini_color}). 

We also note that the various late-time dynamical DE models studied in this work yield values of the Hubble constant that are in strong tension with the local distance-ladder measurement from the SH0ES Team \cite{Riess:2021jrx}. This is not unexpected, since it is by now well established that late-time DE dynamics alone cannot provide a viable resolution to the Hubble tension (see, e.g., \cite{Sola:2017znb,Knox:2019rjx,Krishnan:2021dyb,Lee:2022cyh,Keeley:2022ojz,Gomez-Valent:2023uof,Gonzalez-Fuentes:2025lei,Gonzalez-Fuentes:2026rgu,Pedrotti:2025ccw,Bansal:2026axl,Sabogal:2026ipu}), at least when anisotropic BAO data are included in the fitting analysis\footnote{Angular (or 2D) BAO are in strong tension with anisotropic BAO data, as quantified in \cite{Favale:2024sdq,Pantos:2026rpe,Sapone:2026try}. Data on 2D BAO still leave room for a late-time resolution of the Hubble tension. See, e.g., \cite{Bernui:2023byc,Akarsu:2023mfb,Gomez-Valent:2023uof,Gomez-Valent:2024tdb,Dwivedi:2024okk,Gomez-Valent:2024ejh}.}. In fact, model-agnostic studies incorporating BAO and SNIa data, together with the SH0ES calibration and a prior on the CMB acoustic scale, indicate that resolving the Hubble tension while providing a good fit to the low-$z$ data requires not only new physics before recombination, but also a very large value of $\omega_m=\Omega_m^0h^2$ (i.e., an anomalously high matter energy density at low redshifts), larger than those typically obtained in early DE models. These analyses further indicate that, in the light of the Hubble tension, crossing the phantom divide in the late Universe is not necessary, although there may still be some non-null preference for quintessence \cite{Poulin:2024ken,Gonzalez-Fuentes:2026rgu,Schoneberg:2026buf}. Therefore, the inclusion of the Hubble tension into the discussion can have a very important impact on the results. We state this explicitly to emphasize that, if the Hubble tension persists in the future, our results, as well as those of previous analyses, e.g., those in \cite{Ye:2024ywg,Wolf:2025jed}, will need to be revisited.


\section{Conclusions}\label{sec:conclusions}
The increasing amount and constraining power of cosmological data has set the stage for exploring the phenomenological properties of dark energy, with the hope of achieving a theoretical understanding of the dark sector. One of the most basic and fundamental questions regarding the nature of dark energy is whether its associated energy density evolves with the cosmic expansion, in contrast to the immutable energy density of the cosmological constant that underlies the standard model of cosmology. Hints of dark energy dynamics have emerged intermittently over the last decade, after the advent of Planck and well before DESI (see, e.g., \cite{Sahni:2014ooa,Salvatelli:2014zta,Sola:2015wwa,Sola:2016jky,SolaPeracaula:2016qlq,Zhao:2017cud,SolaPeracaula:2017esw,Sola:2017znb,SolaPeracaula:2018wwm,Park:2024vrw,Gomez-Valent:2024ejh}), often sparked by the numerous anomalies and tensions between the data and $\Lambda\mathrm{CDM}$ \cite{Perivolaropoulos:2021jda,CosmoVerseNetwork:2025alb}. Recently, the BAO observations from DESI, together with data from several SNIa teams and CMB collaborations, have provided new insights into the dark energy puzzle, pointing to the possibility that the effective dark energy density crossed the phantom divide, with a transition from phantom- to quintessence-like behavior occurring at $z\sim 0.5$ \cite{DESI:2025zgx}.

Scalar fields non-minimally coupled to gravity stand as compelling candidates arising naturally from a scalar-tensor approach to modified gravity. In this work, we have studied the case of a non-minimal coupling of the form $F(\varphi)=1+ \alpha\varphi^2$ and a linear potential in the action \eqref{eq:action}. We have presented a detailed study of the analytical solutions to the background equations in the $|\alpha|\varphi^2\ll1$ limit, closing a gap that has been overlooked in the literature, to the best of our knowledge. In this manner, we have also justified our choice of initial conditions. Regardless of the sign of $\alpha$, these models manage to reproduce a phantom crossing close to that required to fully alleviate the BAO-CMB distance tension within $\Lambda$CDM. We have emphasized that in the extensively studied $\alpha<0$ case, this is achieved at the expense of violating stringent local constraints or invoking a screening mechanism that is absent from the action formulation of the model. Even if such a screening mechanism existed, the model would produce large-scale structure at a higher rate than is seemingly preferred by current data from redshift-space distortions and peculiar velocities. Future galaxy survey data below redshift $z\sim 1$ will be decisive in further scrutinizing the viability of this model. Of particular interest will be the analysis of this model using upcoming data releases from DESI \cite{DESI:2023dwi} and Euclid \cite{Amendola:2016saw}. At present, in light of the CMB+SNIa+BAO data set employed in this work, we find positive evidence in its favor, with $\Lambda$CDM excluded at the 2.50$\sigma$ CL.

In contrast, our results with $\alpha>0$ show that a scalar-tensor theory compatible with both cosmological and Solar System observations is physically viable without requiring a screening mechanism. As a result, we manage to keep the action simpler -- i.e., we are not forced to add higher-order derivatives or additional couplings that have to be carefully chosen to suppress the fifth-force effects at local scales. This comes at the price of fine-tuning the initial value of the scalar field, so there is a trade-off between the two approaches. Nonetheless, one may argue that for each combination of parameters we simply choose the only trajectory that is compatible with observations. Of course, a more satisfactory answer would involve constructing a physically-motivated mechanism to justify the scalar field initial conditions and the corresponding values of the model parameters, $\alpha$ and $\beta$. This is beyond the scope of this paper, although we have argued that small values of the scalar field deep in the radiation-dominated epoch could arise from a spontaneous symmetry-breaking mechanism in which the effective potential has a minimum at the origin prior to symmetry breaking. The model with $\alpha>0$+LC is only weakly preferred over the standard model, at $\sim 2\sigma$ CL.

In view of the results obtained in this paper, we deem it prudent to temper our enthusiasm regarding the performance of these models, not only because of the various issues affecting them and the moderate statistical evidence in their favor, but also because the inclusion of the Hubble tension in the analysis could prove to be a game-changer.

Finally, let us briefly discuss some possible directions for future research. This work can be extended and improved in several ways. For instance, modified-gravity effects could be incorporated into the computation of non-linear corrections to the matter power spectrum, with a possible impact on the CMB lensing, while large-scale structure data could be included to obtain tighter constraints. Concrete screening mechanisms could also be investigated, including for the case $\alpha>0$, where they could help alleviate the fine-tuning. Additionally, mapping the non-minimally coupled model onto an interacting DM-DE scenario could be of interest, as screening would not be necessary in this case either. Another interesting avenue would be to study the Hubble tension in the context of these models, as well as their interplay with new physics prior to recombination. We leave some of these investigations for future work.

%
\acknowledgments JAP is supported by grant FPU25/00603 and AGF by grant FPU24/01241, both from the Spanish Ministry of Science, Innovation and Universities (MICIU). AGV is funded by a Ramón y Cajal contract from MICIU with Ref. RYC2024-049138-I, and is also supported by projects PID2022-136224NB-C21 (MICIU), 2021-SGR-00249 (Generalitat de Catalunya) and CEX2024\discretionary{}{}{}-001451-M (ICCUB). He also acknowledges the funding from “la Caixa” Foundation (ID 100010434) and the European Union’s Horizon 2020 research and innovation programme under the Marie Skłodowska-Curie grant agreement No. 847648, with fellowship code LCF/BQ/\discretionary{}{}{}PI23/11970027. AGF and AGV acknowledge the participation in the COST Action CA21136 “Addressing observational tensions in cosmology with systematics and fundamental physics” (CosmoVerse).


\appendix

\section{Geometrical quantities to first order in perturbations}\label{sec:geom}

The Christoffel symbols, Ricci tensor, Ricci scalar and Einstein tensor to first order in the perturbed metric $h_{\mu\nu}$ \eqref{eq:pert} read, respectively:
                               
\begin{equation}\label{eq:Christoffel1}
\Gamma^\alpha_{\beta\kappa}(h) = \frac{\eta^{\alpha\mu}}{2}\left(h_{\mu\beta,\kappa}+h_{\mu\kappa,\beta}-h_{\beta\kappa,\mu}\right)\,,
\end{equation}

\begin{equation}
R_{\mu\nu}(h)=\frac{1}{2}\left[\partial^\alpha\partial_\nu h_{\alpha\mu}+\partial^\alpha\partial_\mu h_{\alpha\nu}-\partial_{\mu}\partial_\nu h-\eta^{\alpha\beta}\partial_\alpha\partial_\beta h_{\mu\nu}\right]\,,
\end{equation}

\begin{equation}
R(h)=\partial^\alpha\partial^\mu h_{\alpha\mu}-\eta^{\alpha\mu}\partial_\alpha\partial_\mu h\,,
\end{equation}

\begin{equation}\label{eq:ET1}
G_{\mu\nu}(h)=\frac{1}{2}\left[\partial^\alpha\partial_\nu h_{\alpha\mu}+\partial^\alpha\partial_\mu h_{\alpha\nu}-\partial_{\mu}\partial_\nu h-\eta^{\alpha\beta}\partial_\alpha\partial_\beta h_{\mu\nu}+\eta_{\mu\nu}(\eta^{\alpha\beta}\partial_\alpha\partial_\beta h-\partial^\alpha\partial^\beta h_{\alpha\beta})\right]\,,
\end{equation}
with $h\equiv\eta^{\mu\nu}h_{\mu\nu}$.


\section{Approximate solutions for the scalar field during radiation and matter domination}\label{sec:background_sol}

In this appendix, we provide all the details of the calculation of the approximate analytical solutions for the scalar field during the RDE and MDE discussed in Sec. \ref{sec:background}, in the limit in which $|\bar{\alpha}|\bar{\varphi}^2\ll 1$, which ensures that the non-minimal coupling  $F$ is close to unity. 
\newline\newline
\noindent {\bf RDE}
\newline
\newline
In that limit, we find that during the RDE the scale factor and Hubble function read 

\begin{equation}
a(t) = C\sqrt{t}\quad {\rm  and}\quad H(t)=\frac{1}{2t}
\end{equation}
in very good approximation, with $C\equiv\left(\frac{4}{3}\kappa^2\rho_r^0\right)^{1/4}$. Substituting these functions in the KG equation \eqref{eq:KG_FLRW} and neglecting the effect of the scalar-field potential term we find, 

\begin{equation}\label{eq:RDE_eq}
\ddot{\varphi}+\frac{3}{2t}\varphi-\frac{A\varphi}{t^{3/2}}=0\,, \quad {\rm with}\quad A\equiv\alpha\rho_m^0C^{-3}\,.
\end{equation}
We perform now the change of variables $\{t,\varphi\}\to\{x,y\}$, 

\begin{equation}
 t=x^r\qquad ;\qquad \varphi = x^py\,,
\end{equation}
where $r$ and $p$ are real numbers. Under this change, Eq. \eqref{eq:RDE_eq} takes the following form, 

\begin{equation}
x^2\frac{d^2y}{dx^2}+x\frac{dy}{dx}\left[1+2p+\frac{r}{2}\right]+y\left[p^2+\frac{pr}{2}-Ar^2x^{r/2}\right]=0\,,
\end{equation}
which is not that far from a Bessel-type differential equation. Indeed, if we set $r=4$ and $p=-1$, it further simplifies, 

\begin{equation}
x^2\frac{d^2y}{dx^2}+x\frac{dy}{dx}-\left[1+16Ax^{2}\right]y=0\,,
\end{equation}
and now it is clear that we just need to define $z\equiv 4\sqrt{A}\, x$ if $\alpha>0$ ($A>0$) or $z\equiv 4\sqrt{|A|}\, x$ if $\alpha<0$ ($A<0$) to finally obtain, 

\begin{equation}
z^2\frac{d^2y}{dz^2}+z\frac{dy}{dz}-\left[z^2+1\right]y=0\quad{\rm for}\,\,\alpha>0\,,
\end{equation}

\begin{equation}
z^2\frac{d^2y}{dz^2}+z\frac{dy}{dz}+\left[z^2-1\right]y=0\quad{\rm for}\,\,\alpha<0\,,
\end{equation}
These equations are the modified and standard Bessel differential equations, respectively. The solution to the former is expressed in terms of the modified Bessel functions of the first and second kind of order one, $I_1(z)$ and $K_1(z)$, whereas the solution to the latter is expressed in terms of the Bessel functions of the first and second kind of order one, $J_1(z)$ and $Y_1(z)$. Therefore, by undoing the changes of variables and expressing the result in terms of the scale factor, we have

\begin{equation}\label{eq:sol_pos}
    \varphi(a)\Big|_{\rm RDE}(a) = \frac{1}{\sqrt{a}}\left[c_0I_1\left(2\sqrt{a}\sqrt{\frac{3\bar{\alpha}\Omega_m^0}{\Omega_r^0}}\right)+c_1K_1\left(2\sqrt{a}\sqrt{\frac{3\bar{\alpha}\Omega_m^0}{\Omega_r^0}}\right)\right]\quad{\rm for}\,\,\alpha>0\,,
\end{equation}

\begin{equation}\label{eq:sol_neg}
\varphi(a)\Big|_{\rm RDE}(a) = \frac{1}{\sqrt{a}}\left[\tilde{c}_0J_1\left(2\sqrt{a}\sqrt{\frac{3\bar{\alpha}\Omega_m^0}{\Omega_r^0}}\right)+\tilde{c}_1Y_1\left(2\sqrt{a}\sqrt{\frac{3\bar{\alpha}\Omega_m^0}{\Omega_r^0}}\right)\right]\quad{\rm for}\,\,\alpha<0\,,
\end{equation}
where $\{c_0,c_1,\tilde{c}_0,\tilde{c}_1\}$ are integration constants to be fixed by the initial conditions. Here, we recall the form of the Bessel functions, 

\begin{equation}
I_1(z)=\frac{z}{2}+\frac{z^3}{2^2\cdot 4}+\frac{z^5}{2^2\cdot 4^2\cdot 6}+\frac{z^7}{2^2\cdot 4^2\cdot 6^2\cdot 8}+...\,,
\end{equation}
\begin{equation}
    K_1(z)=I_1(z)\int\frac{dz}{zI_1^2(z)}\,,
\end{equation}
and 
\begin{equation}
    J_1(x)=\frac{x}{2}-\frac{x^3}{2^2\cdot 4}+\frac{x^5}{2^2\cdot 4^2\cdot 6}-\frac{x^7}{2^2\cdot 4^2\cdot 6^2\cdot 8}+...\,,
\end{equation}
\begin{equation}
Y_1(z)=J_1(z)\int\frac{dz}{zJ_1^2(z)}\,.
\end{equation}
The terms proportional to $K_1(z)$ and $Y_1(z)$ in Eqs. \eqref{eq:sol_pos} and \eqref{eq:sol_neg} are associated to decaying modes and can be neglected, since we assume that the initial conditions are set when they are already very small. Hence, we find, 

\begin{equation}
\varphi(a)\Big|_{\rm RDE} = \frac{\varphi_{\rm ini}}{\sqrt{a}}\sqrt{\frac{\Omega_r^0}{3\bar{\alpha}\,\Omega_m^0}} \,\,I_1\left(2\sqrt{a}\sqrt{\frac{3\bar{\alpha}\Omega_m^0}{\Omega_r^0}}\right)\qquad {\rm for}\,\alpha>0\,,
\end{equation}
\begin{equation}
\varphi(a)\Big|_{\rm RDE} = \frac{\varphi_{\rm ini}}{\sqrt{a}}\sqrt{\frac{\Omega_r^0}{3|\bar{\alpha}|\,\Omega_m^0}} \,\,J_1\left(2\sqrt{a}\sqrt{\frac{3|\bar{\alpha}|\Omega_m^0}{\Omega_r^0}}\right)\qquad {\rm for}\,\alpha<0\,.
\end{equation}
\noindent
{\bf MDE}
\newline
\newline
\noindent In the MDE, instead, if $|\bar{\alpha}|\varphi^2\ll1$ the background evolution is governed by the following scale factor and Hubble function, 

\begin{equation}
a(t)=Dt^{2/3}\qquad ;\qquad H(t)=\frac{2}{3t} \,,
\end{equation}
with $D\equiv \left(\frac{3}{4}\kappa^2\rho_m^0\right)^{1/3}$. Again, we can replace these functions in Eq. \eqref{eq:KG_FLRW} and obtain,

\begin{equation}\label{eq:KGmatter}
    \ddot{\varphi}+\frac{2}{t}\dot{\varphi}-\frac{4\bar{\alpha}}{3t^2}\varphi=0\,,
\end{equation}
whose solution for both positive and negative $\alpha$ reads, 

\begin{equation}
\varphi(t)\Big|_{\rm MDE}=\frac{1}{\sqrt{t}}\left[b_0t^{\frac{1}{2}\sqrt{1+\frac{16}{3}\bar{\alpha}}}+b_1t^{-\frac{1}{2}\sqrt{1+\frac{16}{3}\bar{\alpha}}}\right]\,,
\end{equation}
or, alternatively, in terms of the scale factor, 

\begin{equation}\label{eq:sol_MDE}
\varphi(a)\Big|_{\rm MDE}=a^{-3/4}\left[\tilde{b}_0a^{\frac{3}{4}\sqrt{1+\frac{16}{3}\bar{\alpha}}}+\tilde{b}_1a^{-\frac{3}{4}\sqrt{1+\frac{16}{3}\bar{\alpha}}}\right]\,,
\end{equation}
with $\{b_0,b_1,\tilde{b}_0,\tilde{b}_1\}$ integration constants. For small values of $\bar{\alpha}$, the second term in the solution is a fast decaying mode, which can be again neglected for large enough values of the scale factor, and by Taylor-expanding $\varphi(a)$ we find that it evolves logarithmically with $a$, i.e., 

\begin{equation}\label{eq:log}
\varphi(a)\Big|_{\rm MDE}=b_0\left[1+2\bar{\alpha}\ln (a)+\mathcal{O}(\bar{\alpha}^2)\right]\quad {\rm for}\,\,|\bar{\alpha}|\ll1\,.
\end{equation}
For general values of $\bar{\alpha}>0$,

\begin{equation}\label{eq:solposalpha}
    \varphi(a)\Big|_{\rm MDE}=\varphi(a_*)\left(\frac{a}{a_*}\right)^{\frac{3}{4}\left[-1+\sqrt{1+\frac{16}{3}\bar{\alpha}}\right]}\quad {\rm for}\,\,\bar{\alpha}>0\,,
\end{equation}
 since the second term in Eq. \eqref{eq:sol_MDE} always behaves as a decaying mode. Here, $a_*$ is a scale factor deep in the MDE.
 
 For $\bar{\alpha}<0$, instead, it is important to keep that term in Eq. \eqref{eq:sol_MDE}. The solution will take the following form, depending on whether $\bar{\alpha}$ is larger or smaller than $-3/16$,

\begin{equation}\label{eq:solnegalpha1}
  \varphi(a)\Big|_{\rm MDE}=a^{-3/4}\left[d_0\cosh\left(\frac{3\xi}{4}\,\ln(a)\right)+d_1\sinh\left(\frac{3\xi}{4}\,\ln(a)\right)\right] \quad {\rm if}\,\,-\frac{3}{16}<\bar{\alpha}<0\,,  
\end{equation}

\begin{equation}\label{eq:solnegalpha2}
  \varphi(a)\Big|_{\rm MDE}=a^{-3/4}\left[d_0\cos\left(\frac{3\xi}{4}\,\ln(a)\right)+d_1\sin\left(\frac{3\xi}{4}\,\ln(a)\right)\right] \quad {\rm if}\,\,\bar{\alpha}<-\frac{3}{16}\,, 
\end{equation}
with $\xi(\bar{\alpha})\equiv\sqrt{|1+\frac{16}{3}\bar{\alpha}|}$. Here, we have used the identity $a^x = e^{x\ln(a)}$, together with the fact that exponential functions with real exponents (found when $\bar{\alpha}>-3/16$) can be written in terms of hyperbolic functions, whereas those with imaginary exponents (found when $\bar{\alpha}<-3/16$) can be written in terms of ordinary trigonometric functions.

 Including the effect of the linear potential \eqref{eq:main_functions} on the scalar field in the last stages of the MDE is quite simple. One just needs to consider the particular solution to equation 

\begin{equation}
    \ddot{\varphi}+\frac{2}{t}\dot{\varphi}-\frac{4\bar{\alpha}}{3t^2}\varphi=-\beta\,,
\end{equation}
on top of the solutions to the homogeneous Eq. \eqref{eq:KGmatter}, in  \eqref{eq:solposalpha} and \eqref{eq:solnegalpha1}-\eqref{eq:solnegalpha2}. For instance, for the positive-$\bar{\alpha}$ case, we have

\begin{equation}\label{eq:pos_late_mde}
\bar{\varphi}(a)=\bar{\varphi}(a_*)\left(\frac{a}{a_*}\right)^{\frac{3}{4}\left[-1+\sqrt{1+\frac{16}{3}\bar{\alpha}}\right]}+\frac{2\bar{\beta}}{3\Omega_m^0(2\bar{\alpha}-9)}\left[a^3-a_*^3 \left(\frac{a}{a_*}\right)^{\frac{3}{4}\left[-1+\sqrt{1+\frac{16}{3}\bar{\alpha}}\right]}\right]\,.
\end{equation}
If $\bar{\alpha},\bar{\beta}>0$, the effective potential has a maximum at $\bar{\varphi}_{\rm max}(a)=\bar{\beta}a^3/(3\Omega_m^0\bar{\alpha})$. The scalar field can be slowed down (if $\bar{\varphi}<\bar{\varphi}_{\rm max}$ at some point) and eventually reach a maximum only if $\bar{\alpha}<9/2$, a limiting value that is, in any case, well above those preferred by current data. This is necessary for the field to evolve towards the origin, which is required to satisfy the local constraints discussed in Sec.~\ref{sec:Geff}. Reaching that maximum is also crucial for the model to explain the phantom behavior before crossing the phantom divide (see Sec. \ref{sec:background}). When these conditions are not met, the potential cannot counteract the force exerted by the non-minimal coupling during matter domination. As a result, the field continues to grow until the assumption of small field excursions breaks down, thereby invalidating the linear approximation to the potential and contravening the local constraints.

If $\bar{\alpha}<0$, instead, the effective potential has a minimum at $\bar{\varphi}_{\rm min}(a)=-\bar{\beta}a^3/(3\Omega_m^0|\bar{\alpha}|)$, which is extremely close to the origin deep in the MDE (cf. Fig. \ref{fig:potential}). In fact, for sufficiently large values of $\bar{\alpha}$ the scalar field reaches the minimum of the effective potential even before the onset of matter domination (see Sec. \ref{sec:background} for details). The scalar field sits close to the origin until the linear term of the potential becomes sizable, in the last stages of the MDE. When this happens, the field moves towards negative values, eventually behaving as quintessence.


\bibliographystyle{JHEP}
\bibliography{biblio.bib}

\end{document}